\documentclass{article}

\usepackage{arxiv}

\usepackage[utf8]{inputenc} 
\usepackage[T1]{fontenc}    
\usepackage{hyperref}       
\usepackage{url}            
\usepackage{booktabs}       
\usepackage{amsfonts}       
\usepackage{nicefrac}       
\usepackage{microtype}      
\usepackage{cleveref}       
\usepackage{graphicx}
\usepackage{doi}

\usepackage[mathlines]{lineno}
\usepackage{appendix}
\usepackage{makecell} 

\title{Gross polluters and vehicles' emissions reduction}

\date{}

\author{Matteo B\"{o}hm \\
    Sapienza University of Rome, Italy \\
    \texttt{bohm@diag.uniroma1.it} \\
	\And
	Mirco Nanni \\
    ISTI-CNR, Italy \\
    \texttt{mirco.nanni@isti.cnr.it}
   \And 
   Luca Pappalardo\\
    ISTI-CNR, Italy \\
    \texttt{luca.pappalardo@isti.cnr.it}

}

\begin{document}
\maketitle

\begin{abstract}
	Vehicles' emissions produce a significant share of cities' air pollution, with a substantial impact on the environment and human health.
    Traditional emission estimation methods use remote sensing stations, missing vehicles' full driving cycle, or focus on a few vehicles.
    We use GPS traces and a microscopic model to analyse the emissions of four air pollutants from thousands of private vehicles in three European cities.
    We find that the emissions across the vehicles and roads are well approximated by heavy-tailed distributions and thus discover the existence of gross polluters, vehicles responsible for the greatest quantity of emissions, and grossly polluted roads, which suffer the greatest amount of emissions.
    Our simulations show that emissions reduction policies targeting gross polluters are way more effective than those limiting circulation based on a non-informed choice of vehicles.
    Our study contributes to shaping the discussion on how to measure emissions with digital data.
\end{abstract}


\vfill

\section*{Introduction}
\label{intro}
Estimating air pollutants' distribution over space and time is a significant challenge concerning climate change and human health.
In urban environments, air pollution generated from vehicles' emissions has become more and more evident, to the point that a temporary interruption of regular traffic during the COVID-19's lockdown resulted in a tremendous decrease in CO$_2$ emission~\cite{venter2020, laquere2020, he2020short, liu2020near}.
Even if this brief period's impact on the epochal challenge of climate change is negligible~\cite{forster2020}, it helps outline the impact of the emissions related to transportation on our everyday life.
Greenhouse gas (GHG) emissions from this sector have doubled since 1970, and, in 2016, $11.9\%$ of global GHG emissions were from road transport ($60\%$ of which was from passenger travel)~\cite{ipcc_wg3, ritchie2020sector}.
Moreover, the transport sector emits non-CO$_2$ pollutants such as nitrogen oxides, ozone, particulate matter, and volatile organic compounds, which play a fundamental role in changing climate and are dangerous for human health~\cite{ipcc_wg3}. 
Among the Sustainable Development Goals to be reached by 2030~\cite{assembly2015sustainable}, the United Nations pose an urgent call for action to reduce ``the adverse per capita environmental impact of cities, paying particular attention to air quality''~\cite{assembly2015sustainable}.
In this regard, measuring the vehicles' emissions is primary in designing policies to reduce transportation emissions.

Based on the data at disposal, existing methods to quantify vehicles' emissions range between two extremes. On the one hand, some approaches rely on measurements performed on small samples of vehicles (usually less than 10) but with high spatio-temporal resolutions, such as those coming from particulate sensors~\cite{desouza2020} or portable emissions measurement systems (PEMS)~\cite{chong2020, lujan2018}. These sensors measure emissions in real-world driving conditions, producing accurate estimates but hardly generalisable patterns due to the limited sample size. For example, two studies~\cite{lujan2018, chong2020} analyse emissions from PEMS of one and three light-duty vehicles, finding that the highest emissions are associated with the urban part of the route, flat roads, and low speed.

On the other hand, some studies cover a region's almost entire fleet, such as those using odometer readings from annual safety inspections.
These data describe each vehicle's age, fuel type, engine volume, and mileage, used in macroscopic models to estimate annual emissions.
For example, two studies~\cite{chatterton2015use, diao2014vehicle} use odometer readings to compute mean annual emissions for UK postcode areas and explore the built-environment effects (e.g., work accessibility) on the vehicles' annual miles travelled in Boston.
Unfortunately, odometer readings miss critical information such as instantaneous speed and acceleration~\cite{kancharla2018incorporating, choudhary2016urban, ferreira2015impact, zheng2017influence}, making it challenging to track emissions over time and map them to suburban areas.

GPS traces generated by in-vehicle devices stand as a trade-off between these two extremes. 
Depending on the provider's market penetration, they can cover a representative fraction of the vehicle fleet~\cite{pappalardo2013understanding} and allow computing instantaneous speed and acceleration, which are used within microscopic models to obtain emissions estimates in high spatio-temporal resolution.
GPS traces describe human mobility in great detail~\cite{pappalardo2015returners, gallotti2012towards, luca2020deep, barbosa2018human} and offer an unprecedented tool to implement strategies such as reducing congestion, improving vehicle efficiency, and shifting to lower-carbon options~\cite{ccolak2016understanding, lwin2015estimation, stipancic2019measuring, camargo2020estimating, jenn2020emissions, liang2019air, rolnick2019tackling}.
Given these peculiarities, several studies use GPS traces to analyse the vehicles' emissions at different spatio-temporal scales~\cite{nyhan2016, liu2019}, investigate the relationship between emissions and the urban environment~\cite{reznik2018}, vehicle miles travelled and fuel consumption~\cite{wang2014using}, or trip rates and travel mode choice~\cite{cervero1997travel}. 
Other studies concentrate on congestion-related emissions~\cite{gately2017urban} or braking~\cite{chen2020}, emissions associated with ride-hailing~\cite{sui2019} and bus stops' positioning~\cite{yu2020}, the impact of urban policies~\cite{rahman2017}, methods for emission modelling~\cite{zhu2020, aziz2018}, and air quality monitoring~\cite{desouza2020}.

Despite the depicted variety of literature, it remains unclear what statistical patterns characterise the distribution of emissions per vehicle and road, how these distributions change in time and space, and how we can exploit this information to simulate emission reduction scenarios.
For example, although it is reported that the distribution of emissions from on-road remote sensing sites across the vehicles is skewed~\cite{guenther1994emissions, brand2008taming, huang2018remote}, this finding has been questioned given the inherent limitations of this type of measurement~\cite{huang2020reevaluating}.

This paper analyses the estimated emissions of several air pollutants from thousands of private vehicles moving in different European cities.
We use trajectories produced by onboard GPS devices to compute the vehicles' emissions and match the obtained emissions to the cities' road networks.
We then study how the emissions distribute across vehicles and roads to discover the statistical patterns that characterise emissions and investigate the relationships between emissions, human mobility, and the road network's characteristics.
Finally, we simulate two emission reduction scenarios in which a share of vehicles become zero-emissions or limit their mobility, identifying strategies to drastically reduce emissions over a city while minimising the share of vehicles targeted.

Our framework, which applies to any city provided the availability of vehicles' GPS trajectories and road networks data, may represent practical support for decision-makers to implement strategies to reduce emissions, improve citizens' well-being and design more sustainable cities~\cite{batty2012smart, kitchin2014real, voukelatou2020measuring}.



\section*{Results}
\label{results}

\subsection*{Computation of emissions.}
We use anonymous GPS trajectories describing 423,018 trips from 16,715 private light-duty vehicles moving in Greater London, Rome, and Florence throughout January 2017 (see Table \ref{tab:traj}). 
The spatio-temporal patterns of the vehicles' trajectories are stable across the cities and the seasons of the year (see Supplementary Note 1).

\begin{figure}
    \centering
    \includegraphics[width=\textwidth]{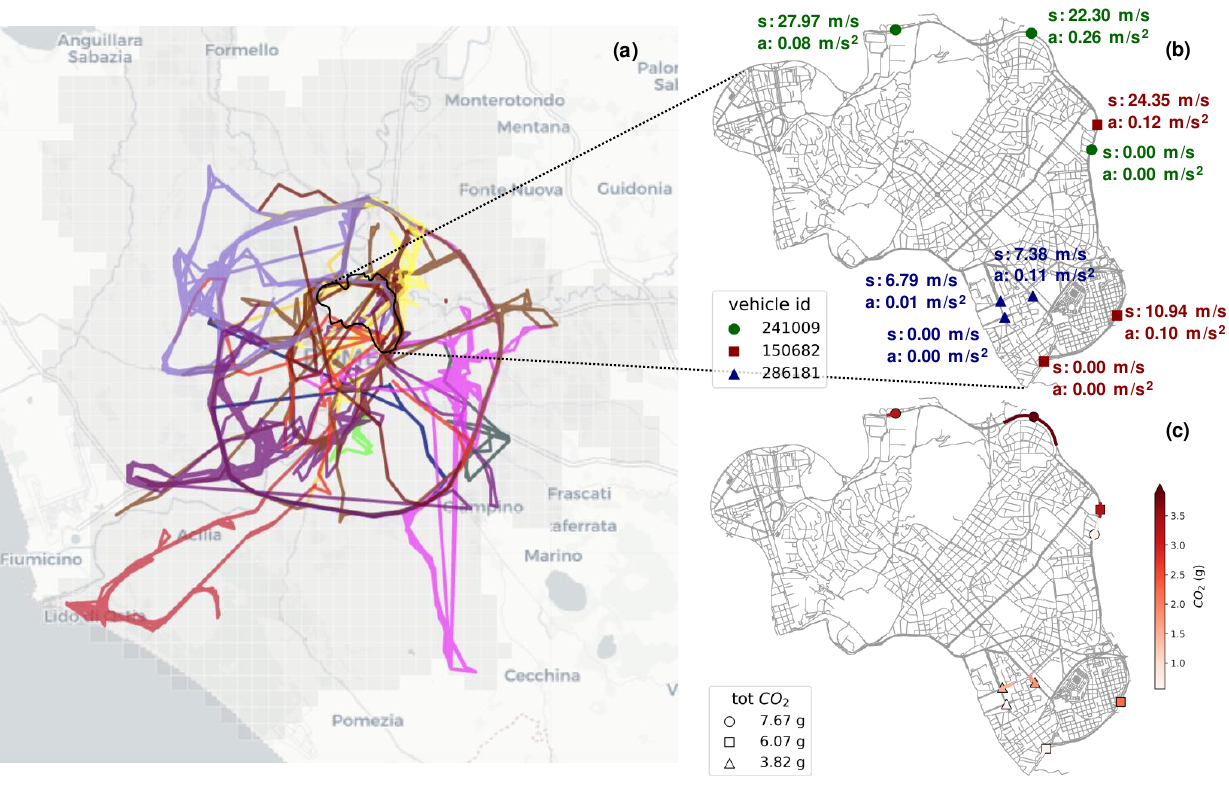}
    \caption{\textbf{Computation of emissions from GPS trajectories.} 
    (a) Visualisation of the trajectories of 20 vehicles travelling during January 2017 in Rome, Italy. 
    Each colour indicates a different trajectory.
    The grey area indicates the territory of the municipality of Rome.
    Plot generated with Python library scikit-mobility~\cite{pappalardo2019scikitmobility}.
    (b) Visualisation of the GPS points of three vehicles passing through a neighbourhood in Rome. 
    Each symbol and colour indicates a different vehicle; we show the corresponding instantaneous speed and acceleration for each point. 
    The point with zero speed and acceleration is the first point of the trajectory sample.
    (c) The instantaneous emissions of CO$_2$ of each GPS point and for each road crossed. 
    The points and the roads are coloured in a gradient from white (low emission) to red (high emission). 
    The legend shows the overall quantity of emissions of each vehicle.
    The road networks in panels (b)-(c) are plotted with the Python library OSMnx~\cite{boeing2017}.}
    \label{fig:Figure1}
\end{figure}

The trajectories are produced by onboard GPS devices, which automatically turn on when the vehicle starts, transmitting a point every minute to the server via a GPRS connection~\cite{pappalardo2013understanding, pappalardo2015returners, gallotti2012towards}.
When the vehicle stops, no points are logged or sent.
The GPS traces are collected by a company that provides a data collection service for insurance companies.
The market penetration of this service is variable, but, in general, it covers at least $2\%$ of the total registered vehicles, and it is representative of the overall amount of vehicles circulating in a city~\cite{pappalardo2013understanding}.
Figure \ref{fig:Figure1}a shows a sample of trajectories for 20 vehicles in Rome.

We define a methodological framework to compute vehicles' emissions from their raw GPS trajectories (see Supplementary Figure \ref{fig:framework}).
We filter the GPS trajectories so that the time between consecutive points is lower than a certain threshold (see Methods and Supplementary Note 2).
For each vehicle, we estimate the instantaneous speed and acceleration in each point of its trajectory and filter out points with unrealistic values (see Methods and Figure \ref{fig:Figure1}b).
We use a Nearest-Neighbour algorithm to assign the points to the city's roads based on road networks downloaded from OpenStreetMap (see Methods).

The three cities are heterogeneous in their road networks: Rome is large but with the sparsest network; London is huge but with the densest network; Florence is small ($\approx1/12$ of Rome and $\approx~1/15$ of London in terms of land area) but with a dense road network (see Supplementary Note 3 for details).

We use a microscopic emissions model~\cite{nyhan2016} that uses speed, acceleration, and fuel type to estimate the vehicles' instantaneous emissions of carbon dioxide (CO$_2$), nitrogen oxides (NO$_x$), particulate matter (PM), and volatile organic compounds (VOC) (see Methods).
Finally, we compute each vehicle's overall emissions as the sum of all its instantaneous emissions during the study period.
Analogously, we compute the overall amount of air pollutants on each road by summing up all the instantaneous emissions from any vehicle passing through that road during the same period (Figure \ref{fig:Figure1}c).

\begin{table}[b]
    \centering
    \begin{tabular}{c | c | c | c | c |}
    \cline{2-5}
         & \multicolumn{4}{c|}{\bf original GPS trajectories} \\
        \cline{2-5}
        
         &  vehicles & trips & points & avg sampling rate (std) \\
         \cline{2-5}
         
        London &  2,745 &  117,930 &  2,978,989 &  72.9 sec (144) \\
        
        Rome &  9,188 &  254,088 &  2,221,206 &  207.7 sec (245.6) \\
        
        Florence &  4,782 &  51,000 &  291,598 &  261.3 sec (400.4) \\
        \cline{2-5}
            \cline{2-5}
         & \multicolumn{4}{c|}{\bf GPS trajectories after filtering} \\
        \cline{2-5}
        
         &  vehicles & trips & points & avg sampling rate (std) \\
         \cline{2-5}
         
        London &  2,721 &  233,627 &  2,936,512 &  58.3 sec (18) \\
        
        Rome &  9,069 &  216,083 &  1,033,487 &  76.8 sec (27.7) \\
        
        Florence &  4,471 &  35,145 &  86,187 &  75.8 sec (31.2) \\
        \cline{2-5}
    \end{tabular}
    \caption{\textbf{Summary statistics of the GPS data.} 
    The number of vehicles, trips, and points, and their mean sampling rate (and standard deviation), for Greater London, Rome, and Florence before and after the filtering step.}
    \label{tab:traj}
\end{table}

\subsection*{Patterns of emissions.}
\label{sec:distribution}
We find that, for all three cities, emissions distribute across vehicles in a heterogeneous way: a few vehicles, that we call gross polluters, are responsible for a tremendous amount of emissions. 
At the same time, most vehicles emit significantly less (Figure \ref{fig:Figure2}).
The distribution of emissions per vehicle is associated with a Gini coefficient higher than 0.55, for all the cities and pollutants (see Supplementary Note 4).
In line with previous studies~\cite{guenther1994emissions, brand2008taming, huang2018remote}, we find that the top 10\% of gross polluters in Florence, Rome, and London are responsible for 47.5\%, 50.5\%, and 38.5\% of the total CO$_2$ emitted during the month, respectively.
The distributions of CO$_2$ emissions per vehicle of Rome and Florence are well approximated by a truncated power law
$p(x) \propto x^{-\alpha} e^{-\lambda x}$, 
with $\alpha = 1.13$, $\lambda=1.04 \times 10^{-3}$ (Rome, Figure \ref{fig:Figure2}e), and $\alpha = 2.12$, $\lambda=1.45 \times 10^{-3}$ (Florence, Figure \ref{fig:Figure2}h). 
Similarly, London's distribution is well approximated by a stretched exponential $p(x) \propto x^{\beta-1} e^{-\lambda x^{\beta}}$, with parameters $\lambda = 5.7 \times 10^{-4}$ and $\beta = 1.26$ (Figure \ref{fig:Figure2}b). 
These results are consistent with those we find for the other three pollutants (NO$_x$, PM, VOC): a truncated power-law well approximates the distribution for Rome and Florence, and a stretched exponential well approximates that for London (see Supplementary Notes 4 and 5 for details).

The picture is similar when considering the distribution of emissions per road: a few grossly polluted roads suffer from a significant quantity of emissions, most of them suffer significantly fewer emissions. 
This distribution is associated with a Gini coefficient higher than 0.64 (see Supplementary Note 4) for all the cities and pollutants, and it is well approximated by a truncated power law, with exponents $\alpha = 1.55$ and $\lambda = 1.08 \times 10^{-4}$ (Rome, Figure \ref{fig:Figure2}f), $\alpha = 1.52$ and $\lambda = 1.30 \times 10^{-4}$ (Florence, Figure \ref{fig:Figure2}i), $\alpha = 2.59$ and $\lambda = 2.88 \times 10^{-4}$ (London, Figure \ref{fig:Figure2}c).
London has both the exponents $\alpha$ and $\lambda$ significantly higher than Rome and Florence, denoting a fairer distribution of the emissions per road (see Supplementary Figure \ref{fig:all_fits_per_road}). 
In Florence and Rome, the top 10\% of grossly polluted roads are associated with more than 90\% of the CO$_2$ emitted during the period. 
In London, this quantity is lower (56.7\%) but still above half of the city's total emissions of CO$_2$.
Again, we find similar results for the other pollutants (see Supplementary Note 4).

The above results hold when changing the year's season (see Supplementary Note 1).
Also, the sample size of the dataset and the choice of the filtering parameter $\theta$ do not affect the significance of our results: the shape of the distributions holds even if we significantly reduce the sample size or change $\theta$ (see Supplementary Note 6).

\begin{figure}
    \centering
    \includegraphics[width=0.85\textwidth]{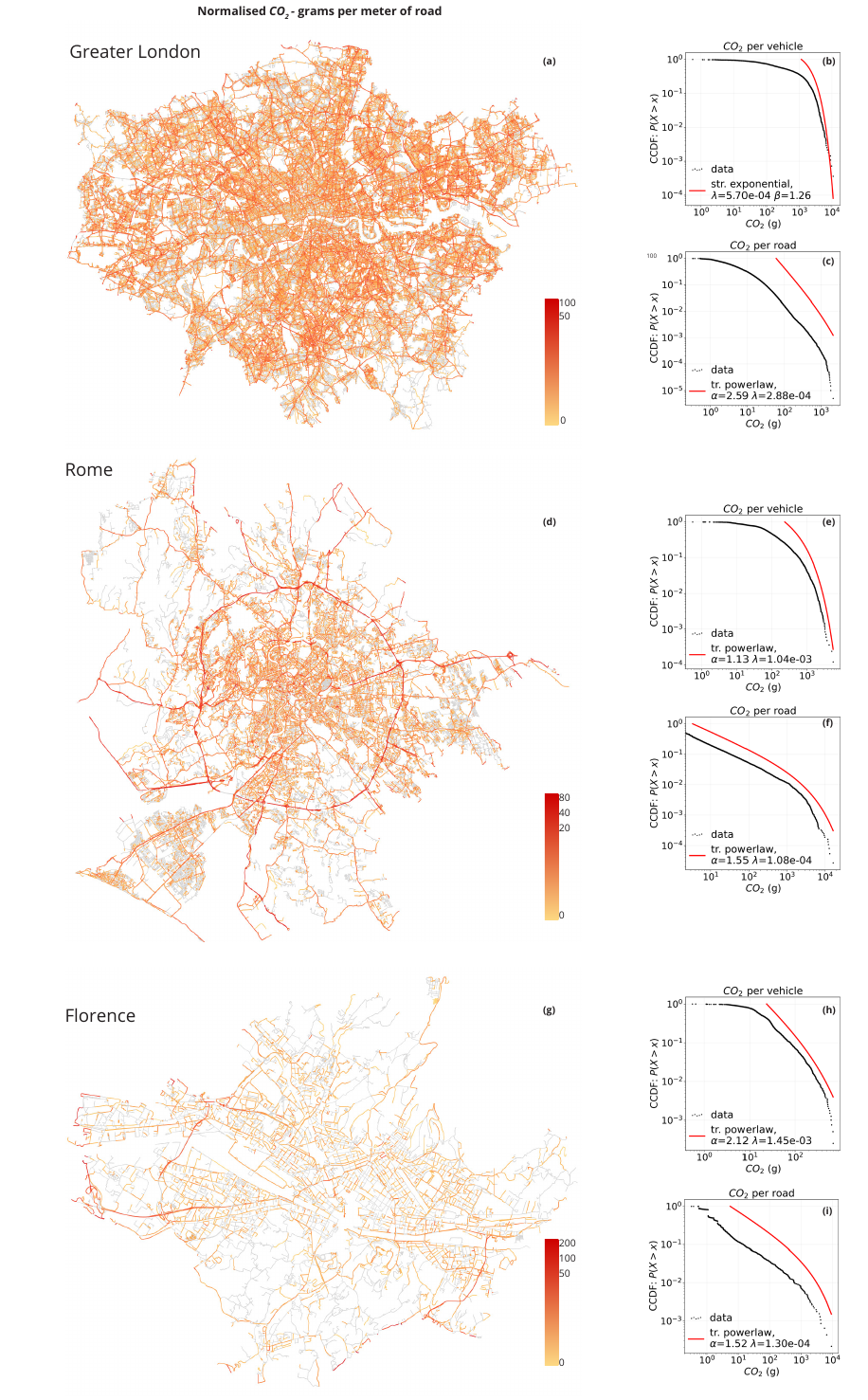}
\end{figure}

\begin{figure}
    \centering
    \caption{\textbf{Distributions of emissions.} 
    The amount of CO$_2$ emissions (expressed as grams per meter of road emitted during January 2017) on the road networks of Greater London (a), Rome (d), and Florence (g).
    Roads are coloured in a gradient from yellow (low emission) to red (high emission). 
    Panels (b), (e), and (h) show in log-log scale the Complementary Cumulative Distribution Function (CCDF, black dots) of the CO$_2$ emissions per vehicle, together with the best fit (red curve) in Greater London, Rome, and Florence, respectively. 
    Panels (c), (f), and (i) show the CCDF (black dots) of the CO$_2$ emissions per road, together with the best fit (red curve), in Greater London, Rome, and Florence, respectively.}
    \label{fig:Figure2}
\end{figure}

\subsection*{Relations with mobility and roads' features.}
\label{sec:correlations}
To investigate the relationship between a vehicle's emissions and mobility patterns, we compute Spearman's correlation coefficient between the emissions and four mobility quantities (see Methods). 
They are the radius of gyration, indicating the characteristic distance travelled by an individual~\cite{gonzalez08understanding, pappalardo2019scikitmobility}; the mobility entropy~\cite{song2010limits, eagle2009eigenbehaviors, pappalardo2016analytical}, characterising the predictability of their visitation patterns; and the total travel time of the vehicles, a principal factor governing emissions.
The travel time shows positive correlations whose strength varies little from city to city. 
The radius correlates positively with the vehicle's emissions, while the entropy correlates negatively.
In London, the travel time has a strong positive correlation ($0.98$) with the emissions (see Table \ref{tab:corrs}), the radius has an almost null correlation ($0.09$), and the entropy has a strong negative correlation ($-0.72$). 
In Rome, the strength of the correlation with the travel time is high ($0.8$), and the radius ($0.58$) and the entropy ($-0.54$) are respectively positively and negatively correlated with the emissions. 
In Florence, the correlation coefficient of the travel time decreases ($0.48$), and those of the radius and entropy show similar behaviour as in Rome: the first is positive ($0.30$), the second is negative ($-0.29$).

As one could expect, the more a vehicle travels, the more emissions it produces. However, vehicles with more regular and predictable behaviour generate the highest emissions, not those with more erratic behaviour.
Indeed, mobility entropy is low when a vehicle performs a high number of recurring trips, indicating predictable travelling patterns. 
In contrast, it is high when the vehicle performs trips from various origins and destinations, denoting a more unpredictable travelling behaviour.
The observed negative correlations suggest that gross polluters are more regular and predictable than low-emitting vehicles.

To deepen these relationships, we use a Generalised Additive Model (GAM)~\cite{hastie1986generalized} to express the emissions as a non-linear combination of the three mobility measures (see Supplementary Note 7). 
We find that the radius and the entropy contribute in an opposite way to determining a vehicle's emissions.
On the one hand, for Rome and Florence, the greater a vehicle's radius of gyration, the greater its emissions (see Supplementary Figures \ref{fig:GAM_rome} and \ref{fig:GAM_florence}). 
For London, the radius of gyration's marginal contribution to the emissions is constant for values $> 7$  km (see Supplementary Figure \ref{fig:GAM_london}).
On the other hand, for Rome and Florence, the greater a vehicle's entropy, the lower its emissions (Supplementary Figures \ref{fig:GAM_rome} and \ref{fig:GAM_florence}). 
For London, the negative marginal contribution of the entropy to the emissions only holds for entropy $> 0.7$ (see Supplementary Figure \ref{fig:GAM_rome}).
We also perform a cluster analysis to group the vehicles based on their radius of gyration, mobility entropy and travel time. We find two clusters, namely the predictable and the erratic drivers, and find that the former emits typically more than the latter (see Supplementary Note 7 and Supplementary Figure \ref{fig:kmeans}).

The interpretation of these results would imply further analyses involving additional data providing information about the motivations behind each vehicle’s trip (e.g., drivers' mobility diaries). For example, the erratic vehicles could emit less because they are primarily used for sporadic excursions towards unknown locations.
In contrast, predictable drivers could be forced to use private vehicles because they live or work in neighbourhoods poorly served by public transportation. 
Also, the heterogeneity in emissions’ distributions could be related to socio-economic inequalities and the centre-periphery divide.
For example, in Rome, ``new low-density settlements often take on the character of sprawl and rely exclusively on private transportation''~\cite{lelo2019socio}.

We also investigate the relationship between the emissions suffered by a road and network features such as the road length and the betweenness centrality of the edge representing the road.
The betweenness centrality is based on how frequently a road falls on the shortest paths connecting two crossroads (see Methods). 
Therefore, a road with a high centrality is more likely to be crossed. 
We find a positive correlation between the betweenness centrality and the emissions in the roads for all three cities (see Table \ref{tab:corrs}). 
This result confirms that our emissions estimates are consistent with the roads' characteristics, as roads central in the network are more likely to host greater emissions. 
Similarly, we find a positive correlation with the road length (see Table \ref{tab:corrs}), as longer roads capture more points and thus more emissions.
Similar results hold for the emissions of the other three pollutants in all the three cities (see Supplementary Note 7). 
Figure \ref{fig:Figure2}a,d,g shows the entire road networks of Greater London, Rome, and Florence, respectively, with the emissions on each road, normalised by the length of the road to highlight the differences between the roads better.

\begin{table}
    \centering
    \begin{tabular}{c | c | c | c || c | c |}
    \cline{2-6}
         & \multicolumn{3}{ c||}{\bf mobility metrics} & \multicolumn{2}{c|}{\bf roads' features} \\
        \cline{2-6}
        
         & radius & entropy & travel time & betweenness centrality & length \\
         \cline{2-6}
         
        London &  0.09 &  -0.72 &  0.98 &   0.27 &  0.22 \\
        
        Rome &  0.58 &  -0.54 &  0.80 &  0.30 &  0.35 \\
        
        Florence &  0.30 &  -0.29 &  0.48 &  0.10 &  0.25 \\
        \cline{2-6}
    \end{tabular}
    \caption{\textbf{Correlations between vehicles' emissions, mobility metrics, and road features.} Spearman's correlation coefficients between (left) CO$_2$ emissions per vehicle and vehicles' mobility metrics (their radius of gyration, uncorrelated entropy, and travel time), and (right) CO$_2$ emissions per road and roads' features (betweenness centrality and length).}
    \label{tab:corrs}
\end{table}

\subsection*{Simulation scenarios.}
\label{sec:simulation}
Reducing emissions is a growing concern for cities, and it is crucial to estimate the impact of policies targeting vehicles to reduce their footprint on the city's environment. 
We investigate the impact of the vehicles' electrification on the total amount of emissions and the distribution of emissions across the roads.
In particular, we study how the electrification of a certain share of vehicles would change the emissions on the three cities' roads.
In this setting, even if a vehicle's electrification would change its driver's mobility behaviour, the vehicle would not create any emissions. 

\begin{figure}
    \centering
    \includegraphics[width=\textwidth]{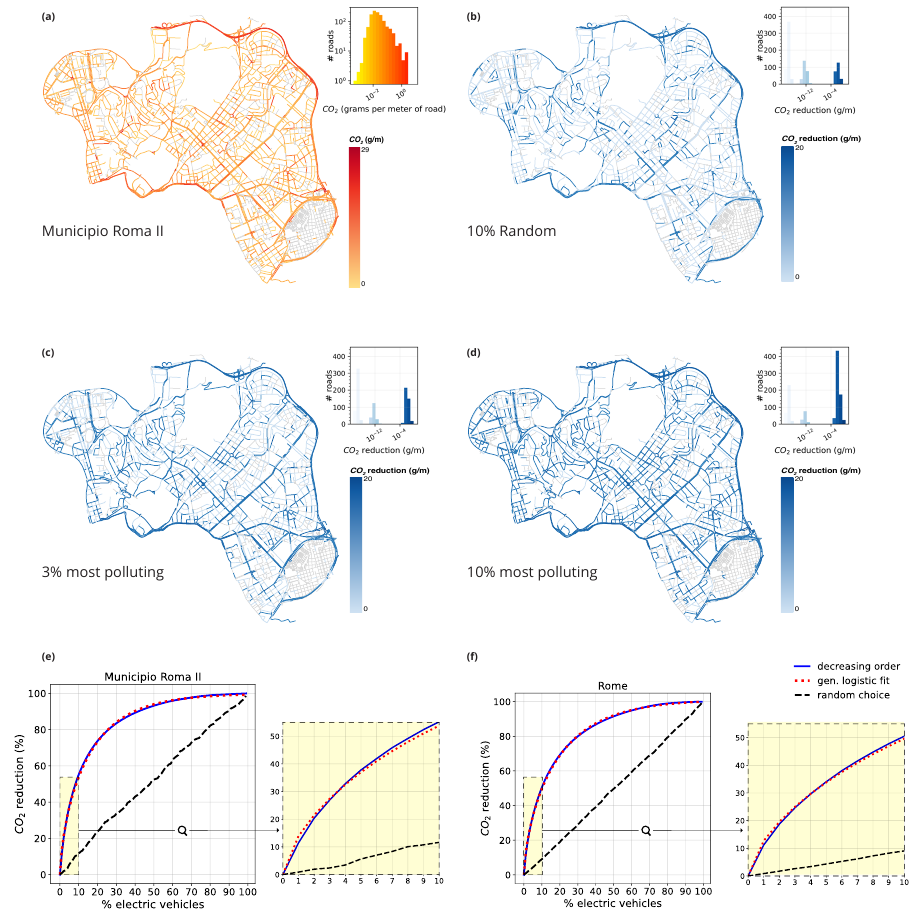}
    \caption{\textbf{Simulation of electrification.} 
    Impact on CO$_2$ emissions of the electrification of the vehicles moving within a neighbourhood (the Second Municipality of Rome), as well as within the whole city of Rome, during January 2017. 
    Panel (a) shows the distribution of emissions of CO$_2$ across the roads of the neighbourhood, and panels (b), (c), (d) show the reduction of emissions experienced by the same roads if a random 10\% of the vehicles, the top 3\% gross polluters, or the top 10\% gross polluters, respectively, shift to electric engines. 
    The roads are coloured in a gradient from light blue (low reduction) to dark blue (high reduction), and the histogram in each panel shows the distribution of the emissions' reduction across the roads, with the bars coloured with the same gradient.
    Panel (e) shows the percentage reduction of CO$_2$ emissions corresponding to a share (0-100\%) of electric vehicles for the Second Municipality of Rome.
    The inset plots zoom in on the first tenth share of electric vehicles.
    The solid blue line is when the vehicles to be electrified are chosen from the most polluting to the least polluting.
    The black dashed line indicates the emission reduction when the vehicles to be electrified are chosen at random.
    The dotted red line is the Generalised Logistic Function (Richard's curve) fit: 
    $f(x) = \frac{\alpha}{(1 + \beta e^{-rx})^{1/\nu}}$, where $\alpha$ represents the upper asymptote, $\beta$ the growth range, $r$ the growth rate, and $\nu$ the slope of the curve.
    Panel (f) shows the same plot for the whole city of Rome.
    The road networks in panels (a)-(d) are plotted with the Python library OSMnx~\cite{boeing2017}.}
    \label{fig:Figure3}
\end{figure}

We find that the electrification of just the top 1\% of gross polluters would reduce emissions as much as electrifying 10\% of random vehicles.
In Figure \ref{fig:Figure3}, we show a case study for the entire city of Rome as well as a single neighbourhood to investigate the impact of massive electrification on emissions. 
We provide results for London and Florence in Supplementary Note 8.
As the share of gross polluters that shift to electric engines grows, the impact on the roads in reducing emissions becomes more evident. 
In particular, if the top 10\% of gross polluters shift to an electric engine, 107 roads have a significant reduction in the grams of CO$_2$ per meter (at least equal to 0.01 $g/m$), see Table \ref{tab:emissions_reduction} and Figure \ref{fig:Figure3}d.
In contrast, if 10\% of the vehicles that shift to electric engines are chosen at random, only 18 roads have a significant reduction of emissions (see Figure \ref{fig:Figure3}b). 
These results hold for both single neighbourhoods and the entire city (see Figure \ref{fig:Figure3}e,f for Rome and Supplementary Note 8 for London and Florence).

The percentage reduction of the overall emissions grows almost linearly when the share of electric vehicles is chosen at random.
In contrast, a Generalised Logistic Function (GLF, also known as Richard's curve~\cite{richards1959flexible, fekedulegn1999parameter}) approximates the growth rate when the gross polluters are electrified first.
We use non-linear least squares to fit the GLF, $f(x) = \frac{\alpha}{(1 + \beta e^{-rx})^{1/\nu}}$, where $\alpha$ represents the upper asymptote, $\beta$ the growth range, $r$ the growth rate, and $\nu$ the slope of the curve. 
The model gives $R^2 = 0.99$ both for the selected neighbourhood (Figure \ref{fig:Figure3}e) and the whole city of Rome (Figure \ref{fig:Figure3}f). 
The estimated growth rate $r$ of the curve is $4.84 \times 10^{-2}$ for the neighbourhood and $3.96 \times 10^{-2}$ for the entire city of Rome. Its slope $\nu$ is almost the same if we move from the neighbourhood to the entire city ($-1.55$ and $-1.56$, respectively). 
$\alpha$ and $\beta$ are $\approx 100$ and $-1$ for both the neighbourhood and the city.
Similar results hold for Florence (Supplementary Figure \ref{fig:electrification}b, Supplementary Table \ref{tab:logistic_fit_electrification}).
In Greater London (Supplementary Figure \ref{fig:electrification}a), the growth starts slowly ($\nu = -0.86$): 
there are fewer vehicles with high emissions levels, and electrifying the most polluting vehicles is slightly less effective in reducing emissions than in the other two cities (Supplementary Table \ref{tab:logistic_fit_electrification}).

Given the increasing importance of remote working, especially since the COVID-19 pandemic~\cite{vyas2020impact, nagel2020influence}, we simulate the impact of a massive shift to remote working on reducing vehicles' emissions. 
Even if this working style may affect individual mobility patterns, we assume that it eliminates commuting trips. 
Indeed, if an individual works from home, the most straightforward implication is the removal of the commuting trips from their mobility habits.
We identify the vehicles' home and work locations (see Methods) and study the emissions generated from their commuting patterns.
We perform a simulation in which a growing share of these commuters become home workers, i.e., they do not travel anymore between their home and work locations. 

We find that emissions reduction is more effective when the home workers are gross polluters. 
In this case, remote working for the top $1\%$ gross polluters leads to the same reduction reached if they were $\approx4\%$ random vehicles (see Supplementary Figure \ref{fig:homeworking}).
Again, a GLF well fits emissions reduction when the gross polluters become home workers. 
In particular, we obtain estimates for $\nu$ (the slope of the curve) that are similar for Rome and Florence ($-1.30$ and $-1.35$, respectively) and lower for London ($-0.72$); see Supplementary Note 8 and Supplementary Figure \ref{fig:homeworking} for details.

Overall, these results demonstrate that targeting specific profiles of vehicles can significantly improve emission reduction policies.


\begin{table}
    \centering
    \begin{tabular}{c | c | c | c |}
        \cline{2-4}
        
         CO$_2$ reduction (g/m) &  10\% random & 3\% most polluting & 10\% most polluting \\
         \cline{2-4}
         
        $\ge 10^{-4}$ &  232 &  385 & 665 \\
        
        $\ge 10^{-3}$ &  137 &  258 & 469 \\
        
        $\ge 10^{-2}$ &  18 &  53 & 107 \\
        
        $\ge 10^{-1}$ &  0 &  6 & 10 \\
        \cline{2-4}
    \end{tabular}
    \caption{\textbf{The roads and their CO$_2$ reduction in three different scenarios.} Number of roads that experience certain levels of CO$_2$ reduction in three different scenarios of vehicles' electrification: (i) when electrifying $10\%$ random vehicles (first column), (ii) when electrifying the top $3\%$ most polluting ones (second column), and (iii) when electrifying the top $10\%$ most polluting ones (third column).}
    \label{tab:emissions_reduction}
\end{table}

\section*{Discussion}
\label{discussion}
Using GPS data to estimate the emissions from thousands of vehicles in three European cities with different sizes and characteristics, we show the existence of gross polluters, i.e., vehicles responsible for the greatest quantity of emissions. 
The existence of gross polluters has been reported in previous studies using measurements from on-road remote sensing sites~\cite{guenther1994emissions, brand2008taming, huang2018remote}.
Nevertheless, these studies have been questioned because measurements from on-road sites cannot represent a vehicle's overall emission level~\cite{huang2020reevaluating}.
Our study contributes to reshaping this discussion since our findings are based on a microscopic emission model that captures the vehicles' instantaneous emissions in great detail.
We add new elements to this debate, discovering that gross polluters exist in different cities and for different pollutants (CO$_2$, NO$_x$, PM, VOC) and reporting the existence of grossly polluted roads that suffer the greatest amount of emissions.

The heterogeneous patterns governing the distribution of emissions across vehicles and roads are well approximated by heavy-tailed distributions, with exponents that vary from city to city and from pollutant to pollutant.
These peculiar exponents may depend on the characteristics of the city's road network and people's commuting behaviour.
For example, London has a vast and dense road network, and people use private vehicles less intensively~\cite{londonmobility2019} than in Rome and Florence. 
Thus, one can argue that mobility behaviour in London leads to a distribution of emissions per vehicle that is fairer than people's behaviour in Rome, which is characterised by a vast and sparse road network and intensive use of private vehicles~\cite{romemobility2019}. 

Our study can be reproduced with any city provided the availability of vehicles' GPS trajectories and road networks data and may help find more effective strategies to reduce emissions.
For example, our study demonstrates that blocking the circulation based on a non-informed choice (e.g., blocking vehicles with odd or even number plates) has less impact on reducing emissions than identifying and targeting a small share of gross polluters.
Moreover, we provide a precise model to estimate the overall reduction of emissions generated by the electrification of a particular share of vehicles or by reducing the number of commuting patterns travelled by the vehicles (e.g., caused by a transition to the home working of their drivers).

There are several directions in which this study can be extended.
For example, since it focuses on light-duty vehicles, all the results we show are valid for this fleet of vehicles only. 
Although they are the vast majority of the vehicles circulating in a city, we are aware that the absence of other vehicles such as heavy-duty vehicles (e.g., buses, trucks) generates an incomplete mosaic of the emissions within the urban environment.
We, therefore, hope for a more comprehensive study that may include different types of vehicles.

Also, our analysis can be extended by investigating how the emission patterns change between weekdays and weekends or weather conditions and considering more sophisticated simulation scenarios. 
For example, it would be interesting to investigate the impact of policies that aim to improve walking, transit, or cycling on the distribution of emissions, the number of gross polluters, and grossly polluted roads.
Finally, the relation between the vehicles' emissions and mobility patterns may be examined more deeply to investigate whether the observed heterogeneous distributions originate from other inequalities (e.g., socio-economic inequalities and centre-periphery divide).

Meanwhile, our study may shape the discussion on measuring emissions with digital data and how to use such measurements to simulate emission reduction scenarios. 
If we learn how to use such a resource, we have the potential of monitoring in real-time the level of emission in our urban environments and taking immediate, informed actions when they overcome a certain tolerance threshold. 
This fact is crucial because policymakers' decisions depend on what we measure, how good our measurements are, and how promptly we react to these measurements.



\section*{Methods}
\label{methods}

\paragraph{Data Filtering.}
\label{sec:preprocessing}
In our GPS dataset, each trajectory point is associated with a vehicle identifier, a trajectory identifier, a timestamp, and a latitude/longitude pair. 
The sampling rate of the trajectory points may affect the estimate of instantaneous speed and acceleration. 
Since the mean time interval between trajectory points varies from city to city (it is about one minute for London and four minutes for Rome and Florence, see Table \ref{tab:traj}), we perform a pre-processing step to align them.
For each trajectory, we retain only those sub-trajectories (i.e., disjoint subsets of points) that satisfy two constraints: (i) there are at least two points, and (ii) the time interval between consecutive points is less than $\theta = 120$ seconds.
The filtering step causes a drop in the number of points and, by consequence, of vehicles. 
We analyse the trends of both the number of vehicles and points resulting from the filtering step varying the filtering parameter $\theta$ from 1 second to 300 seconds (see Supplementary Note 1).
By choosing $\theta = 120$ seconds, we lose 53.5\% of points in Rome, 1.4\% in London, and 70.4\% in Florence. 
Consequently, we discard 1.3\% of vehicles in Rome, 0.9\% in London, and 6.5\% in Florence (see Table \ref{tab:traj}).

Different works use different values for $\theta$ (e.g., the time interval between the points is set to 1 second \cite{chen2020}, 3 seconds \cite{sui2019}, 5 seconds \cite{nyhan2016}, 5 to 50 seconds \cite{liu2019}).
Our choice derives from our data sampling rate and is a trade-off between the reliability of the results and the data coverage.
As the last step, we compute, for each vehicle, the speed and acceleration in each point, and retain only points whose speed is lower than 300 km/h and whose acceleration is in the range $[-10, +10]$ m/s$^2$ (as suggested by Nyhan et al.\cite{nyhan2016}).

\paragraph{Road network and GPS points snapping.}
\label{sec:matching}
The road network of each city is extracted from OpenStreetMap (OSM) \cite{OpenStreetMap}, a collaborative project to create a free editable map of the world and provide the geodata underlying the map.
In particular, the road network is provided as a multigraph $G = (V, E)$, with $V$ being the set of nodes $v$ and $E$ being the multiset of edges $e$, in which the edges represent public roads accessible to vehicles (including service roads). 
Each edge $e\in E$ is a pair of two identifiers, indicating the starting and ending OSM nodes, plus a key that discriminates between parallel edges (if present). Moreover, it carries some information about the road it represents, such as its name, length, and type (e.g., whether it is a motorway or residential street).
In this context, we call crossroads the nodes $v\in V$ that have at least two roads (i.e. edges $e\in E$) incident on them.
To download, compute statistics, and visualise the road networks, we conceived methods based on the Python library OSMnx~\cite{boeing2017}.
Our matching step consists of a ball tree Nearest-Neighbour algorithm that assigns each point of a GPS trajectory to its nearest edge in the road network.
This point snapping step allows us to assign the vehicles' emissions to the roads onto which they are produced (see Figure \ref{fig:Figure1}c).
Note that the Nearest-Neighbour algorithm used does not affect the quality of the emissions estimated with the microscopic model. 
Indeed, the instantaneous emissions are computed based on the points' speed and acceleration computed from the GPS data (not matched yet with the road network). 
Moreover, our GPS data come with a quality index that ranges from 1 to 3 and indicates the precision of each GPS point's location. 
A point's maximum quality (i.e., index = 3) indicates that at least four satellites' signals received by the GPS device are used for the trilateration of the point's location. 
Our study only uses the points with the highest quality (i.e., index = 3). 
Given this accuracy in the points' locations, the reconciliation problem is straightforward~\cite{white2000some}, and we assign the location obtained from the GPS receiver to the nearest edge in the network.

\paragraph{Computing emissions.}
We implement a microscopic emissions model~\cite{nyhan2016} to compute the instantaneous emissions associated with each trajectory point $p$. 
We denote the quantity of pollutant $j \in \{\textrm{CO}_2, \textrm{NO}_x, \textrm{PM}, \textrm{VOC}\}$ emitted at point $p$ from vehicle $u$ as $E^{j,u}_p$ and the instantaneous speed and acceleration of the vehicle in $p$ as $v_p$ and $a_p$, respectively.
Information about its engine type (whether it is petrol, diesel, or LPG) is available for each vehicle.
This information determines, together with the type of pollutant, the emission factors $f_i$. 
We use the following equation to compute the instantaneous emissions $E^{j,u}_p$ of pollutant $j$ from vehicle $u$ in point $p$:
\begin{linenomath*}
    \begin{equation}
    E^{j,u}_p = f_1^{j,u} + f_2^{j,u} v_p + f_3^{j,u} v_p^2 + f_4^{j,u} a_p + f_5^{j,u} a_p^2 + f_6^{j,u} v_p a_p
    \end{equation}
\end{linenomath*}
where for NO$_x$ and VOC emissions the factors $f_1,...,f_6$ change with acceleration (based on whether $a_p \ge -0.5 \;$ m/s$^2$ or $a_p < -0.5 \;$ m/s$^2$).
In Supplementary Table \ref{tab:emission_factors}, we show the value of factors $f_1, \dots, f_6$ varying the vehicle's type and acceleration.

\paragraph{Mobility measures and roads' centrality.}
We use four quantities to describe the mobility of a vehicle $u$: 
\begin{itemize}
\item the radius of gyration~\cite{gonzalez08understanding, pappalardo2015returners, pappalardo2013understanding} $r_g(u) = \sqrt{\frac{1}{n} \sum_{i \in P} dist(\mathbf{r}_i(u) - \mathbf{r}_{cm}(u))^2}$, where $P$ is the set of $n$ points recorded for $u$, $\mathbf{r}_i(u)$ indicates the coordinates of trajectory point $i \in P$, $\mathbf{r}_{cm}(u)$ is the centre of mass of $u$, and \emph{dist} the haversine distance between two points on earth; 
\item the temporal-uncorrelated entropy~\cite{song2010limits, eagle2009eigenbehaviors, pappalardo2016analytical} $S(u) = -\sum_{i=1}^{N_u} p_u(i) \log_2 p_u(i)$, where $N_u$ is the number of distinct locations visited by $u$, and $p_u(i)$ is the probability that $u$ visits location $i$;
\item the travel time of $u$, computed as the sum of all the travel times of its trajectories.
\end{itemize}
We measure the centrality of a road, a proxy of its traffic volume in the city, as its betweenness centrality.
In network science, the betweenness centrality of an edge $e$ (i.e., a road in our case) is defined as $C_b(e) = \sum_{s,t \in V} \frac{\sigma(s,t|e)}{\sigma(s,t)}$, where $V$ is the set of nodes in the network, $\sigma(s,t)$ is the number of shortest paths between $s$ and $t$, and $\sigma(s,t|e)$ is the number of those shortest paths passing through edge $e$.

\paragraph{Home and work locations.}
The first step when identifying an individual's home and work locations is the selection of the starting and ending points of their trajectories \cite{pappalardo2020individuallevel}.
The position of the starting (or ending) points of the trajectories that start from (or end to) the same semantic location may not coincide. 
This may happen because: (a) a driver can park the vehicle within a certain radius from the location; (b) since the first point sent by the GPS device often lacks precision and is discarded, the second point sent is taken as the starting point of the trajectory.
For the above reason, we spatially cluster these points within a radius of 250 meters and take the centroid of each cluster as the vehicle's stop location. 

To identify a vehicle's home and work locations, we use a principle commonly adopted in the literature~\cite{pappalardo2020individuallevel}: the home location is the stop location corresponding to the most frequent cluster; the work location is the stop location corresponding to the second most frequent cluster.
We discard the vehicles for which it is impossible to identify the most frequent stop location(s) (for example, the vehicle visited each location only once). 
We successfully identify home and work locations for $55\%$, $31\%$, and $16\%$ of the vehicles moving in London, Rome, and Florence, respectively.
There are two main reasons we cannot identify the home and work locations of many vehicles in Florence. First, the average number of trajectories per vehicle (10.7) is much lower than in the other two cities (27.6 in Rome and 43 in London) (see Table \ref{tab:traj}). The fewer trajectories a vehicle has, the more difficult it is to identify its home and work locations. Second, while a relatively small city, Florence is an essential hub for the surroundings. Thus, many users could live outside the city. Moreover, even if they live inside the city, given various restricted traffic areas in the city’s historic centre, many could reach the workplace by public transportation or walking. 
This leads to a low number of commuting trajectories inside the city.


\section*{Acknowledgements}
This work is partially funded by the H2020 projects Track \& Know (grant agreement \#780754, M.N.), SoBigData++ (grant agreement \#871042, M.B.), and HumanE-AI-Net (grant agreement \#952026, L.P.).
The authors thank Giuliano Cornacchia, Vasiliki Voukelatou, Massimiliano Luca, and Giovanni Mauro for the useful suggestions, Daniele Fadda for the precious support with data visualization. 
M.B. and L.P. also thank Francesco Totti for the inspiration.

\paragraph{Data availability statement} The data that support the findings of this study are not publicly available due to privacy restrictions and were used under license for the current study. Aggregated source data for figures are available from the authors upon reasonable request.
\paragraph{Code availability statement} The Python code to reproduce the analyses in the study from public GPS data (taxi trips) is publicly available in GitHub (\url{https://github.com/matteoboh/mobility_emissions}) and Zenodo repository \cite{bohm2022zenodo}.

\paragraph{Author contributions} M.B. designed the study, performed the data preprocessing and cleaning, developed the code for and performed the statistical analyses, the experiments, the simulations and the plots, contributed to the interpretation of the results, and wrote the paper. 
M.N. provided the data, contributed to the work methodology and interpretation of the results, and revised the writing of the paper. 
L.P. developed the concept and designed the study, contributed to the work methodology and interpretation of the results, wrote the paper and coordinated the study. All authors read, edited, and approved the final version of the paper. 

\paragraph{Competing interests} The authors declare no competing interests.

\paragraph{Corresponding authors} The corresponding authors are M.B. (bohm@diag.uniroma1.it) and L.P. (luca.pappalardo@isti.cnr.it)

\bibliographystyle{abbrv} 
\bibliography{biblio}  






\clearpage
\appendix
\appendixpage

\section{Supplementary Information}
\label{supp:info}

\subsection*{Supplementary Note 1: vehicles and emissions throughout the year.}
\label{supp:seasonality}
We use a whole year of GPS data to investigate (i) the spatio-temporal patterns of the vehicles' trajectories, and (ii) the robustness of our results regarding the seasonal variations of the emissions.

Regarding point (i), our sample for this analysis is represented by all the vehicles that made at least one trip in each season of the year, for each city (2374 vehicles in London, 4936 in Rome, 2906 in Florence).
Supplementary Figures \ref{fig:traj}c,d,e show the spatial distribution of the GPS points forming the trajectories describing these vehicles' trips throughout the entire year, in Greater London, Rome and Florence. There are no areas in the three cities without GPS points, with the exception of green areas where cars are not allowed (e.g., Richmond Park and Hampstead Heath in London, Tenuta di Castelporziano and Parco Regionale dell'Appia Antica in Rome, Parco delle Cascine and Giardino di Boboli in Firenze).
To investigate whether these spatial patterns change throughout the year, we consider the number of different vehicles passing on each road and show its distribution in each season of the year (Supplementary Figures \ref{fig:traj}a,b,c). The number of distinct vehicles per road in London (Supplementary Figure \ref{fig:traj}a) varies more than the other two cities. This may be explained by the lower number of vehicles and the higher number of roads in London as compared to Rome and Florence (Supplementary Table \ref{tab:cities}). Regardless of this peculiarity, the mean number of vehicles per road is reasonably stable across the seasons and the cities.

Regarding point (ii), we compute the Gini coefficient of all the emissions’ distributions obtained for the four seasons of 2017 (winter, spring, summer, autumn). We find that the unequal distributions not only hold when changing the city but also when changing the season of the year (Supplementary Figures \ref{fig:seasons_gini_london}, \ref{fig:seasons_gini_rome}, \ref{fig:seasons_gini_florence}). 
Indeed, the Gini coefficients are similar across the seasons for all the distributions and cities: the maximum value of their coefficient of variation is $2.5\%$, obtained for the distributions of CO$_2$ emissions across London's vehicles (Supplementary Figure \ref{fig:seasons_gini_london}b). Even if this variation in the seasonal Gini coefficients is small, summer and spring are the seasons with the highest degree of inequality in the emissions. 
Also, we show the distribution of seasonal variations in the quantities emitted across the vehicles and roads in Supplementary Figures \ref{fig:seasons_boxplots_london}, \ref{fig:seasons_boxplots_rome}, \ref{fig:seasons_boxplots_florence}. 
While the median and the quartiles of the distributions are similar across the seasons, their mean value changes because it is more sensitive to the points in the tail. 
Thus, the highest means are always obtained in summer and spring, whose distributions’ tails are slightly heavier, coherently with what we observe for the seasonal Gini coefficients.

In conclusion, even if we obtain slightly more skewed emissions' distributions in summer and spring, this analysis shows that these differences are narrow, supporting the robustness of our results.

\subsection*{Supplementary Note 2: trajectory pre-processing.}
\label{supp:filtering}
The distribution of the sampling rate (i.e., the time interval between consecutive points belonging to the same trajectory) differs among the three cities (Supplementary Table \ref{tab:stats_sampling_rate}). 
Due to differences in the GPS data acquisition process, London’s sampling rate distribution is concentrated around the median of $60$ seconds, with $90\%$ of the points being sampled with no more than $90$ seconds from each other. On the contrary, the sampling rates for Rome and Florence are higher on average, and their distribution has a higher variance, with $10\%$ of the points being distant more than, respectively, $425$ and $469$ seconds from each other.
High sampling rates are not desirable for estimating instantaneous speed and acceleration and hence instantaneous emissions. 
Thus, for each vehicle, we retain only those sub-trajectories (i.e., disjoint subsets of points) that satisfy two constraints: (1) they are composed of at least two points, and (2) the time interval between consecutive points is less than $\theta$ seconds.
This filtering step causes a drop in the number of points and, by consequence, of vehicles. We analyse the trends in both the number of vehicles and points resulting from the filtering step when varying the filtering parameter $\theta$ from $1$ second to $300$ seconds (Supplementary Figures \ref{fig:exp_sampling_rate}a,b,c).
The number of vehicles and points surviving the filtering process generally grows with $\theta$. 
This growth is similar in Rome (Supplementary Figure \ref{fig:exp_sampling_rate}b) and Florence (Supplementary Figure \ref{fig:exp_sampling_rate}c). In London, it is steeper due to the lower variability in the sampling rate of London’s data: for $\theta > 60$ seconds, only $1\%$ of the vehicles are discarded by the filtering process. For Rome and Florence, we find acceleration in the growth of the two curves for $\theta > 60$, even if not as steep as for London. 
This is explained by their modal value of sampling rate, which is equal to $70$ seconds for Rome and $65$ seconds for Florence (Supplementary Table \ref{tab:stats_sampling_rate}) ($11\%$ and $5,4\%$ of the values lie in the interval $[60, 75]$ seconds). 
Thus, many points and vehicles are discarded from the filtering for values of $\theta$ that overcome these limits.
By choosing $\theta = 120$ seconds, we incur a loss of points equal to $53.5\%$ in Rome, $1.4\%$ in London, and $70.4\%$ in Florence. 
Consequently, the vehicles that are discarded are $1.3\%$ in Rome, $0.9\%$ in London, and $6.5\%$ in Florence.

We also look at the average sampling rate when applying the filtering and its trend when increasing $\theta$. In Supplementary Figure \ref{fig:exp_sampling_rate}d, we show the average sampling rate of the points after the filtering step in the three cities for $\theta$ in $[1, 300]$ seconds. When $\theta = 120$ seconds, we reach an average sample rate equal to $76.8$, $58.3$, and $75.8$ seconds in Rome, London, and Florence. Although this seems counterintuitive, note that $\theta$ is an upper bound to the maximum time interval between a trajectory's two consecutive points: for a specific value of $\theta = \theta'$ we obtain, after the filtering, trajectories in which the points are distant no more than $\theta'$ seconds and, thus, on average, less than $\theta'$ seconds.

After this filtering process, one can argue that, still, both London's number of trajectories per vehicle and number of points per trajectory ($\approx86$ and $\approx13$, respectively) are considerably bigger than the Roman ($\approx24$ and $\approx9$, respectively) and Florentine ($\approx8$ and $\approx6$, respectively) ones.
As the empirical Complementary Cumulative Distribution Functions (CCDF) of the data show (Supplementary Figures \ref{fig:ccdf_vehicle} and \ref{fig:ccdf_road}), this fact does not influence the shape of the distributions, which is reasonably similar across the three cities.

\subsection*{Supplementary Note 3: road networks.}
\label{supp:networks}
The three cities we analyse (Greater London, Rome and Florence) are heterogeneous concerning their road networks: Rome is huge but with the sparsest network; London is huge but with the densest network; Florence is small ($\approx1/12$ of Rome and $\approx1/15$ of London in terms of land area) but has a dense road network too.
London has about three times the number of roads than Rome and a smaller average road length (see Supplementary Table \ref{tab:cities}). 
In terms of the number of crossroads per km$^2$, Florence ($\approx73$) is between Rome ($\approx39$) and London ($\approx98$). 
Also, in terms of road density (total length of all the roads divided by the land area of the city), Florence (10,483) is $\approx1.6$ times denser than Rome (6,759) and not far from London's density (12,367).
In Rome, 0.4\% of roads are motorways (usually long and large roads without traffic signals), while in London and Florence, they are only 0.02\%.

\subsection*{Supplementary Note 4: distribution of emissions.}
\label{supp:distributions}
Emissions distribute across both vehicles and roads in a heterogeneous way. 
In Supplementary Figure \ref{fig:lorenz}, we show the Lorenz curves representing the distributions of the four pollutants per vehicle (Supplementary Figures \ref{fig:lorenz}b,d,f,h) and road (Supplementary Figures \ref{fig:lorenz}a,c,e,g). 
Together with the associated Gini coefficients, these curves highlight the inequality of our distributions.
Each Lorenz curve shows the $y\%$ of the total emissions ($y$-axis) from the bottom $x\%$ of vehicles ($x$-axis). 
The more distant the curve from the line of perfect equality (i.e. the case in which the $x\%$ of vehicles emit the $x\%$ of emissions), the more is unequal the distribution. 
The Gini coefficient is the ratio of the area between the Lorenz curve and the line of perfect equality to the total area under this line. It varies between 0 and 1, with high values corresponding to unequal distributions.
For all the cities and distributions, we obtain a Gini coefficient that is $> 0.55$.
We observe the highest Gini coefficient for Rome and Florence (Supplementary Figures \ref{fig:lorenz}a,c,e,g, Gini $>0.8$). 
For London, the Gini coefficient ranges between $0.65$ and $0.70$.
The emissions of CO$_2$ (Supplementary Figure \ref{fig:lorenz}a) and VOC (Supplementary Figure \ref{fig:lorenz}h) per vehicle are distributed in a similar way across the cities (with a maximum difference between the Gini coefficients equal to $0.09$). 
This is not the case for the emissions of NO$_x$ (Supplementary Figure \ref{fig:lorenz}d) and PM (Supplementary Figure \ref{fig:lorenz}f), whose inequality in their distributions across the vehicles slightly change with the city. In London, distributions have a lower inequality(Gini coefficient around $0.62$) than in Rome (in the case of NO$_x$) and Florence (in the case of PM).

All the distributions are well approximated by at least one of the models we consider (power law, truncated power law, log-normal, stretched exponential), and at least one of these models is always a better fit than the exponential one.
Considering the distributions of emissions across vehicles (Supplementary Figure \ref{fig:all_fits_per_vehicle}) and London as a first case, they are always well approximated by a stretched exponential model, with $\lambda$ varying with the pollutant, and $\beta$ in the interval $[1.22, 1.30]$ for all the four pollutants (Supplementary Figures \ref{fig:all_fits_per_vehicle}a,d,g,j). 
Moreover, in the case of NO$_x$ and VOC, also the log-normal model is a good fit for the distribution of their emissions across London's vehicles, with parameters respectively ($\mu=0.31$, $\sigma=0.53$) for NO$_x$ emissions and ($\mu=2.02$, $\sigma=0.52$) for VOC emissions (Supplementary Figures \ref{fig:all_fits_per_vehicle}d,j).
Then, considering the case of Rome, the distributions of emissions of both CO$_2$ and VOC are well approximated by a truncated power law, with similar exponents $\alpha$ but different $\lambda$, as they are respectively ($\alpha=1.13$, $\lambda=1.04 \times 10^{-3}$) for CO$_2$ emissions and ($\alpha=1.21$, $\lambda=0.51$) for VOC emissions (Supplementary Figures \ref{fig:all_fits_per_vehicle}b,k). 
For the distributions of NO$_x$ and PM emissions, the log-likelihood ratio tests give no overall best-fitting model (Supplementary Figures \ref{fig:all_fits_per_vehicle}e,h). 
Indeed, in the case of NO$_x$ emissions, the power law, the truncated-power law and the log-normal all are better fits than the exponential and stretched exponential models (Supplementary Table \ref{tab:test_NOx_Rome}). 
However, the tests give no evidence for concluding that one of the three has a better fit than the others. 
For the PM emissions, all the four heavy-tailed models have better fits than the exponential one, but none of them is better than the others in terms of goodness of fit (Supplementary Table \ref{tab:test_PM_Rome}). 
Finally, the distributions of CO$_2$, NO$_x$, PM, VOC emissions per vehicle in the city of Florence are well approximated by a truncated power law (Supplementary Figures \ref{fig:all_fits_per_vehicle}c,f,i,l), with estimated parameters respectively: ($\alpha=2.12$, $\lambda=1.45 \times 10^{-3}$), ($\alpha=1.97$, $\lambda=0.90$), ($\alpha=1.76$, $\lambda=25.20$), ($\alpha=1.73$, $\lambda=1.39$). 

The distributions of emissions across the roads of the three cities are more homogeneous than those across the vehicles w.r.t. their best-fitting model (Supplementary Figure \ref{fig:all_fits_per_road}). 
Indeed, for both Rome and Florence these distributions are all well approximated by a truncated power law with parameters $\alpha$, $\lambda$ that are similar for each pollutant (Supplementary Figures \ref{fig:all_fits_per_road}b,c,e,f,h,i,k,l): for the emissions of CO$_2$ per road they are respectively ($\alpha=1.55$, $\lambda=1.08 \times 10^{-4}$) in Rome and ($\alpha=1.52$, $\lambda=1.30 \times 10^{-4}$) in Florence; for the NO$_x$, ($\alpha=1.65$, $\lambda=0.27$) in Rome and ($\alpha=1.56$, $\lambda=0.29$) in Florence; for the PM, ($\alpha=1.61$, $\lambda=42.25$) in Rome and ($\alpha=1.52$, $\lambda=35.24$) in Florence; for the VOC, ($\alpha=1.66$, $\lambda=0.06$) in Rome and ($\alpha=1.57$, $\lambda=0.10$) in Florence. 
These results reflect the similarity of these distributions in Rome and Florence that we observed in their Lorenz curves (Supplementary Figures \ref{fig:lorenz}a,c,e,g).
In the case of London's distributions of emissions per road, for the CO$_2$ the best fit is again represented by the truncated power law (Supplementary Figure \ref{fig:all_fits_per_road}a), with exponents $\alpha=2.59$ and $\lambda=2.88 \times 10^{-4}$, while for the other three pollutants the comparisons give no overall best-fitting model.
Indeed, regarding NO$_x$, PM and VOC emissions across London's roads (Supplementary Figures \ref{fig:all_fits_per_road}d,g,j), all the four heavy-tailed models have better fits than the exponential one. 
At the same time, the log-normal, the truncated power law and the stretched exponential models all have better fits than the simple power law, with no overall winning model between the three (Supplementary Tables \ref{tab:test_NOx_London}, \ref{tab:test_PM_London} and \ref{tab:test_VOC_London}).

\subsection*{Supplementary Note 5: fitting of distributions.}
\label{supp:fitting}
We fit the distributions of emissions of the four pollutants (CO$_2$, NO$_x$, PM, VOC) across vehicles and roads for the three cities (Florence, Rome, Greater London) using the statistical methods developed in Clauset et al.~\cite{clauset2009powerlaw} and Klaus et al.~\cite{klaus2011statistical} and implemented by Alstott et al. 2014~\cite{alstott2013powerlaw}. 
These methods use a maximum-likelihood fitting and evaluate the goodness-of-fit using the Kolmogorov-Smirnov distance, comparing different models with a log-likelihood ratio test. 
This fitting procedure is used on 4x2x3=24 distributions in total.

We fit five models -- power law, truncated power law, lognormal, exponential, and stretched exponential -- to the data and compare pairwise their goodness-of-fit with a log-likelihood ratio test. 
We give particular attention to the comparison with the exponential model, as it is generally considered the minimum alternative candidate for evaluating the heavy-tailedness of a distribution.
Indeed, if the exponential is the best fit, one should reconsider the data's heavy-tail distribution hypothesis.
For each model, we use log-likelihood ratio tests for comparing its goodness-of-fit to the data with all the others, and we choose the best fitting model as the one (if any) that wins the higher number of comparisons.

As the power-law behaviour of all our distributions starts from a certain value $x_{min}$ (as it happens for the majority of the phenomena that obey power laws), we estimate $x_{min}$ as follows. 
We compute the power-law fit for each possible value of $x_{min}$ in the range of the data and evaluate the goodness of fit with the Kolmogorov-Smirnov statistic ($D$), that is the maximum distance between the empirical Cumulative Distribution Function (CDF) and that of the fitted model (the lower $D$, the better the fit). 
We choose the $x_{min}$ corresponding to the global minimum of $D$. 
However, the choice of $x_{min}$ influences the precision of the estimator we use.
Indeed, the Maximum Likelihood Estimator $\hat{\alpha}$ of the exponent of the power law is asymptotically Gaussian with variance $\sigma = \frac{(\alpha-1)}{n}$ (see Proposition B.4 in Clauset et al.~\cite{clauset2009powerlaw}), where $n$ is the number of data points $x_i \geq x_{min}$.
As $\sigma$ grows with $x_{min}$ (as $n$ decreases), the trade-off is given by the choice of the best power law fit (i.e., the $x_{min}$ with minimum $D$) with a low variability for $\hat{\alpha}$.
As Supplementary Figures \ref{fig:xmin_vehicle} and \ref{fig:xmin_road} show for Rome and Florence, the global minima of $D$ also correspond to low $\sigma$, with the only exception of the distributions of CO$_2$ and VOC emissions per vehicle in Rome. Nevertheless, the local minima of $D$ found when $\sigma < 0.05$ are not far from the global minimum (there is a difference of $\approx 0.02$), so we can choose those $x_{min}$ with a reasonable loss in terms of goodness of fit.
This is not the case for London, for which the global minima for $D$ are reached at high values of $x_{min}$, which correspond to $\sigma$ higher than 0.4. 
At the same time, choosing a lower $\sigma$ would lead to a solution with high values of $D$ (around 1.5). 
We conclude that a power-law fit is not reasonable for London.
Instead, the distributions for London are better approximated by a stretched exponential.

\subsection*{Supplementary Note 6: robustness of the results to sample size and filtering parameter.}
\label{supp:samplesize_theta}
To demonstrate that the sample size we use does not affect the significance of our results, we perform extensive experiments on different sample sizes. 
In particular, we replicate the computation of the Gini coefficient on subsamples of different dimensions: from $10\%$ of the available vehicles to $100\%$ ($10\%$-step). 
For each sample dimension, we compute the Gini coefficient on ten different random samples, obtaining coefficients that are reasonably stable across the different sample dimensions.
For all cities, we show the results obtained for the distributions of all pollutants across vehicles and roads in Supplementary Figure \ref{fig:sample_exp_vehicle} and Supplementary Figure \ref{fig:sample_exp_road}, respectively. 
Despite the higher variance in the coefficients computed on smaller samples, both the mean and the median Gini coefficient (and its entire distribution) remain stable and consistently above $0.5$. 
These experiments demonstrate that the sample size does not affect the shape of the observed distributions, neither for vehicles nor for roads.

We also investigate the robustness of our results given the choice of the filtering parameter $\theta$.
We find that the inequality of the emissions’ distributions holds when changing $\theta$. 
Supplementary Figures \ref{fig:theta_gini_london}, \ref{fig:theta_gini_rome}, and \ref{fig:theta_gini_florence} represent the Lorenz curves and associated Gini coefficients of all the emissions' distributions in the three cities. 
We draw two conclusions: \emph{(i)} the changes in the Gini coefficients depend on the type of pollutant and distribution, not on $\theta$; \emph{(ii)} the Lorenz curves appear reasonably similar across the $\theta$, with relatively small changes in the Gini coefficients. 
Not surprisingly, the most unpredictable behaviour is obtained for $\theta = 30$ seconds, especially for Florence, as a consequence of the much smaller sample dimension (see Supplementary Figure \ref{fig:exp_sampling_rate}a,b,c).

In conclusion, although the estimates of emissions become rougher when the parameter $\theta$ increases, we find that the emissions’ unequal distributions are robust concerning the choice of $\theta$.

\subsection*{Supplementary Note 7: relations with vehicles' mobility and roads' features.}
\label{supp:relations}
The relationship between the emissions of the four pollutants, the vehicles' mobility, and the roads' characteristics is often non-linear (Supplementary Figures \ref{fig:scatter_mobility_london}-\ref{fig:scatter_roads_florence}). 
Thus, we used Spearman's correlation coefficient to investigate these relationships as a first step.
Spearman's correlation measures the direction and strength of the monotonic (even non-linear) relation between two features. More precisely, given two random variables $X_1$ and $X_2$, their Spearman's correlation coefficient $\rho$ is computed as $\rho = \frac{Cov[R_1, R_2]}{Var[R_1]Var[R_2]}$, where $R_1$, $R_2$ are the ranked features obtained from $X_1$ and $X_2$, respectively.

For all pollutants and cities, the strongest (and almost linear) correlation with the emissions is that with the vehicles' travel time ($> 0.6$): the more a vehicle travels, the more emissions it produces (see Supplementary Figures \ref{fig:scatter_mobility_london}, \ref{fig:scatter_mobility_rome}, \ref{fig:scatter_mobility_florence}, and Supplementary Tables \ref{tab:corrs_NOx}, \ref{tab:corrs_PM}, \ref{tab:corrs_VOC}).
In London, the correlation is the highest ($> 0.95$). 
In Rome, it is slightly lower ($\in [0.8, 0.9]$); in Florence, it ranges between $0.48$ (for CO$_2$) and $0.74$ (for PM).
The mobility entropy exhibits negative and strong correlations with the emissions of all pollutants and cities, with the highest coefficients found in London ($-0.76$ for NO$_x$ and VOC) and the lowest in Florence ($-0.29$ for CO$_2$). 
Since a low mobility entropy indicates a predictable mobility behaviour, these negative correlations suggest that gross polluters tend to be more regular and predictable than low-emitting vehicles. 
Finally, the correlations with the radius of gyration are almost null for London (even though the coefficients found for NO$_x$, PM and VOC are not statistically significant). 
They are positive for Rome (CO$_2$ $0.58$, NO$_x$ $0.36$, PM $0.28$, VOC $0.48$) and Florence (CO$_2$ $0.30$, NO$_x$ $0.13$, VOC $0.09$), with the only exception of PM that exhibits a negative correlation ($-0.47$).
In conclusion, with a few exceptions, gross polluters travel the most and have the most predictive travelling behaviour.

We deepen this analysis using a Generalised Additive Model (GAM)~\cite{hastie1986generalized}, a non-parametric regression technique that is a generalisation of the Generalised Linear Models (GLMs).
We choose a GAM because we aim to understand our predictors' effects (the radius of gyration, the entropy, and the travel time) on the response variable (emissions), which may be non-linear. 
Being the GAM an additive model, the interpretation of the marginal effect of a single predictor on the response variable does not depend on the values of the other predictors.
Moreover, a GAM allows control for the smoothness of the predictor functions and, thus, control overfitting.
We can write the GAM as:
\begin{linenomath*}
    \begin{equation}
    g(E[e]) = f_1(r_g) + f_2(S) + f_3(T) + \beta_0
    \end{equation}
\end{linenomath*}
where $e$ is the amount of emissions (and $E[e]$ its expected value), $r_g$ is the radius of gyration, $S$ is the temporal-uncorrelated entropy, $T$ is the travel time, and $\beta_0$ is the intercept. 
$f_1$ and $f_2$ are non-parametric smooth functions (penalised B-splines). 
We assume $f_3$ to be a linear function, as the relation that the emissions have with the travel time can be reasonably assumed to be linear. 
Finally, $g$ is the so-called link function that connects the expected value of the response variable to the predictors.
Essentially, the splines $f_1$, $f_2$ are functions expressed as a linear combination of a finite set of basis functions. 
By controlling for the number of these basis functions, we can control for the smoothness of $f_1$ and $f_2$: the higher the number of basis functions, the smoother the resulting spline. 
Moreover, as we use penalised B-splines, the overfitting can also be controlled through a penalisation parameter. 
Given a certain number of basis functions for the splines, we choose the best parameter through cross-validation.

In Supplementary Figures \ref{fig:GAM_london}, \ref{fig:GAM_rome}, and \ref{fig:GAM_florence} we show, for each city, the partial dependence plots from four different GAMs obtained by reducing the number of basis functions for the smoothers $f_1$, $f_2$ from 20 to 5.
We observe that, at the price of a worse performance, reducing the number of basis functions reduces overfitting and provides better interpretability of each predictor's marginal effect on the response variable. 
Indeed, as both the $GCV$ and $AIC$ increase, the pseudo-$R^2$ slowly decreases.
We also observe that the radius and the entropy contribute in an opposite way to determining a vehicle's emissions.
On the one hand, for Rome and Florence, the greater the typical distance a vehicle travels (i.e., its radius of gyration), the greater its emissions.  
In London, starting from a radius value of around 7, its role in determining the emissions becomes constant (see Supplementary Figure \ref{fig:GAM_london}).
On the other hand, in Rome and Florence, the greater a vehicle's entropy, the lower its emissions: the emissions decrease when the entropy increases, except for the interval $[0.6, 0.8]$, where the opposite thing happens (see Supplementary Figure x). 
In London, the entropy has a negative contribution to emissions (i.e. the more entropy, the fewer emissions) starting from $0.7$; for lower values, it has a slightly positive impact on emissions.

Finally, we perform a cluster analysis of the vehicles with respect to their mobility measures, and observe the level of emissions in each of the resulting clusters. We find two clusters of vehicles, that we name erratic and predictable drivers, and investigate the typical emissions in each cluster (see Supplementary Figure \ref{fig:kmeans}). We use the GPS trajectories describing all the trips made during 2017 by all the vehicles with at least one trip in each season of the year. For each city, we compute each vehicle’s radius of gyration, mobility entropy, and travel time, as well as its total emissions. Then, we use $k$-means~\cite{tan2016introduction} for clustering the vehicles with respect to their radius of gyration, entropy and travel time. We repeat the clustering for values of $k$ ranging from 1 to 10 and select the best $k$ with the "elbow method''~\cite{tan2016introduction}. For all cities, we find that $k=2$ is the best choice (Supplementary Figures \ref{fig:kmeans}a,b,c). We name these two clusters \textit{erratic drivers}, those with high values of mobility entropy (usually higher than 0.7), and \textit{predictable drivers}, whose entropy is lower (Supplementary Figures \ref{fig:kmeans}d,e,f). 
There are 669 \textit{erratic drivers} in London, 2462 in Rome, and 1877 in Florence, while there are 1705 \textit{predictable drivers} in London, 2474 in Rome, and 1027 in Florence.
We find that the level of emissions of the \textit{predictable drivers} is higher in all the cities: the average CO$_2$ emitted by these latter is 6.40 times higher than that of the \textit{erratic drivers} in London, 2.33 in Rome, and 2.41 in Florence (Supplementary Figure \ref{fig:kmeans}g,h,i). We find similar results for the other pollutants (see Supplementary Figures \ref{fig:kmeans}j-\ref{fig:kmeans}r). Thus, the clustering analysis confirms what we find with the Spearman correlations and GAMs: the most erratic vehicles emit less than the more predictable ones.

Concerning the relation between the quantities of air pollutants emitted on the roads, their (betweenness) centrality w.r.t. the entire network, and their length, we find positive correlation coefficients for all air pollutants and cities (Supplementary Figures \ref{fig:scatter_roads_london}, \ref{fig:scatter_roads_rome}, \ref{fig:scatter_roads_florence}, and Supplementary Tables \ref{tab:corrs_NOx}, \ref{tab:corrs_PM}, \ref{tab:corrs_VOC}).
The road's betweenness centrality is the frequency with which it falls on the shortest paths connecting two crossroads in the network, and it can be considered a proxy of the traffic volume it hosts. 
Given this, our results suggest that the grossly polluted roads are more likely to be the longest and most congested in the road network.

\subsection*{Supplementary Note 8: emissions reduction simulations.}
\label{supp:simulations}
Supplementary Figure \ref{fig:electrification} shows the reduction of the CO$_2$ emitted overall by the vehicles in London and Florence when a share of them are zero-emitting vehicles (e.g., electric vehicles).
We compute this reduction in two cases: when the share of vehicles to be electrified are chosen starting from the most polluting ones (and proceeding in decreasing order), and when they are chosen at random. 
In London, the reduction of emissions in the first case is $\approx 6$ times more effective than with the random choice (Supplementary Figure \ref{fig:electrification}a). 
This means that electrifying the top $1\%$ most polluting vehicles leads to a $\approx 6\%$ reduction of the overall emissions of CO$_2$. 
In contrast, a random $1\%$ electric vehicles reduces $\approx 1\%$ in the CO$_2$ emissions.
In Florence, this difference is more considerable: electrifying the top $1\%$ most polluting vehicles is $\approx 10$ times more effective than electrifying $1\%$ vehicles chosen at random (Supplementary Figure \ref{fig:electrification}b).
We find that the growth of this reduction is well described by a Generalised Logistic Function (GLF, or Richard's curve). 
Indeed, we use non-linear least squares to fit a \textit{GLF} $f(x) = \frac{\alpha}{(1 + \beta e^{-rx})^{1/\nu}}$ to the reduction of the overall emissions of CO$_2$ when the electrified vehicles are chosen in decreasing order starting from the most polluting ones. We set the starting values for the parameters following Fekedulegn et al.~\cite{fekedulegn1999parameter}.
The parameters of the curve change with the city (Supplementary Table \ref{tab:logistic_fit_electrification}); in particular, its slope is $-0.86$ for London and $-1.91$ in Florence, as this latter experiences a faster growth in the reduction of overall CO$_2$ emissions obtained by electrifying the most polluting vehicles.

We also simulate the impact of a massive shift to home working on the reduction of vehicles’ emissions in the three cities (Supplementary Figure \ref{fig:homeworking}). 
We find that the reduction of the emissions is more effective when the home workers are gross polluters: in this case, the remote working of the top $1\%$ gross polluters leads to the same reduction reached if they were $\approx4\%$ random vehicles. 
Again, the growth of this reduction is well described by a GLF, whose parameters slightly change with the city (see Supplementary Table \ref{tab:logistic_fit_homeworking}).


\clearpage

\section{Supplementary Figures}
\label{supp:figures}

\begin{figure}
    \centering
    \includegraphics[width=0.9\textwidth]{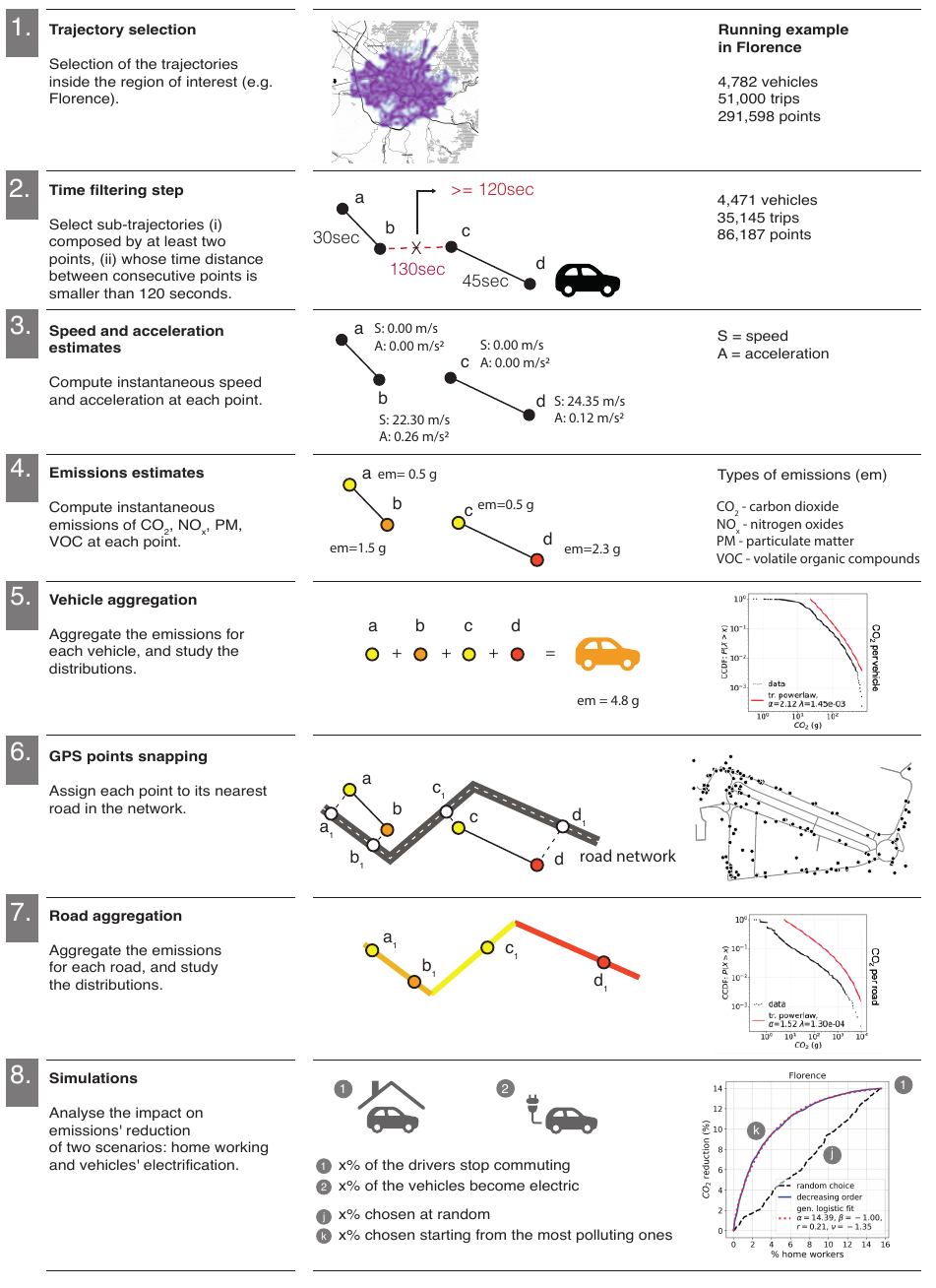}
    \caption{\textbf{Step-by-step procedure for the computation of emissions and results analyses}.
    The left column describes the steps followed starting from the data, passing through the data processing, and ending with the analyses performed. The central column shows a schema of what happens in each step. The right column shows some numbers and results in support of the central column.
    The heatmap in step 1. is plotted with the Python library scikit-mobility~\cite{pappalardo2019scikitmobility}. The small road network in step 7. is plotted with the Python library OSMnx~\cite{boeing2017}.}
    \label{fig:framework}
\end{figure}

\begin{figure}
    \centering
    \includegraphics[width=\textwidth]{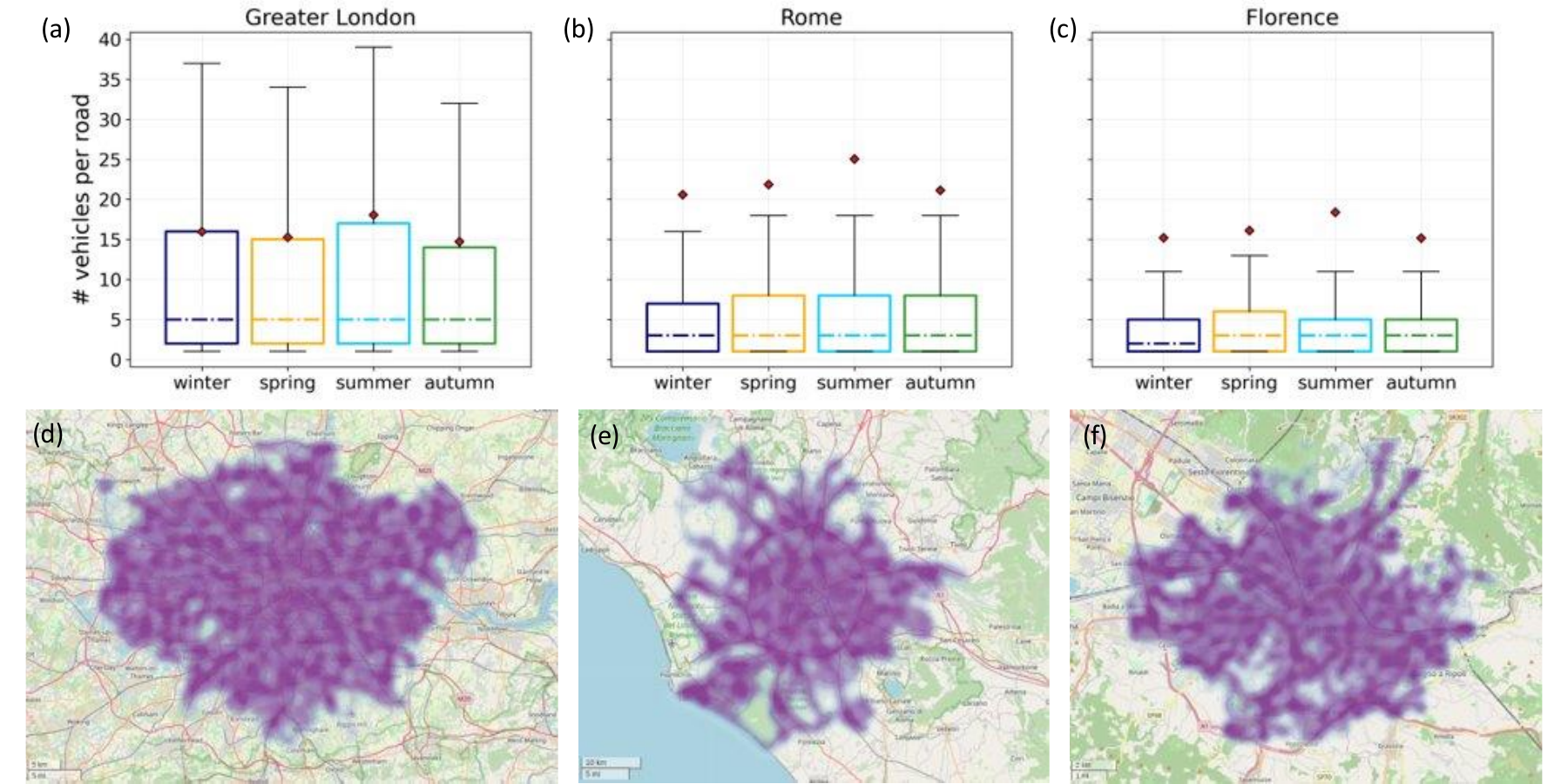}
    \caption{\textbf{Spatio-temporal patterns of the vehicles' trajectories}.
    Panels (a), (b), (c) show the distribution of the number of distinct vehicles passing through a road during the winter (blue), spring (yellow), summer (light blue) and autumn (green) of 2017, for Greater London (a), Rome (b), and Florence (c). The mean of each distribution is indicated with the red diamond, and the scale of the y-axis is the same for the three panels.
    Panels (d), (e), (f) show the spatial distribution of the GPS points forming the trajectories of the vehicles for the three cities in 2017.}
    \label{fig:traj}
\end{figure}


\begin{figure}
    \centering
    \includegraphics[width=0.9\textwidth]{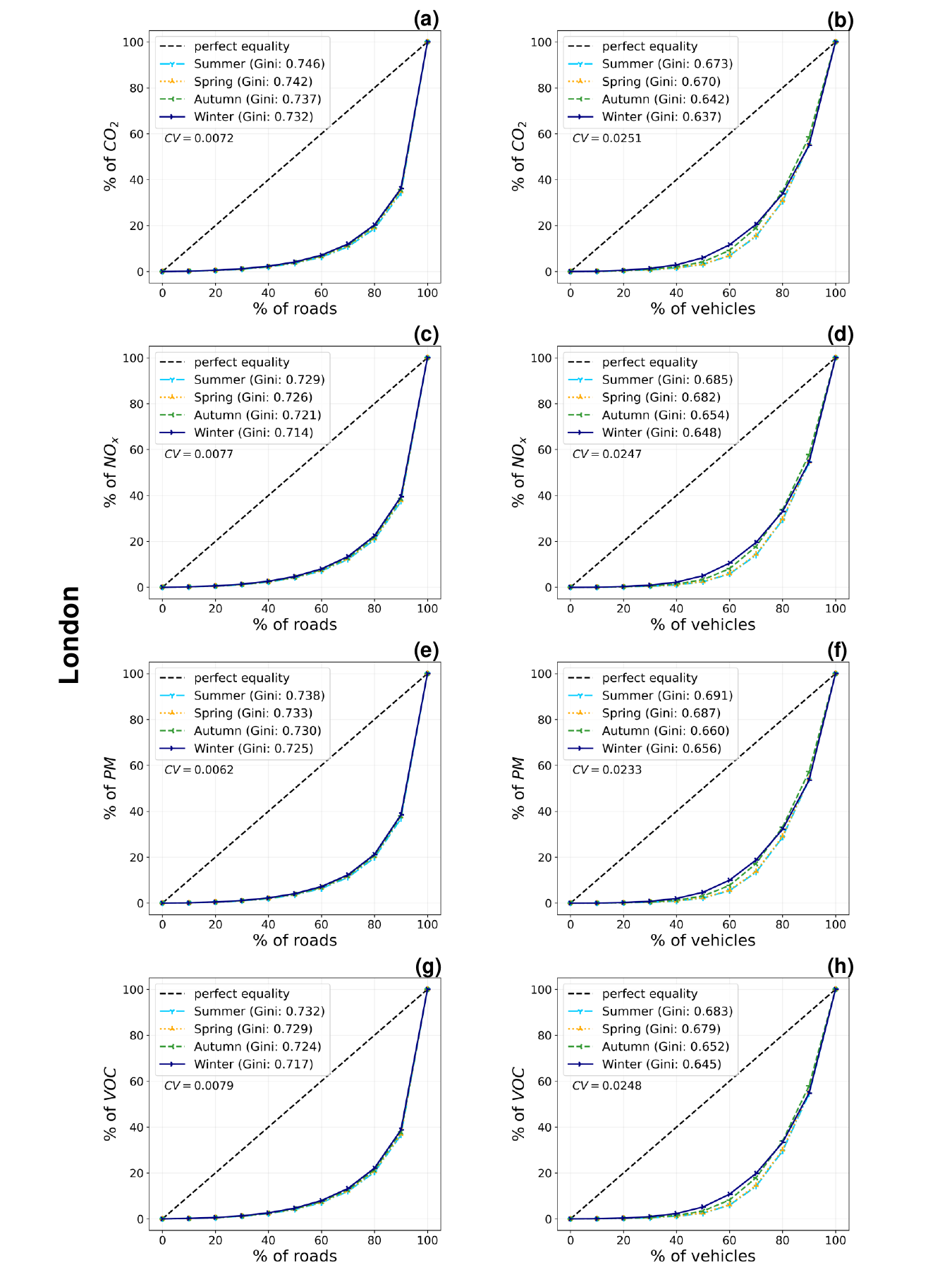}
    \caption{\textbf{Distributions of emissions across vehicles and roads in different seasons of the year (Greater London)}.
    Lorenz curves showing the share of overall emissions associated respectively to the bottom $x\%$ of the roads (left) and to the bottom $x\%$ of the vehicles (right) in Greater London, in spring (yellow curve), summer (light blue curve), autumn (green curve), and winter (blue curve), and for CO$_2$ (a, b), NO$_x$ (c, d), PM (e, f), and VOC (g, h). The black dashed line indicates a uniform distribution. 
    We show in the legend the Gini coefficient for each curve. Above the legend, we also show their coefficient of variation (CV), i.e. the ratio of their standard deviation to their mean.}
    \label{fig:seasons_gini_london}
\end{figure}

\begin{figure}
    \centering
    \includegraphics[width=0.9\textwidth]{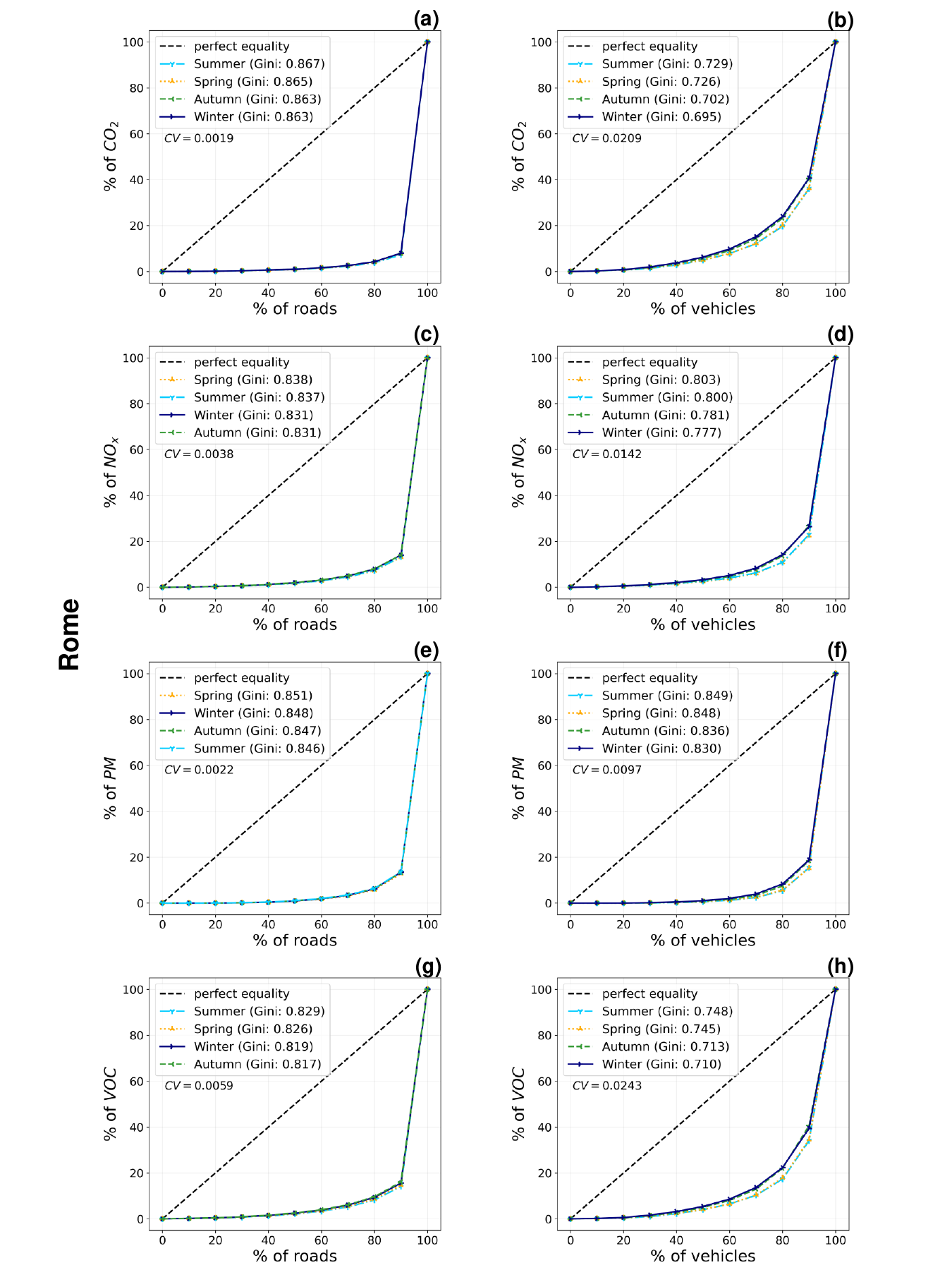}
    \caption{\textbf{Distributions of emissions across vehicles and roads in different seasons of the year (Rome)}.
    Lorenz curves showing the share of overall emissions associated respectively to the bottom $x\%$ of the roads (left) and to the bottom $x\%$ of the vehicles (right) in Rome, in spring (yellow curve), summer (light blue curve), autumn (green curve), and winter (blue curve), and for CO$_2$ (a, b), NO$_x$ (c, d), PM (e, f), and VOC (g, h). The black dashed line indicates a uniform distribution. 
    We show in the legend the Gini coefficient for each curve, and also, above the legend, their coefficient of variation (CV), i.e. the ratio of their standard deviation to their mean.}
    \label{fig:seasons_gini_rome}
\end{figure}

\begin{figure}
    \centering
    \includegraphics[width=0.9\textwidth]{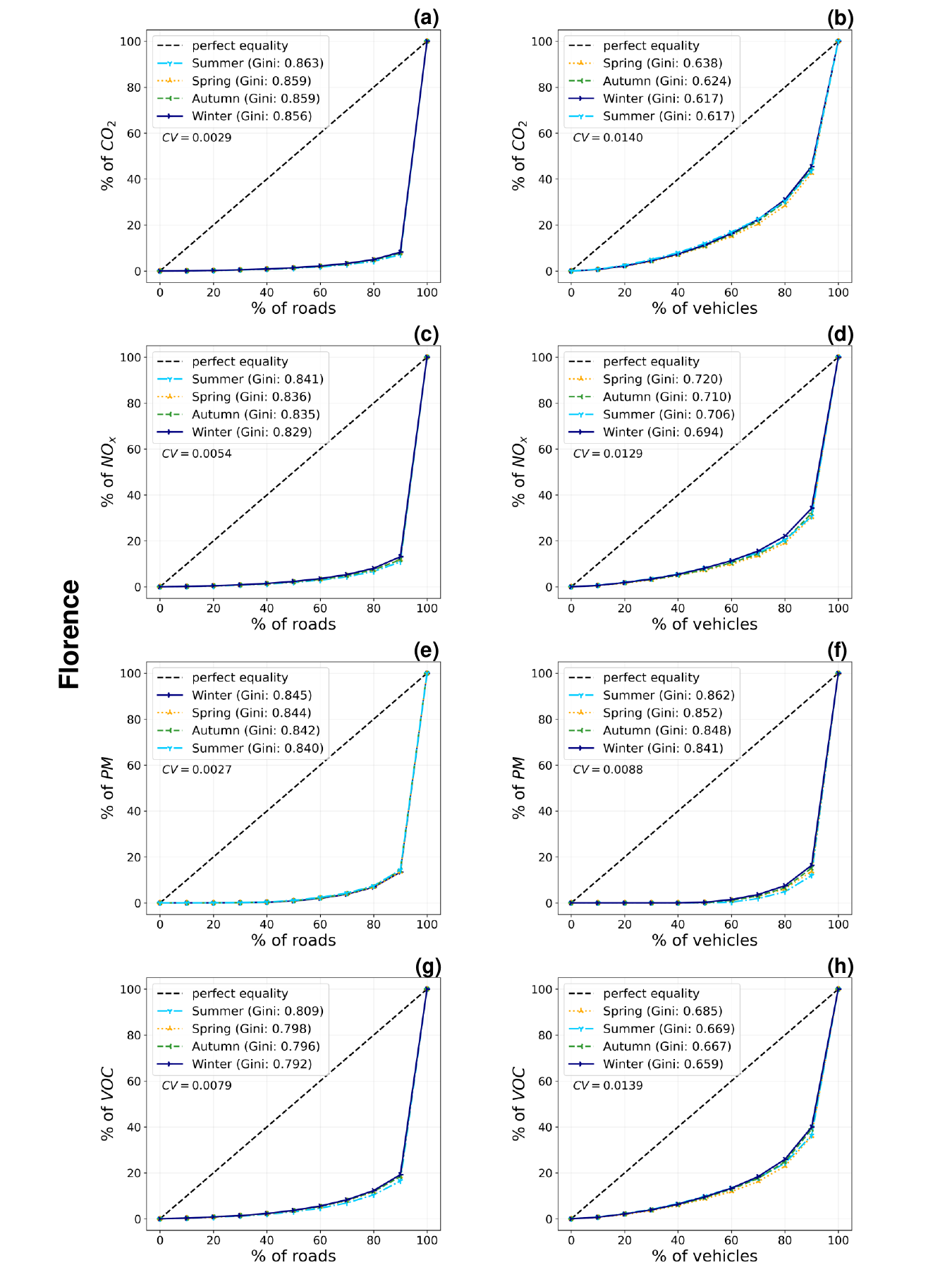}
    \caption{\textbf{Distributions of emissions across vehicles and roads in different seasons of the year (Florence)}.
    Lorenz curves showing the share of overall emissions associated respectively to the bottom $x\%$ of the roads (left) and to the bottom $x\%$ of the vehicles (right) in Florence, in spring (yellow curve), summer (light blue curve), autumn (green curve), and winter (blue curve), and for CO$_2$ (a, b), NO$_x$ (c, d), PM (e, f), and VOC (g, h). The black dashed line indicates a uniform distribution. 
    We show in the legend the Gini coefficient for each curve, and also, above the legend, their coefficient of variation (CV), i.e. the ratio of their standard deviation to their mean.}
    \label{fig:seasons_gini_florence}
\end{figure}

\begin{figure}
    \centering
    \includegraphics[width=\textwidth]{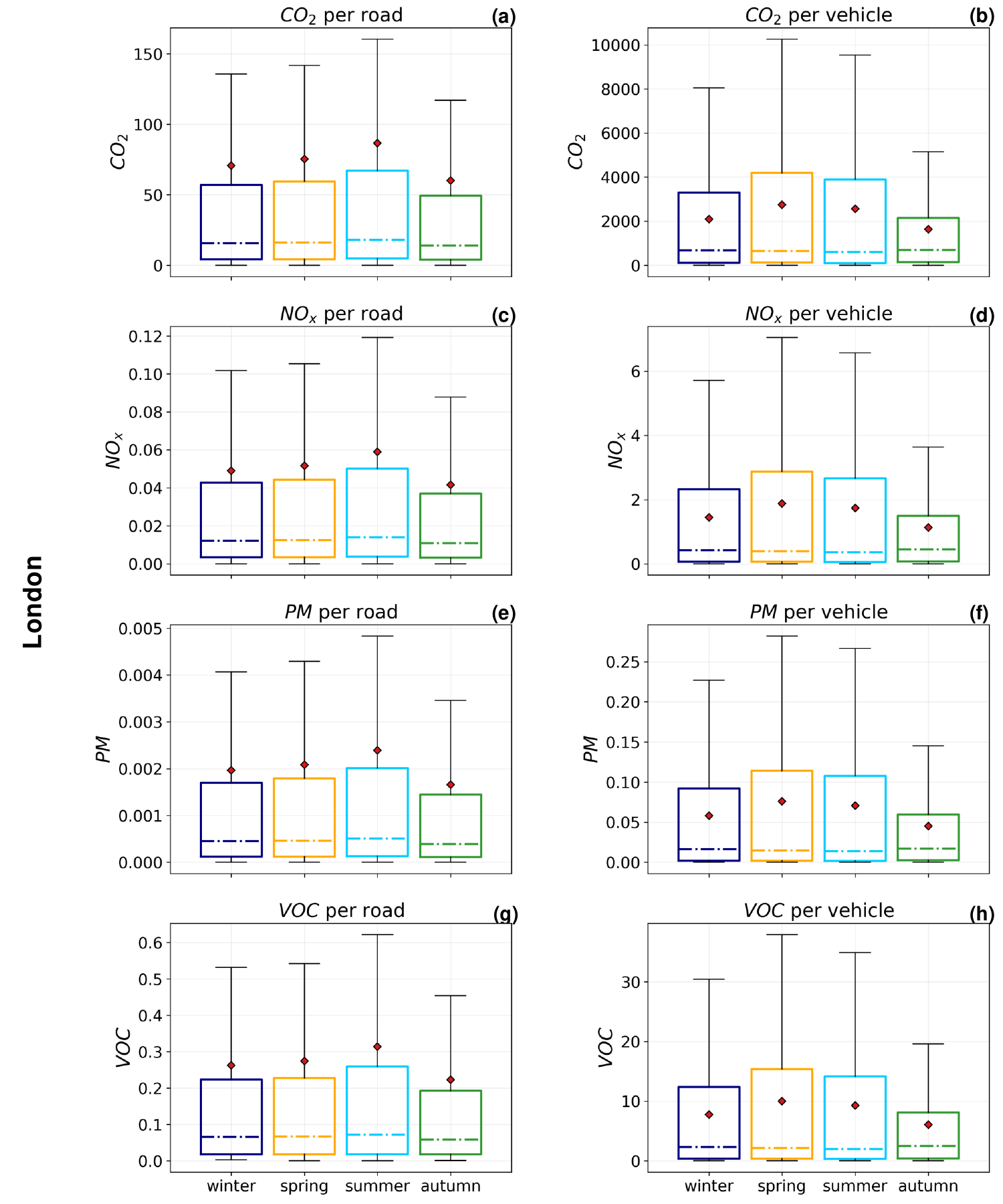}
    \caption{\textbf{Emissions' distributions across the four seasons of the year in Greater London}.
    The boxplots of the CO$_2$ (a, b), NO$_x$ (c, d), PM (e, f) and VOC (g, h) emissions’ distributions across vehicles (right) and roads (left) in the four seasons of the year, for the city of Greater London. The mean of each distribution is indicated with the red diamond. The long tails of the distributions are not shown here as they would flatten the boxplots on the x-axis.}
    \label{fig:seasons_boxplots_london}
\end{figure}

\begin{figure}
    \centering
    \includegraphics[width=\textwidth]{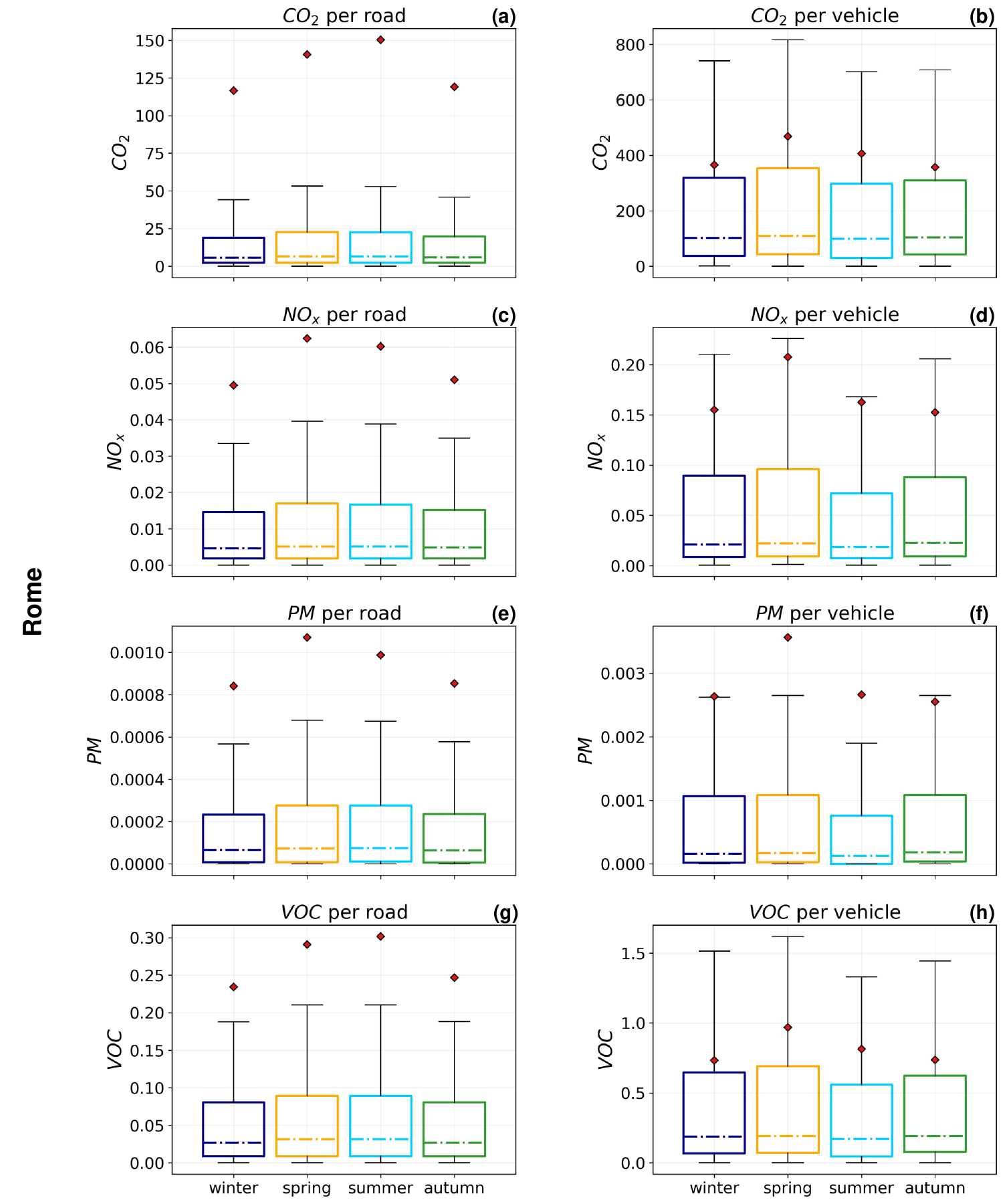}
    \caption{\textbf{Emissions' distributions across the four seasons of the year in Rome}.
    The boxplots of the CO$_2$ (a, b), NO$_x$ (c, d), PM (e, f) and VOC (g, h) emissions’ distributions across vehicles (right) and roads (left) in the four seasons of the year, for the city of Rome. The mean of each distribution is indicated with the red diamond. The long tails of the distributions are not shown here as they would flatten the boxplots on the x-axis.}
    \label{fig:seasons_boxplots_rome}
\end{figure}

\begin{figure}
    \centering
    \includegraphics[width=\textwidth]{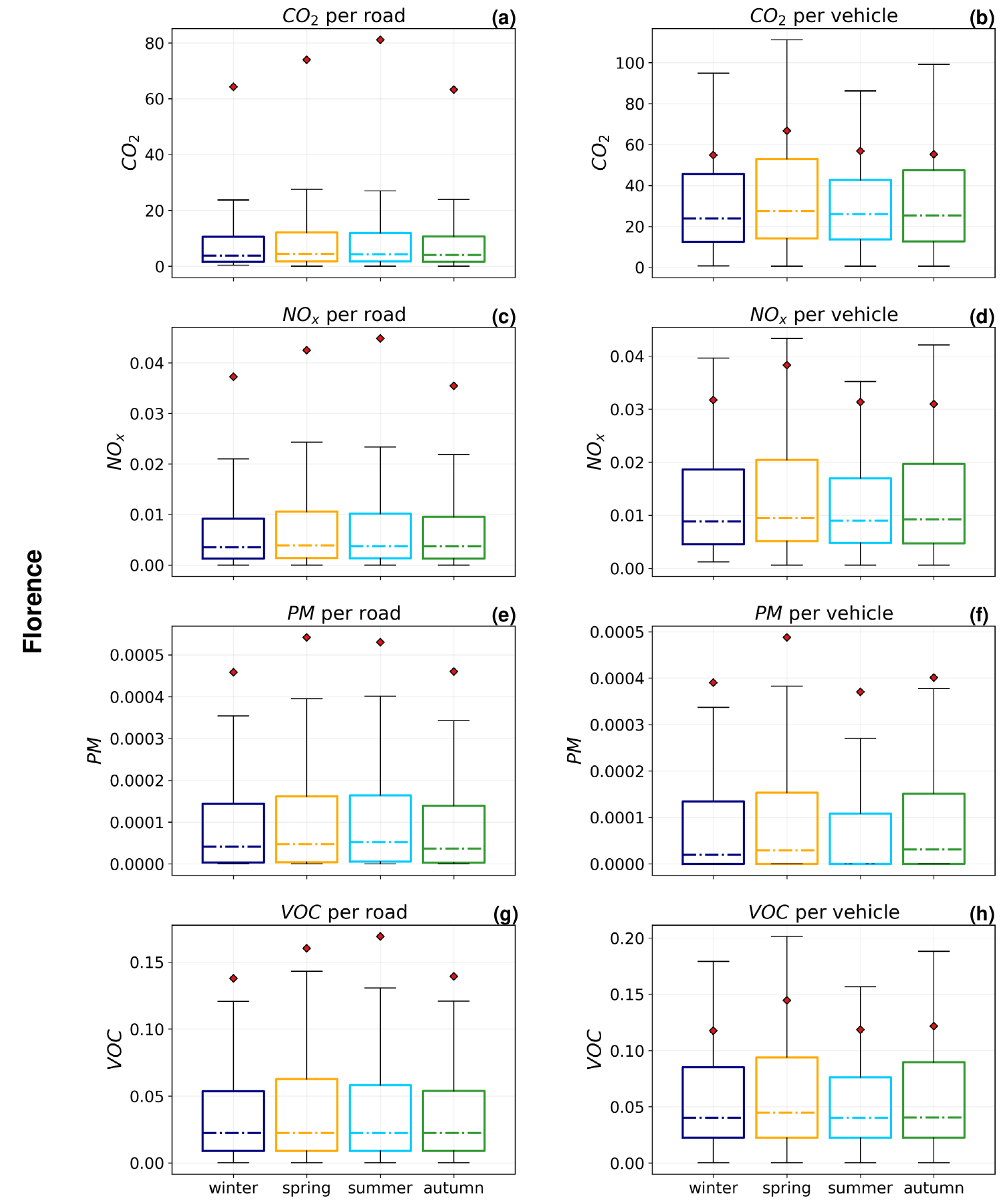}
    \caption{\textbf{Emissions' distributions across the four seasons of the year in Florence}.
   The boxplots of the CO$_2$ (a, b), NO$_x$ (c, d), PM (e, f) and VOC (g, h) emissions’ distributions across vehicles (right) and roads (left) in the four seasons of the year, for the city of Florence. The mean of each distribution is indicated with the red diamond. The long tails of the distributions are not shown here as they would flatten the boxplots on the x-axis.}
    \label{fig:seasons_boxplots_florence}
\end{figure}

\begin{figure}
    \centering
    \includegraphics[width=\textwidth]{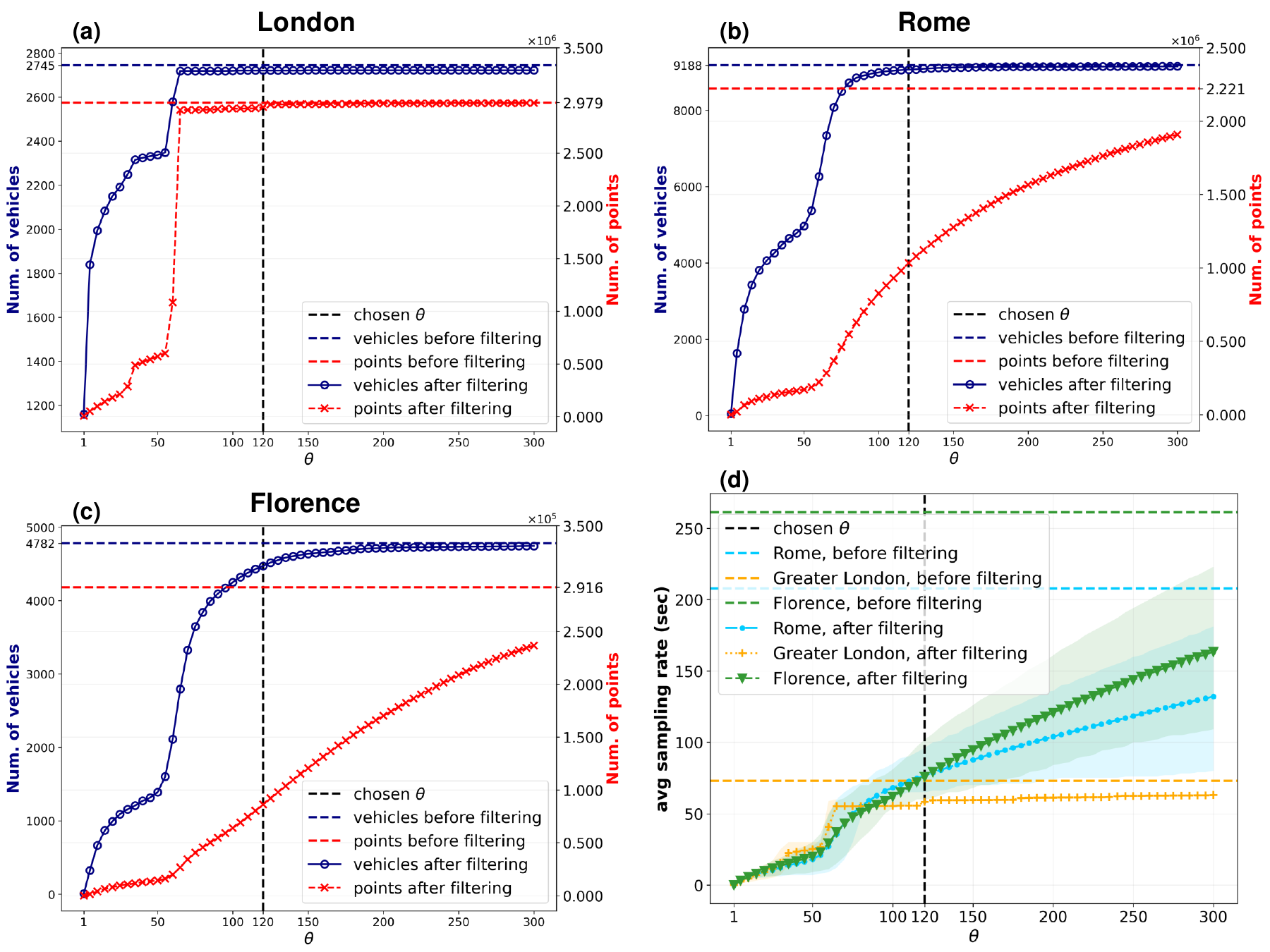}
    \caption{\textbf{Effects of changing the sampling rate parameter $\theta$ in the filtering step}.
    Figures (a), (b), and (c) show the number of vehicles (solid blue line) and points (solid red line) surviving the filtering process when changing $\theta$ (x-axis) from $1$ to $300$ seconds, for Greater London, Rome and Florence, respectively. 
    The horizontal dashed blue and red lines indicate the number of vehicles and points present in each dataset before the filtering. 
    The vertical dashed black line shows the value of $\theta$ used in our experiments.
    (d) Average sampling rate of the points after the filtering process when $\theta$ grows in the same interval $[1,300]$ seconds for the three cities. 
    The horizontal dashed yellow, light blue and green lines indicate the average sampling rate of the points before the filtering for Greater London, Rome, and Florence, respectively. The vertical dashed black line shows the value of $\theta$ used in our experiments.}
    \label{fig:exp_sampling_rate}
\end{figure}

\begin{figure}
    \centering
    \includegraphics[width=0.95\textwidth]{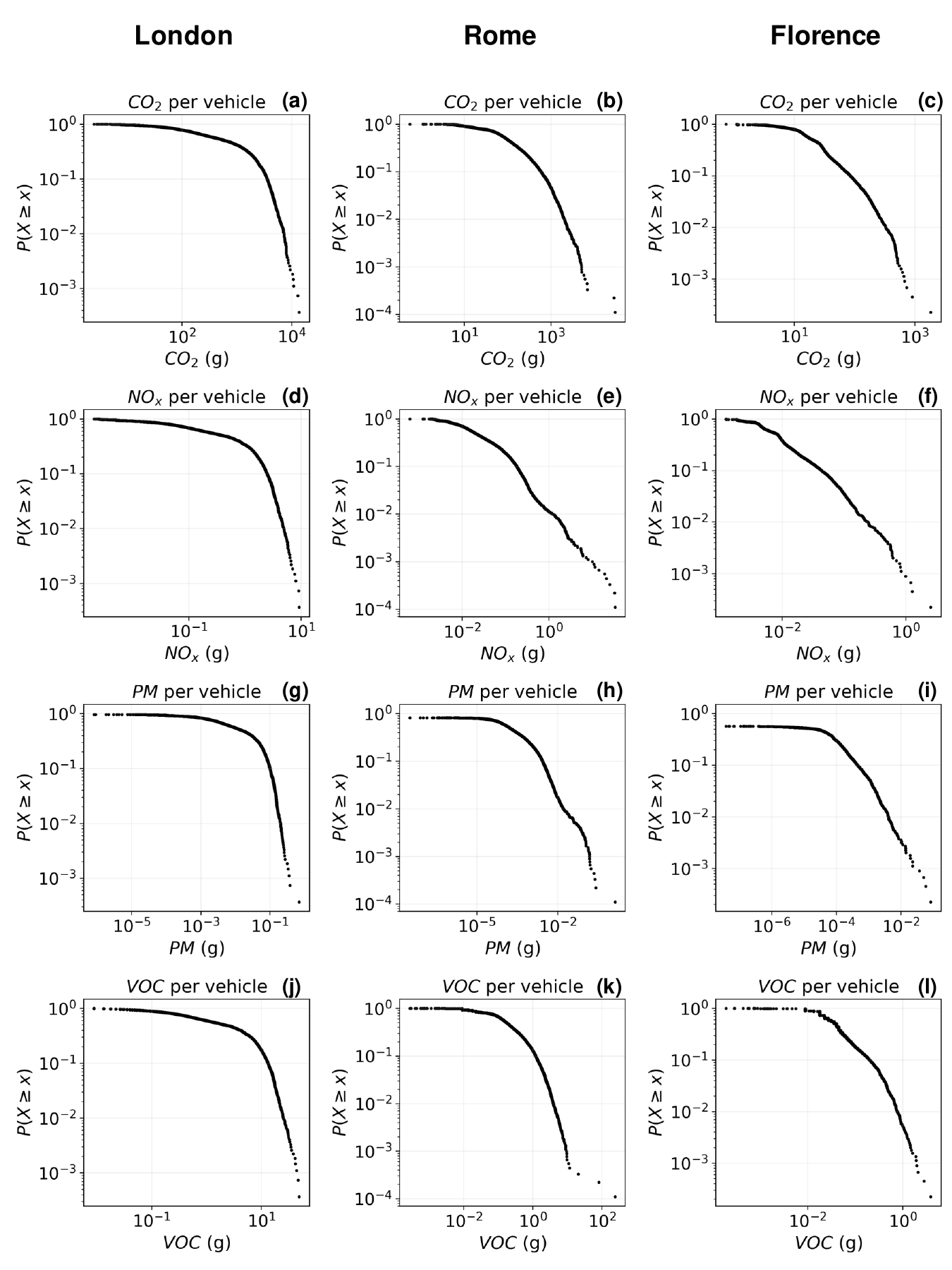}
    \caption{\textbf{Empirical Complementary Cumulative Distribution Function (CCDF) of the emissions per vehicle}.
    For each pollutant and city, we show the empirical CCDF of the data. For each quantity $x$ emitted by a vehicle, the CCDF is the probability of getting a vehicle that emitted more than $x$.
    The scale is logarithmic in both the axes.
    The shape of the distribution is similar across the cities and air pollutants.}
    \label{fig:ccdf_vehicle}
\end{figure}

\begin{figure}
    \centering
    \includegraphics[width=0.95\textwidth]{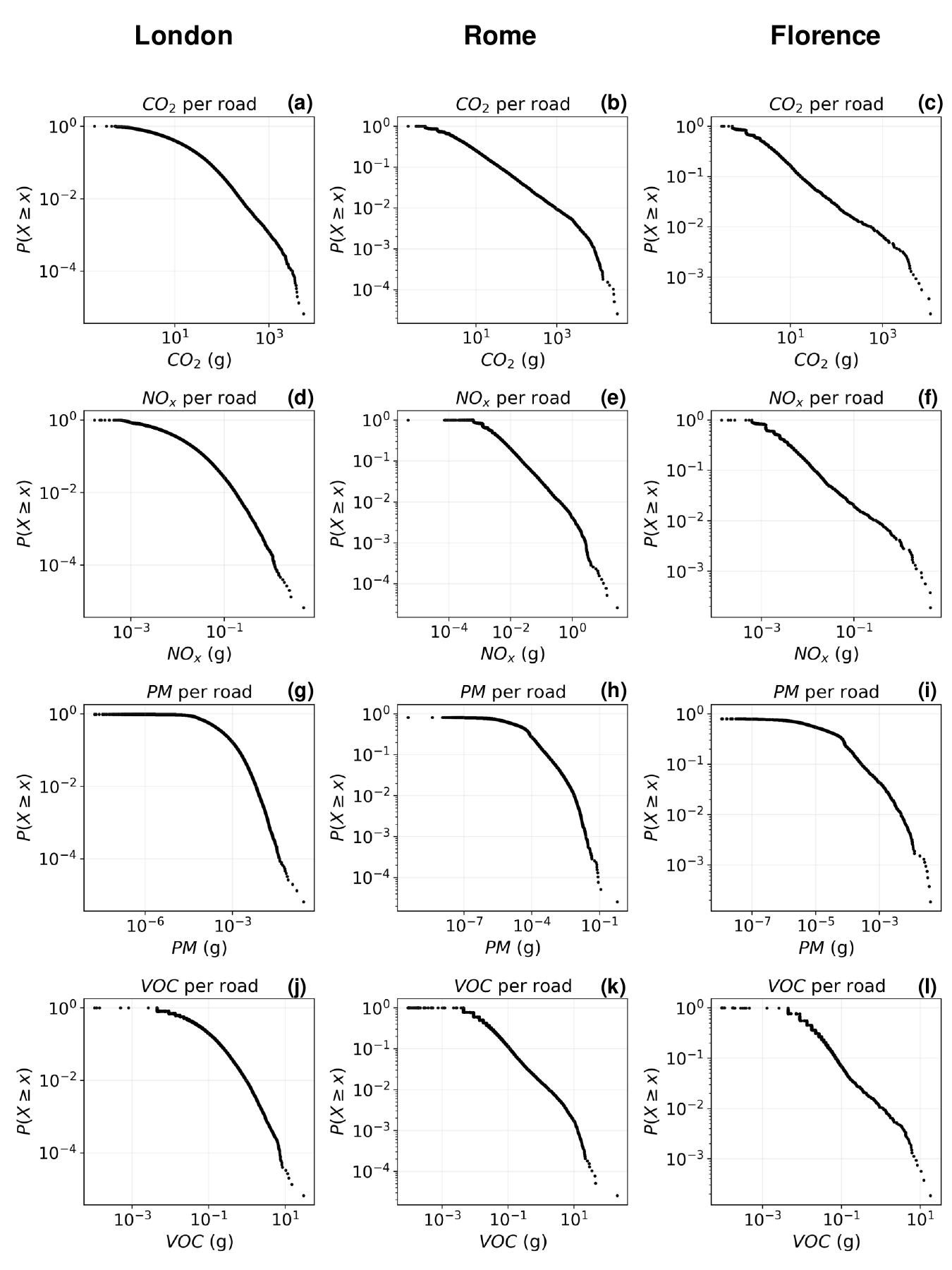}
    \caption{\textbf{Empirical Complementary Cumulative Distribution Function (CCDF) of the emissions per road}.
    For each pollutant and city, we show the empirical CCDF of the data. For each quantity $x$ emitted on a road, the CCDF is the probability of getting a road that has more emissions than $x$.
    The scale is logarithmic in both the axes.
    The shape of the distribution is similar across the cities and air pollutants.}
    \label{fig:ccdf_road}
\end{figure}

\begin{figure}
    \centering
    \includegraphics[width=0.9\textwidth]{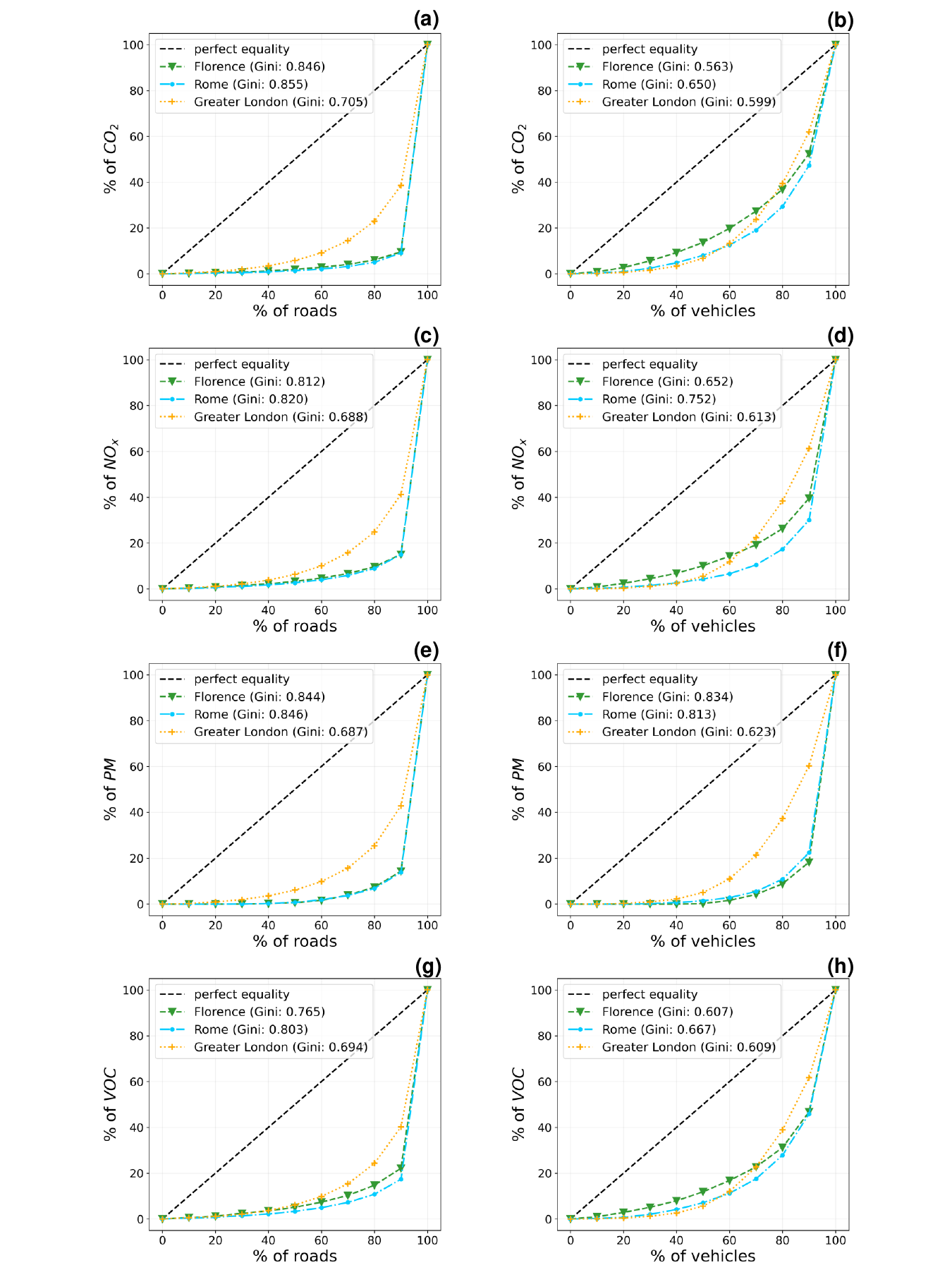}
    \caption{\textbf{Distributions of emissions across vehicles and roads}. 
    Lorenz curves showing the share of overall emissions associated respectively to the bottom $x\%$ of the roads (left) and to the bottom $x\%$ of the vehicles (right), for Greater London (yellow curve), Rome (light blue curve), and Florence (green curve), and for CO$_2$ (a, b), NO$_x$ (c, d), PM (e, f), and VOC (g, h). The black dashed line indicates a uniform distribution. 
    We show in the legend the Gini coefficient for each curve.}
    \label{fig:lorenz}
\end{figure}

\begin{figure}
    \centering
    \includegraphics[width=0.9\textwidth]{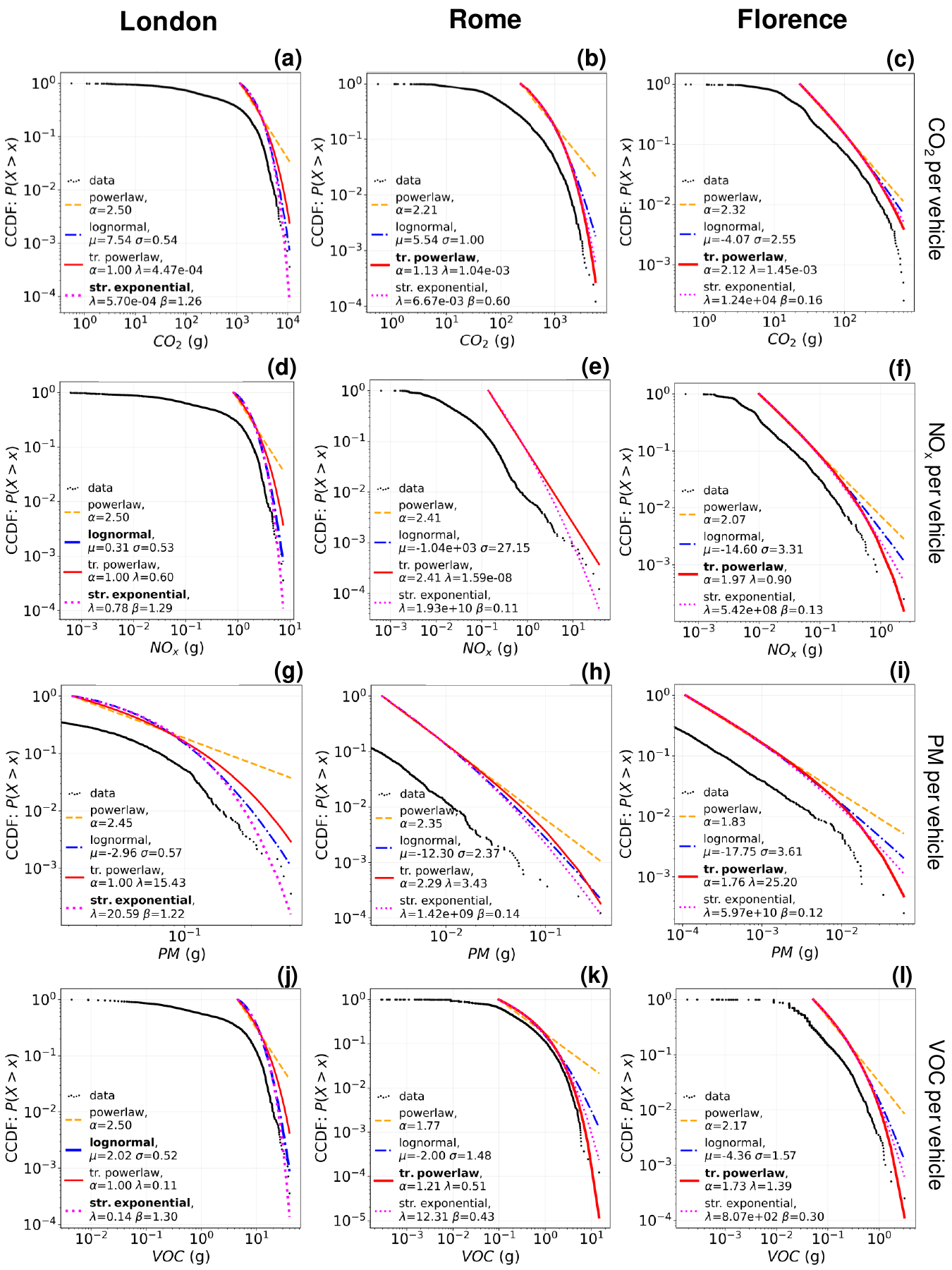}
    \caption{\textbf{Fitting of the distributions of emissions per vehicle}.
    For each pollutant and city, we show the Complementary Cumulative Distribution Function (CCDF) of the emissions per vehicle (black dots), together with the fitted distributions: power law (dashed orange curve), log-normal (dashed blue curve), truncated power law (solid red curve), stretched exponential (dotted purple curve).
    In the legend, we highlight the best fit(s), if any, in bold.
     We use a maximum-likelihood fitting method and evaluate the goodness-of-fit using the Kolmogorov-Smirnov distance, comparing different models with a log-likelihood ratio test.}
    \label{fig:all_fits_per_vehicle}
\end{figure}

\begin{figure}
    \centering
    \includegraphics[width=0.9\textwidth]{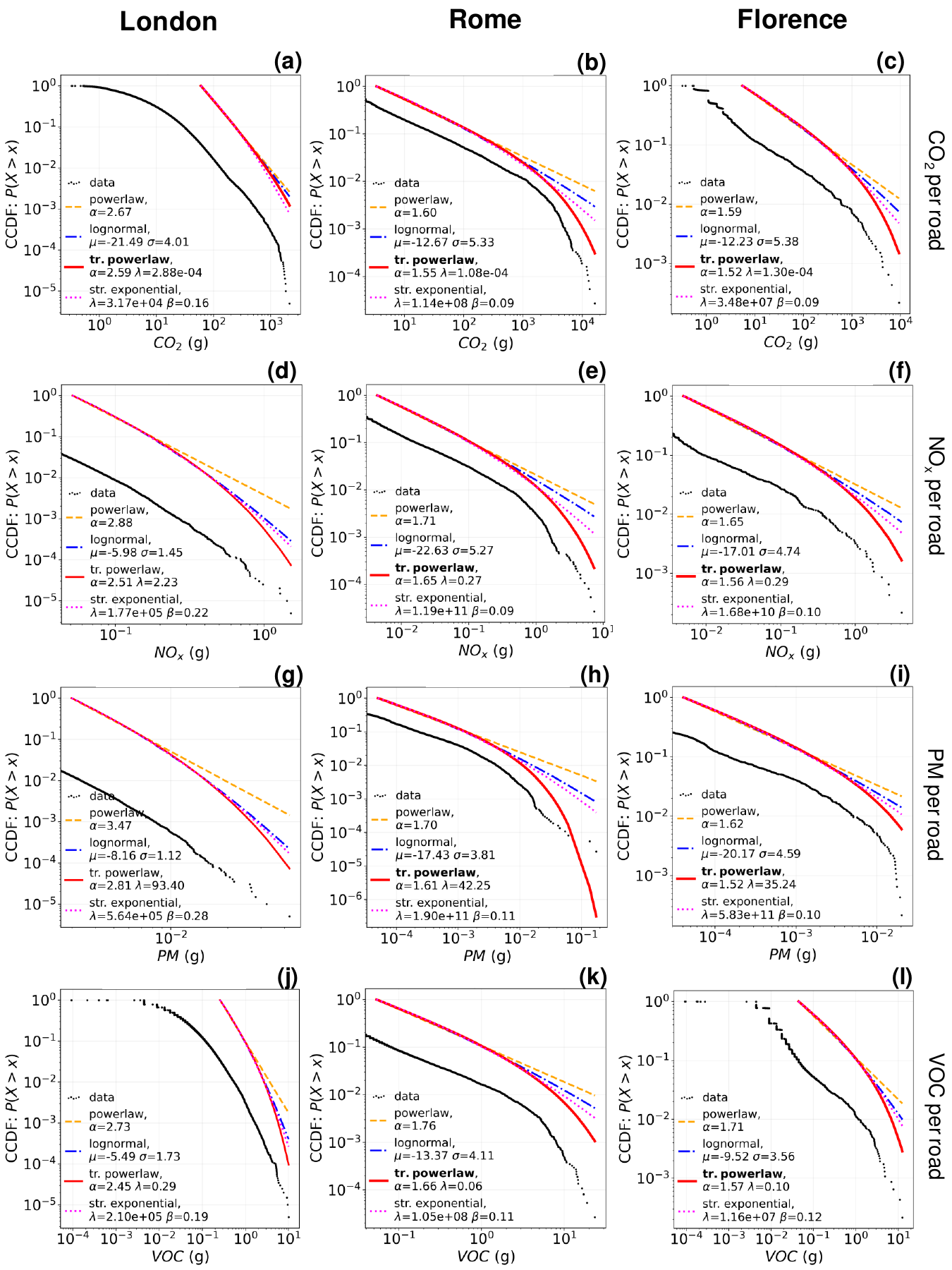}
    \caption{\textbf{Fitting of the distributions of emissions per road}.
    For each pollutant and city, we show the Complementary Cumulative Distribution Function (CCDF) of the emissions per road (black dots), together with the fitted distributions: power law (dashed orange curve), log-normal (dashed blue curve), truncated power law (solid red curve), stretched exponential (dotted purple curve).
    In the legend, we highlight the best fit(s), if any, in bold. 
    We use a maximum-likelihood fitting method and evaluate the goodness-of-fit using the Kolmogorov-Smirnov distance, comparing different models with a log-likelihood ratio test.}
    \label{fig:all_fits_per_road}
\end{figure}

\begin{figure}
    \centering
    \includegraphics[width=\textwidth]{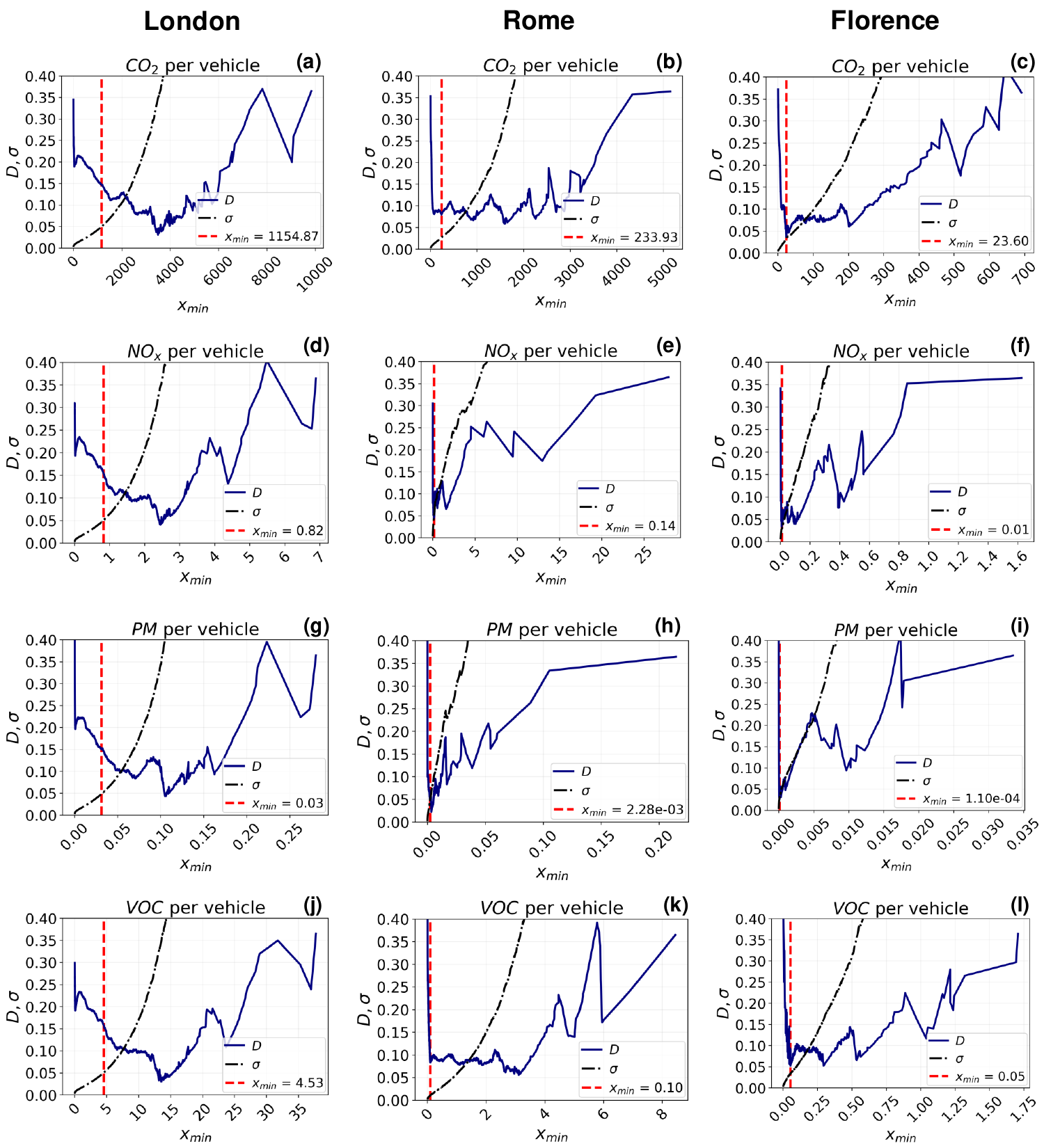}
    \caption{\textbf{Choice of $x_{min}$ when fitting the distributions of emissions per vehicle}.
    For each pollutant and city, we show the values of the Kolmogorov-Smirnov distance $D$ between the data and the power law fit (blue solid line) and the standard error $\sigma$ of the power law exponent (black dashed line) for each possible choice of $x_{min}$. The value of $x_{min}$ that minimises $D$ under the constraint $\sigma < 0.05$ is indicated by the red dashed vertical line.}
    \label{fig:xmin_vehicle}
\end{figure}

\begin{figure}
    \centering
    \includegraphics[width=\textwidth]{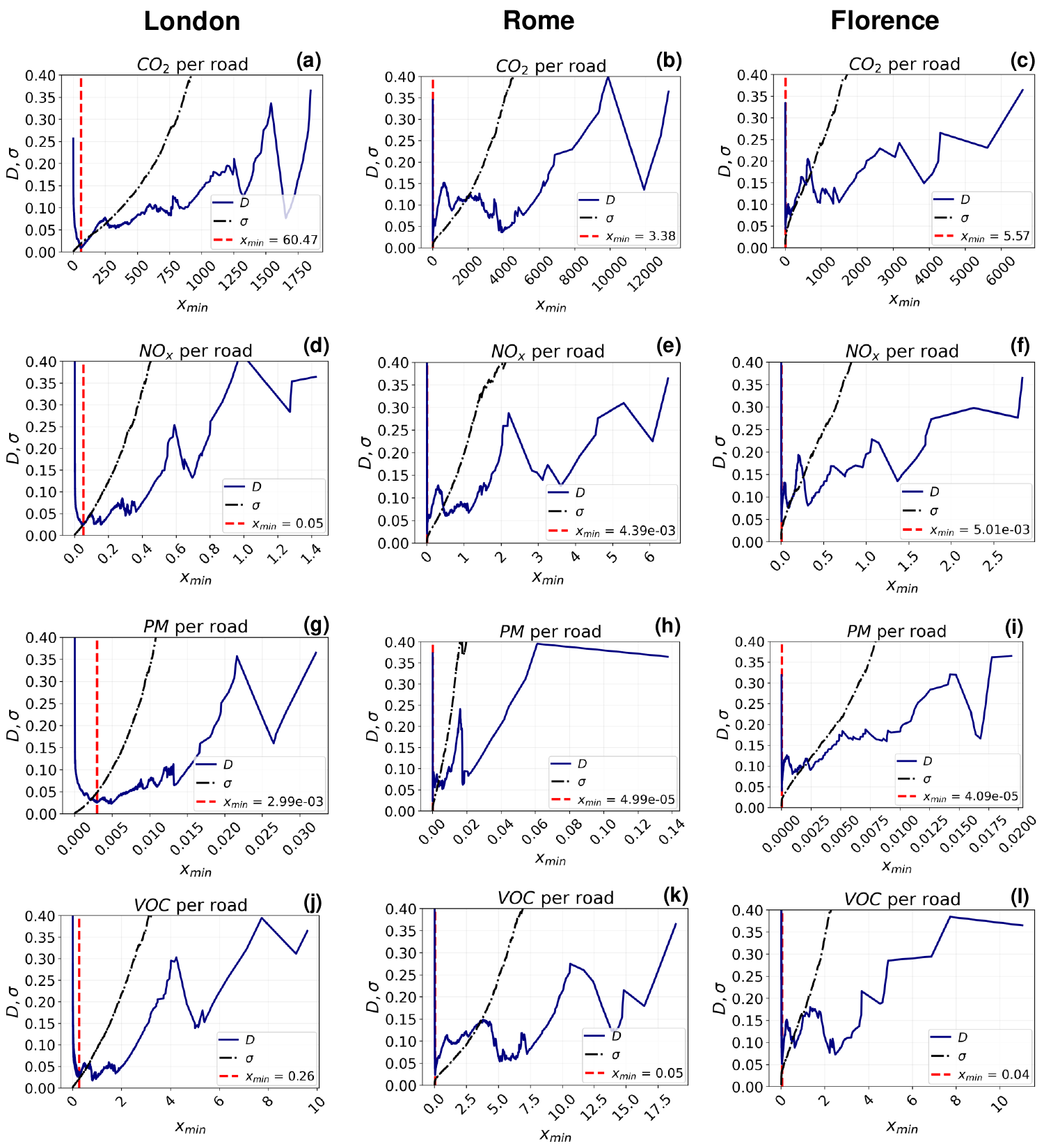}
    \caption{\textbf{Choice of $x_{min}$ when fitting the distributions of emissions per road}.
    For each pollutant and city, we show the values of the Kolmogorov-Smirnov distance $D$ between the data and the power law fit (blue solid line) and the standard error $\sigma$ of the power law exponent (black dashed line) for each possible choice of $x_{min}$. The value of $x_{min}$ that minimises $D$ under the constraint $\sigma < 0.05$ is indicated by the red dashed vertical line.}
    \label{fig:xmin_road}
\end{figure}

\begin{figure}
    \centering
    \includegraphics[width=\textwidth]{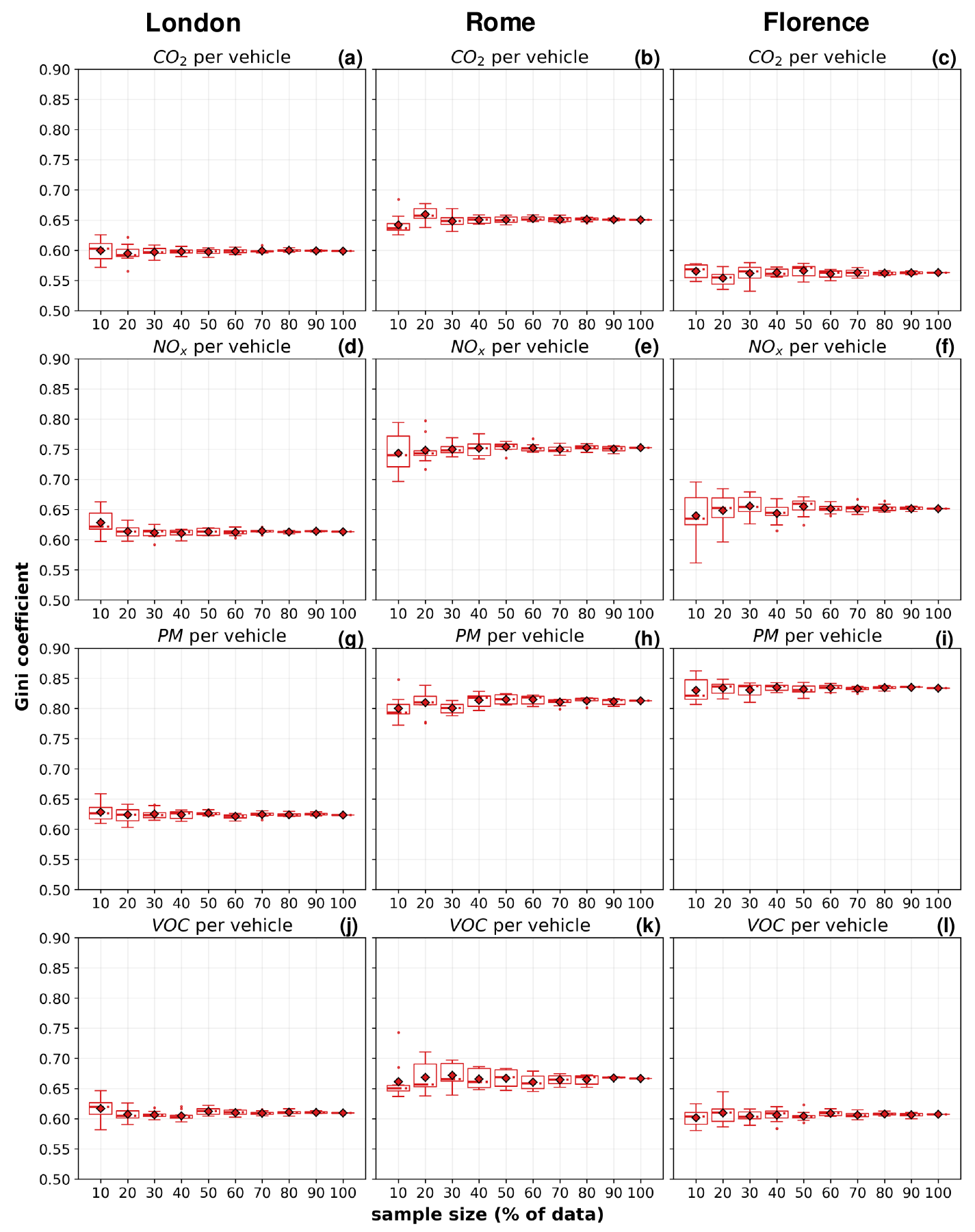}
    \caption{\textbf{Distributions of the Gini coefficient computed on different sample sizes of the emissions across the vehicles}.
    The distribution of the Gini coefficients of the emissions' distributions of CO$_2$ (a ,b ,c), NO$_x$ (d, e, f), PM (g, h, i) and VOC (j, k, l) across the vehicles when the sample dimension grows from the 10\% to 100\% of the available data, for the three cities.}
    \label{fig:sample_exp_vehicle}
\end{figure}

\begin{figure}
    \centering
    \includegraphics[width=\textwidth]{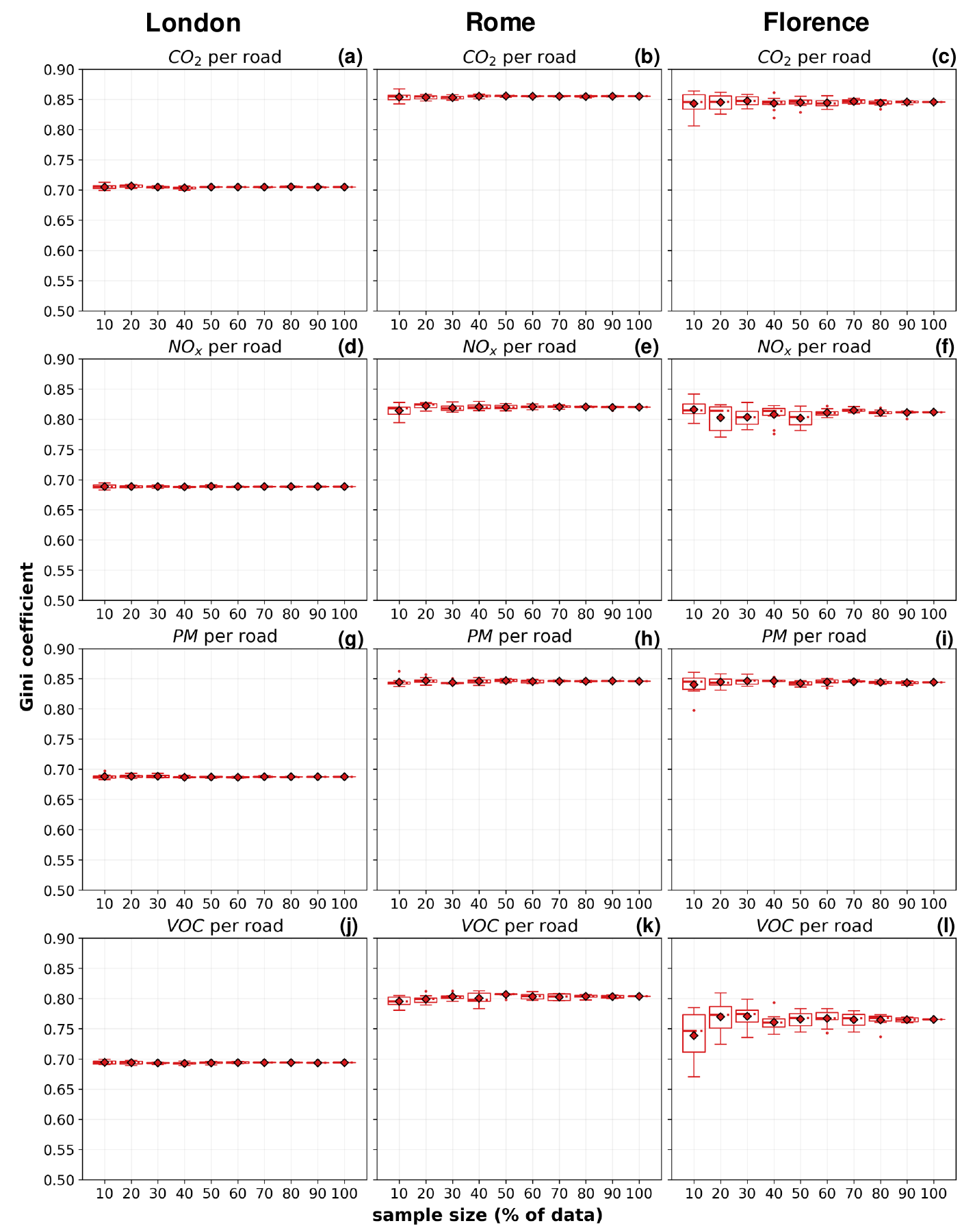}
    \caption{\textbf{Distributions of the Gini coefficient computed on different sample sizes of the emissions across the roads}.
    The distribution of the Gini coefficients of the emissions' distributions of CO$_2$ (a ,b ,c), NO$_x$ (d, e, f), PM (g, h, i) and VOC (j, k, l) across the roads when the sample dimension grows from the 10\% to 100\% of the available data, for the three cities.}
    \label{fig:sample_exp_road}
\end{figure}

\begin{figure}
    \centering
    \includegraphics[width=0.9\textwidth]{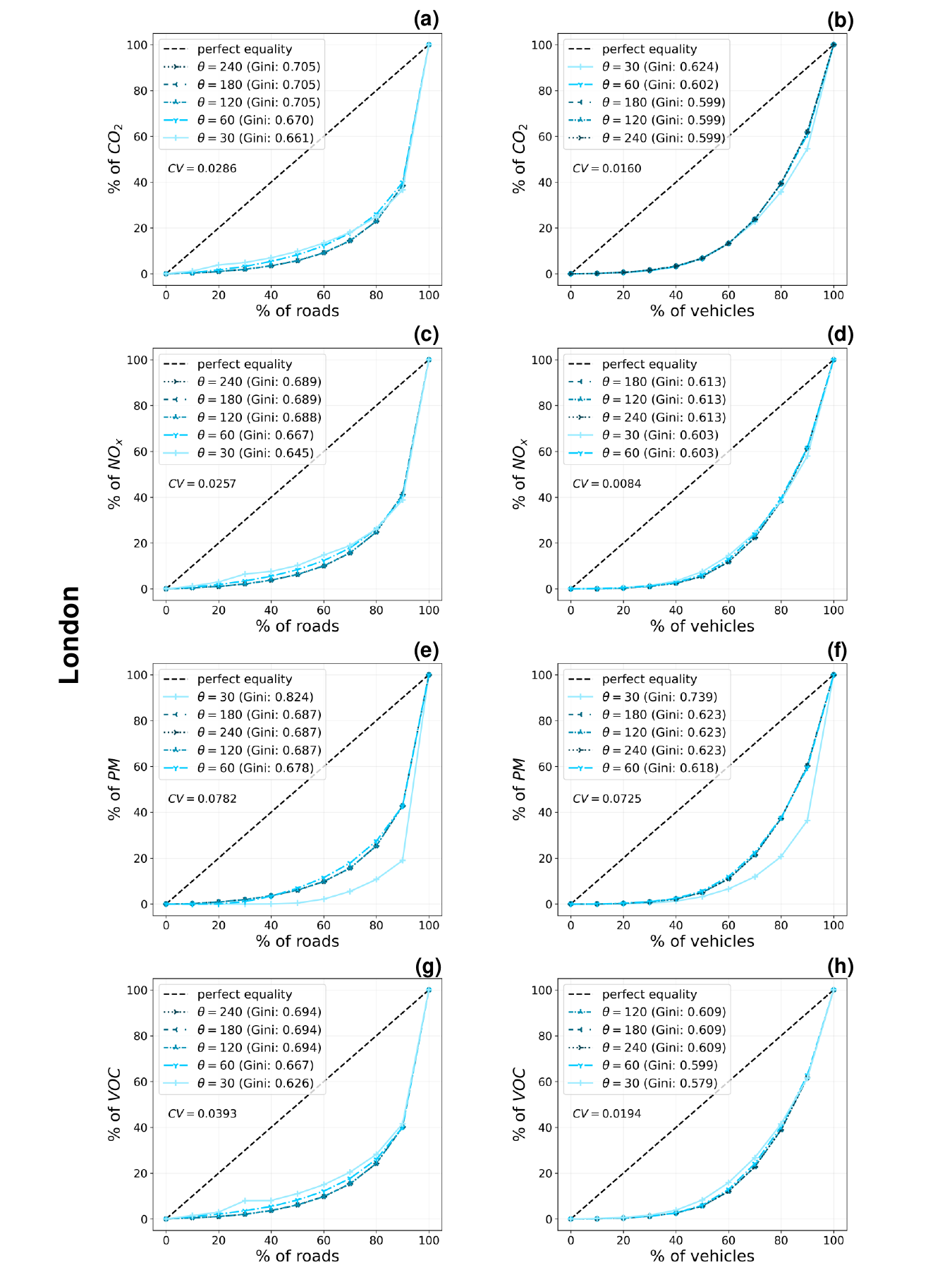}
    \caption{\textbf{Distributions of emissions across vehicles and roads for different choices of $\theta$ (Greater London)}.
    Lorenz curves showing the share of overall emissions associated respectively to the bottom $x\%$ of the roads (left) and to the bottom $x\%$ of the vehicles (right) in Greater London, for $\theta$ equal to 30, 60, 120, 180, and 240 seconds (from the lighter to the darker blue curves, respectively), and CO$_2$ (a, b), NO$_x$ (c, d), PM (e, f), and VOC (g, h). The black dashed line indicates a uniform distribution. 
    We show in the legend the Gini coefficient for each curve. Above the legend, we also show their coefficient of variation (CV), i.e. the ratio of their standard deviation to their mean.}
    \label{fig:theta_gini_london}
\end{figure}

\begin{figure}
    \centering
    \includegraphics[width=0.9\textwidth]{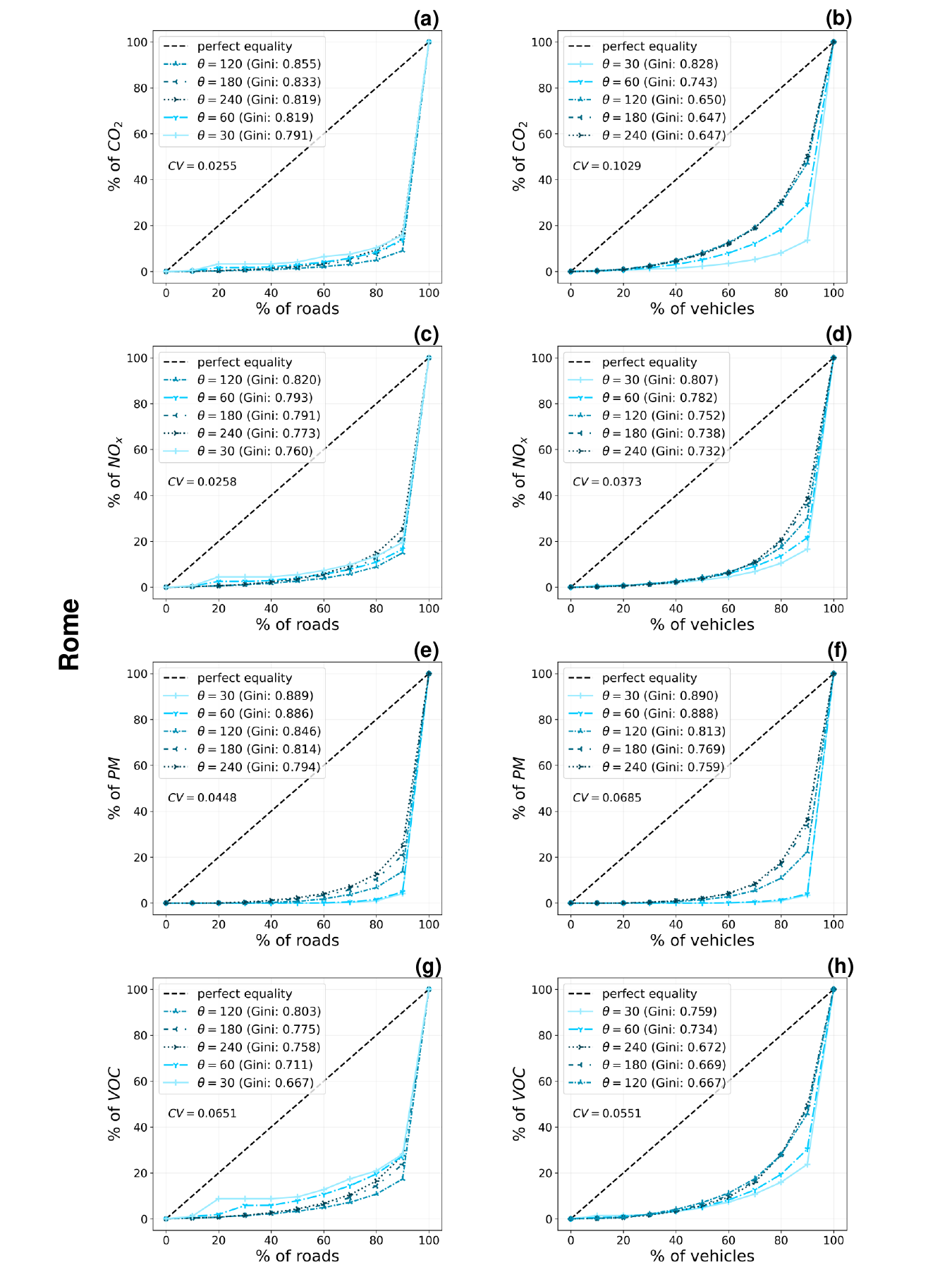}
    \caption{\textbf{Distributions of emissions across vehicles and roads for different choices of $\theta$ (Rome)}.
    Lorenz curves showing the share of overall emissions associated respectively to the bottom $x\%$ of the roads (left) and to the bottom $x\%$ of the vehicles (right) in Rome, for $\theta$ equal to 30, 60, 120, 180, and 240 seconds (from the lighter to the darker blue curves, respectively), and CO$_2$ (a, b), NO$_x$ (c, d), PM (e, f), and VOC (g, h). The black dashed line indicates a uniform distribution. 
    We show in the legend the Gini coefficient for each curve. Above the legend, we also show their coefficient of variation (CV), i.e. the ratio of their standard deviation to their mean.}
    \label{fig:theta_gini_rome}
\end{figure}

\begin{figure}
    \centering
    \includegraphics[width=0.9\textwidth]{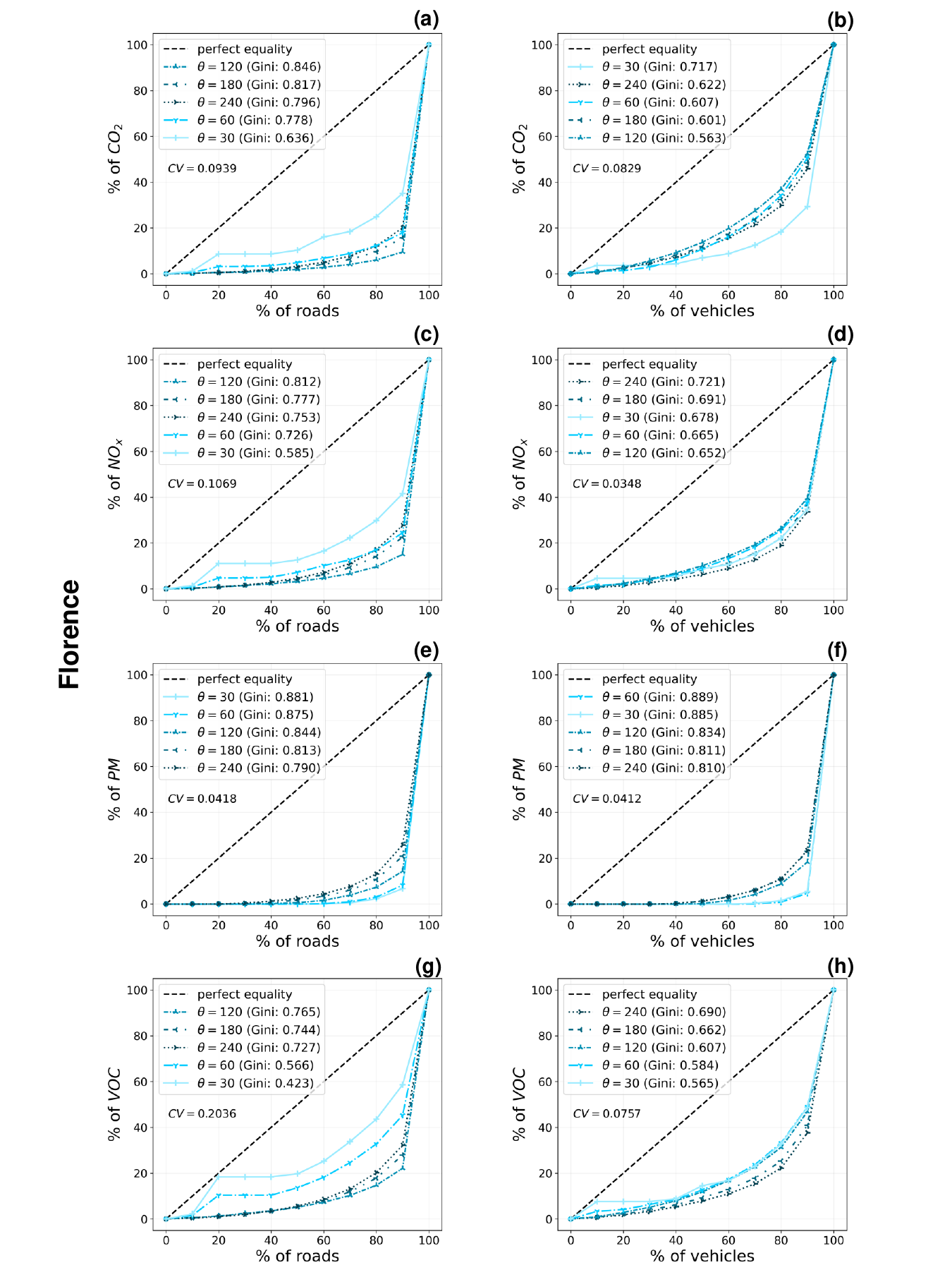}
    \caption{\textbf{Distributions of emissions across vehicles and roads for different choices of $\theta$ (Florence)}.
    Lorenz curves showing the share of overall emissions associated respectively to the bottom $x\%$ of the roads (left) and to the bottom $x\%$ of the vehicles (right) in Florence, for $\theta$ equal to 30, 60, 120, 180, and 240 seconds (from the lighter to the darker blue curves, respectively), and CO$_2$ (a, b), NO$_x$ (c, d), PM (e, f), and VOC (g, h). The black dashed line indicates a uniform distribution. 
    We show in the legend the Gini coefficient for each curve. Above the legend, we also show their coefficient of variation (CV), i.e. the ratio of their standard deviation to their mean.}
    \label{fig:theta_gini_florence}
\end{figure}

\begin{figure}
    \centering
    \includegraphics[width=\textwidth]{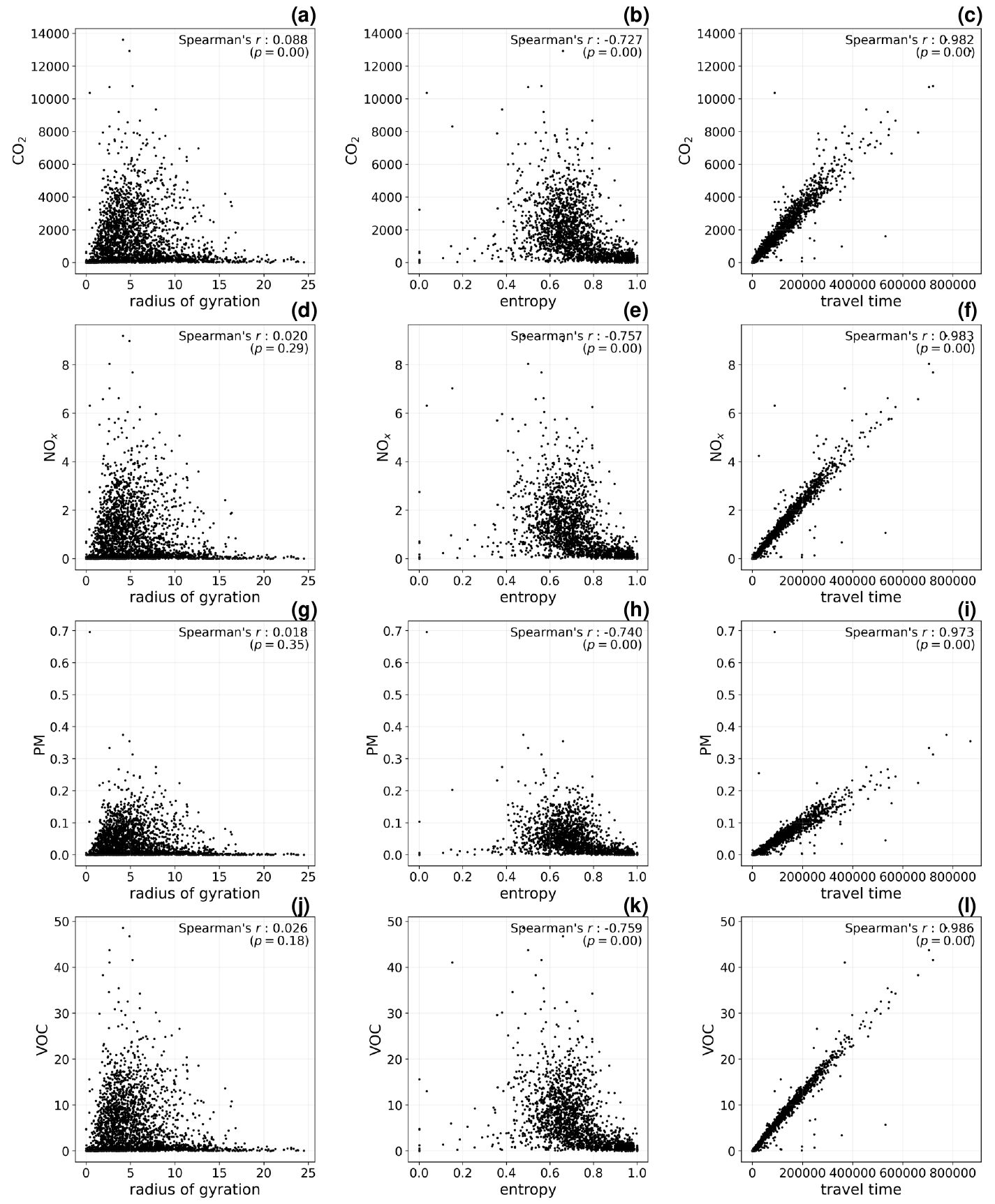}
    \caption{\textbf{Correlations between emissions and mobility metrics for London.} 
    The scatter plots show the relationship between the emissions of the four pollutants (rows) and three mobility metrics of the vehicles (columns): the radius of gyration, temporal-uncorrelated entropy, and travel time.
    On the upper right corner of each figure we show the Spearman's correlation coefficient, and the corresponding $p$-value.}
    \label{fig:scatter_mobility_london}
\end{figure}

\begin{figure}
    \centering
    \includegraphics[width=\textwidth]{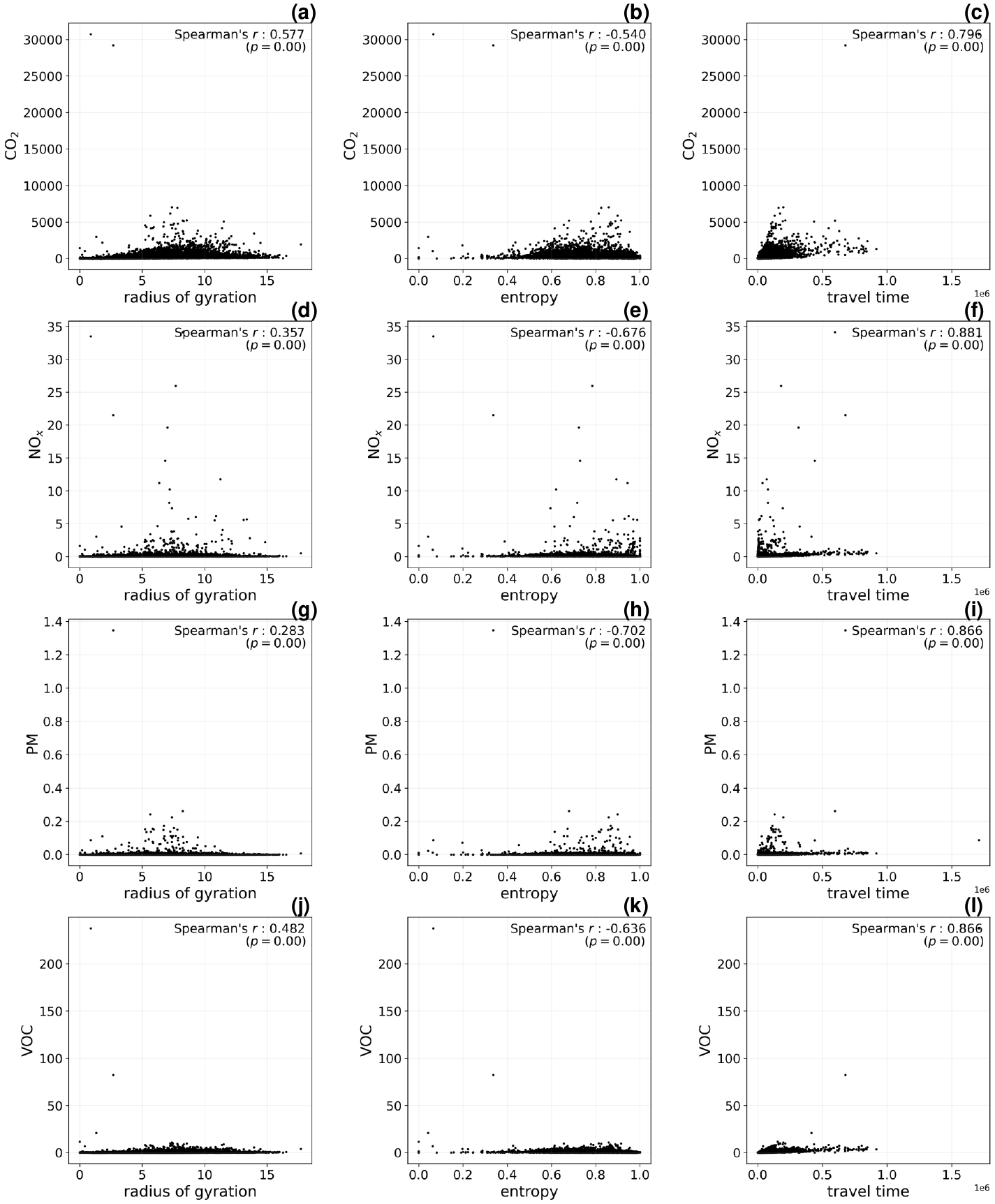}
    \caption{\textbf{Correlations between emissions and mobility metrics for Rome.} 
    The scatter plots show the relation between the emissions of the four pollutants (rows) and three mobility metrics of the vehicles (columns): the radius of gyration, temporal-uncorrelated entropy, and travel time.
    On the upper right corner of each figure we show the Spearman's correlation coefficient, and the corresponding $p$-value.}
    \label{fig:scatter_mobility_rome}
\end{figure}

\begin{figure}
    \centering
    \includegraphics[width=\textwidth]{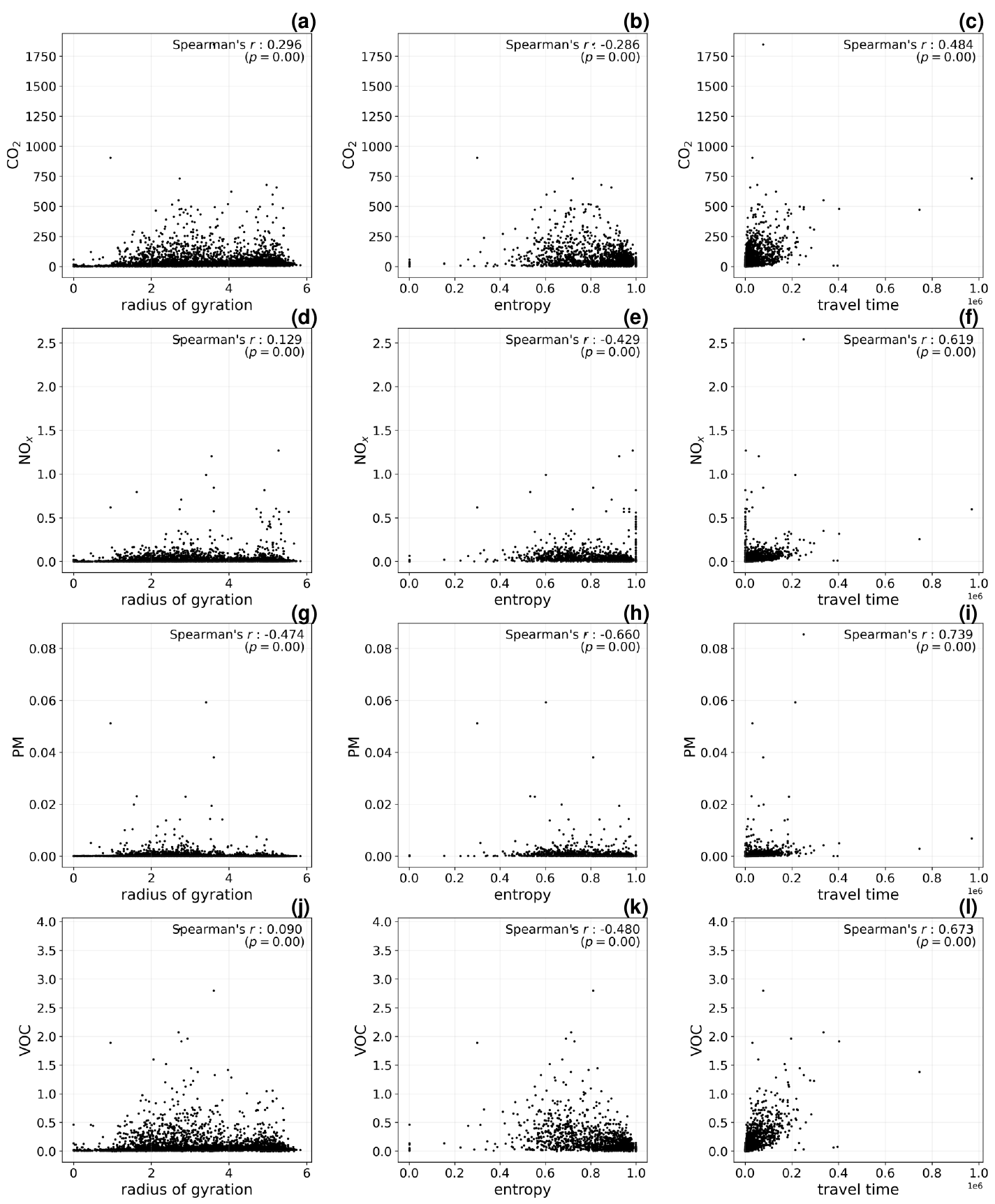}
    \caption{\textbf{Correlations between emissions and mobility metrics for Florence.} 
    The scatter plots show the relation between the emissions of the four pollutants (rows) and three mobility metrics of the vehicles (columns): the radius of gyration, temporal-uncorrelated entropy, and travel time.
    On the upper right corner of each figure we show the Spearman's correlation coefficient, and the corresponding $p$-value.}
    \label{fig:scatter_mobility_florence}
\end{figure}

\begin{figure}
    \centering
    \includegraphics[width=0.85\textwidth]{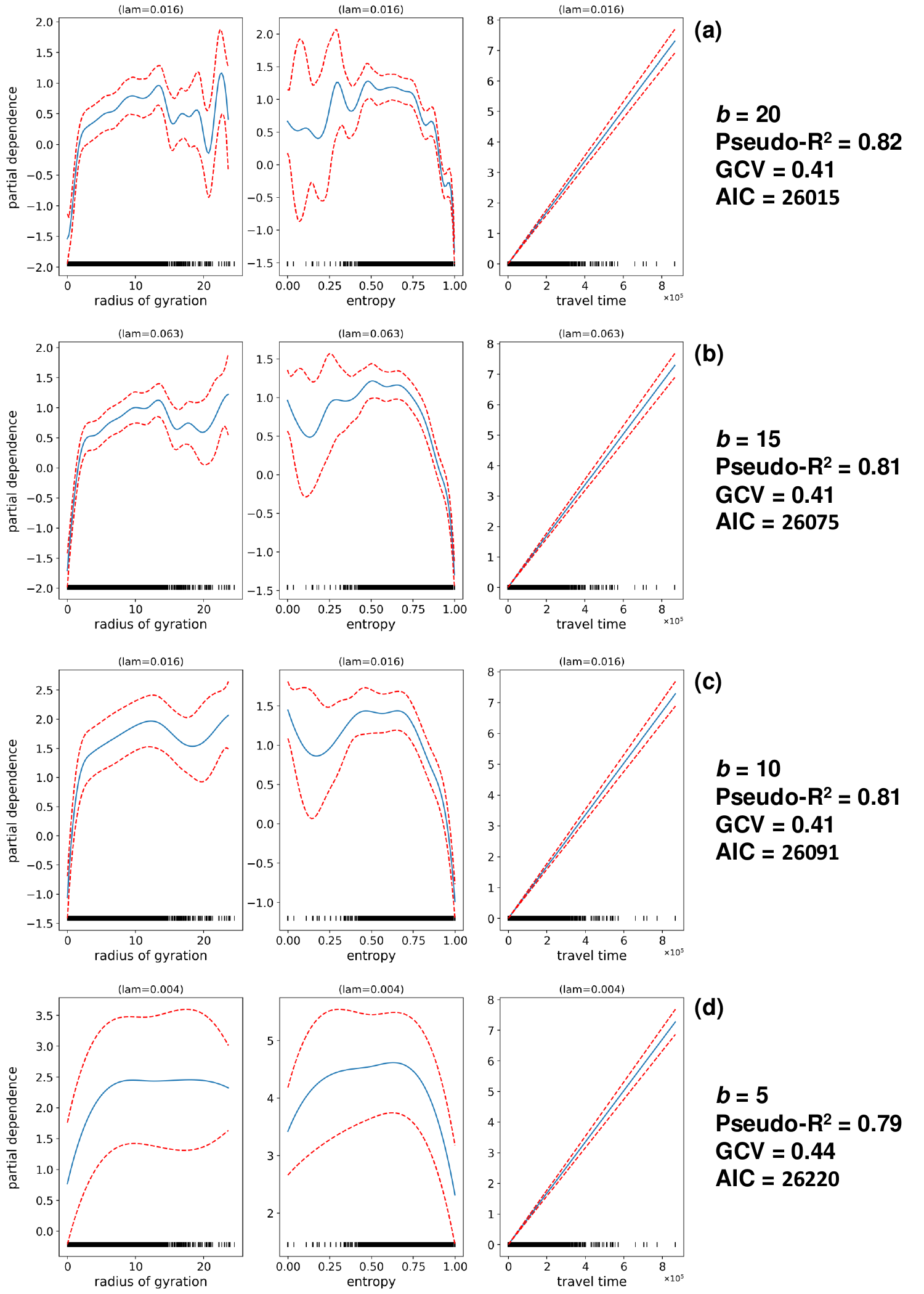}
    \caption{\textbf{Interpretation of the predictors of the GAM for Greater London}. 
    We show, from (a) to (d), the partial dependence plot showing the marginal contribution of each predictor in the model (from left to right: the radius of gyration, the entropy, and the travel time) to the CO$_2$ emissions when changing the number of basis function $b$ for the smoothers from 20, to 15, to 10, and finally to 5. 
    For each of these four choices of $b$, we show the pseudo-$R^2$, $GCV$ (Generalised Cross Validation), and $AIC$ obtained with that model and the best penalisation parameter (on the top of every single plot).
    The black lines at the bottom of each plot indicate the data distribution on the x-axis.
    Reducing the number of basis functions for the smoothers solves overfitting. Also, it leads to better interpretability of the marginal effect each predictor has on the response variable while paying in terms of model performance, as both the $GCV$ and $AIC$ increase, and the pseudo-$R^2$ decreases, even if slowly.}
    \label{fig:GAM_london}
\end{figure}

\begin{figure}
    \centering
    \includegraphics[width=0.85\textwidth]{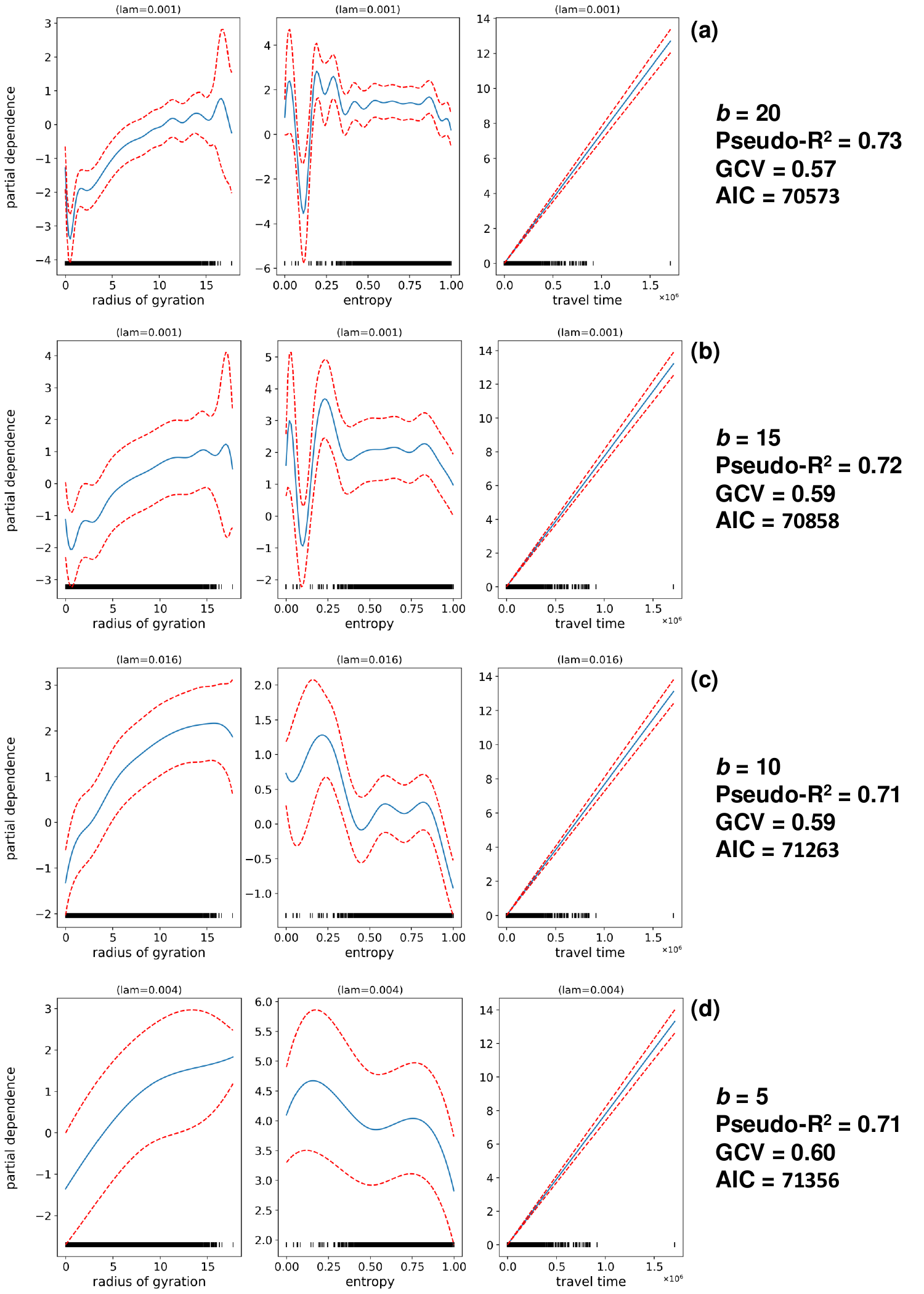}
    \caption{\textbf{Interpretation of the predictors of the GAM for Rome}. 
    We show, from (a) to (d), the partial dependence plot showing the marginal contribution of each predictor in the model (from left to right: the radius of gyration, the entropy, and the travel time) to the CO$_2$ emissions when changing the number of basis function $b$ for the smoothers from 20, to 15, to 10, and finally to 5. Also, for each of these 4 choices of $b$, we show the pseudo-$R^2$, $GCV$ (Generalised Cross Validation), and $AIC$ obtained with that model, as well as the best penalisation parameter (on the top of each single plot).
    The black lines at the bottom of each plot indicate the data distribution on the x-axis.
    Reducing the number of basis functions for the smoothers solves overfitting. 
    It leads to better interpretability of the marginal effect each predictor has on the response variable, while paying in terms of model performance, as both the $GCV$ and $AIC$ increase, and the pseudo-$R^2$ decreases, even if slowly.}
    \label{fig:GAM_rome}
\end{figure}

\begin{figure}
    \centering
    \includegraphics[width=0.85\textwidth]{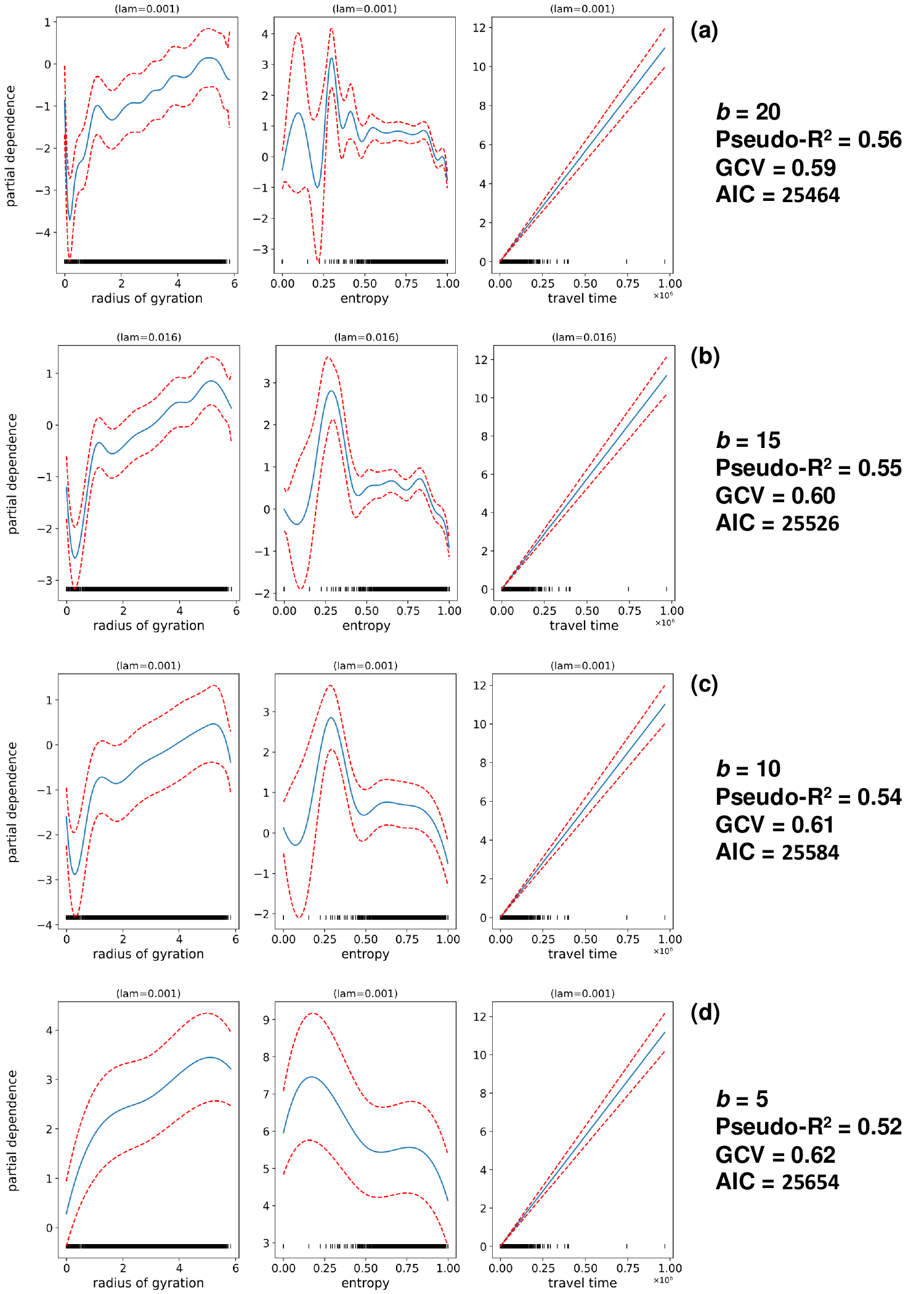}
    \caption{\textbf{Interpretation of the predictors of the GAM for Florence}. 
    We show, from (a) to (d), the partial dependence plot showing the marginal contribution of each predictor in the model (from left to right: the radius of gyration, the entropy, and the travel time) to the CO$_2$ emissions when changing the number of basis function $b$ for the smoothers from 20, to 15, to 10, and finally to 5. Also, for each of these 4 choices of $b$, we show the pseudo-$R^2$, $GCV$ (Generalised Cross Validation), and $AIC$ obtained with that model, as well as the best penalisation parameter (on the top of each single plot).
    The black lines at the bottom of each plot indicate the data distribution on the x-axis.
    Reducing the number of basis functions for the smoothers solves overfitting.
    It also results in better interpretability of the marginal effect each predictor has on the response variable, while paying in terms of model performance, as both the $GCV$ and $AIC$ increase, and the pseudo-$R^2$ decreases, even if slowly.}
    \label{fig:GAM_florence}
\end{figure}

\begin{figure}
    \centering
    \includegraphics[width=0.9\textwidth]{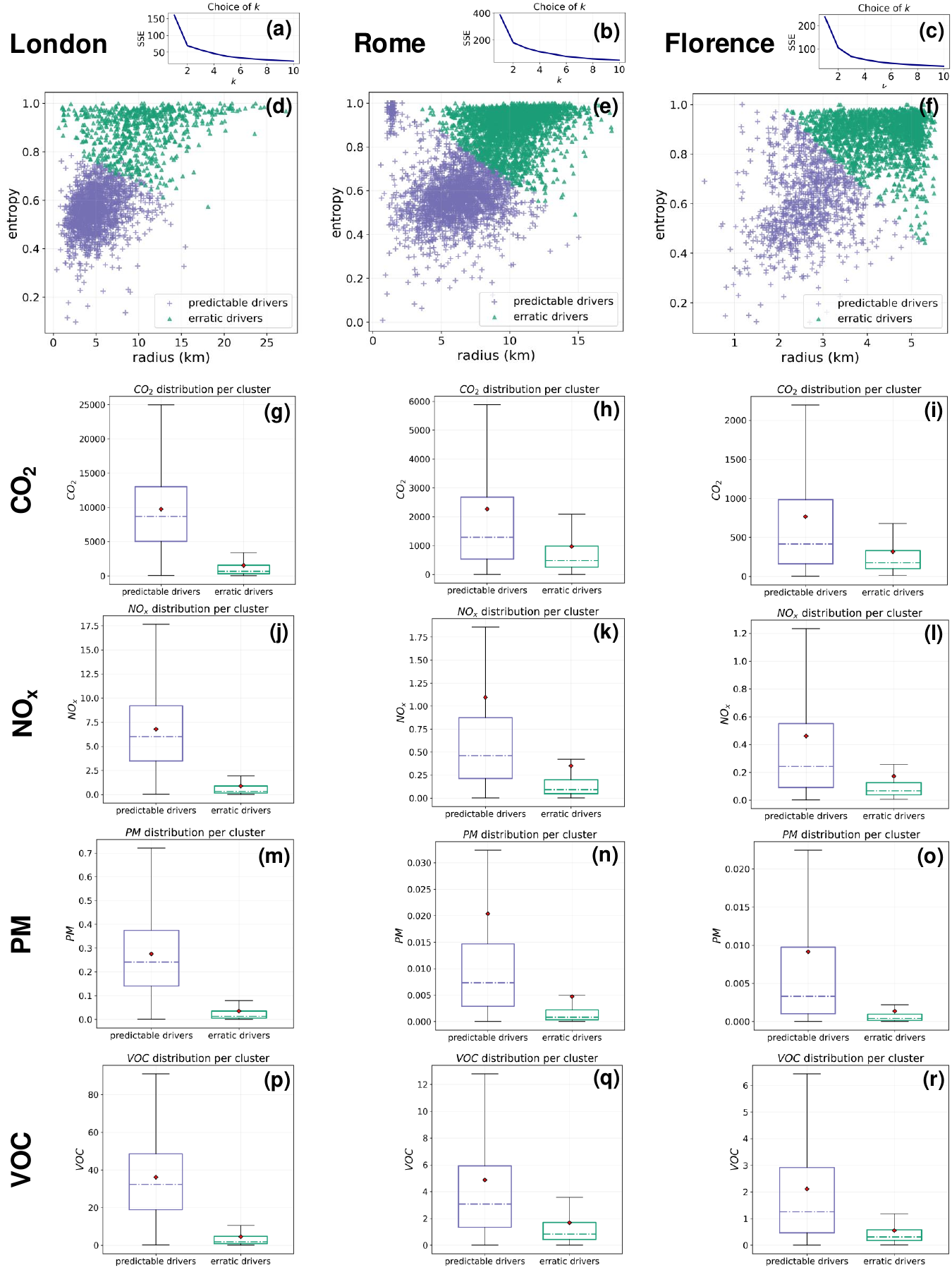}
    \caption{\textbf{$K$-means clustering of the vehicles}.
    Panels (a), (b), (c) show the trend in the sum of squared errors (SSE) when the number of clusters $k$ grows from 1 to 10, in Greater London, Rome and Florence, respectively. 
    Panels (d), (e), (f) show a scatter plot of the vehicles in the space described by their radius of gyration (x-axis) and mobility entropy (y-axis), for each city. For each vehicle, we show the cluster to which it is assigned by a $k$-means algorithm with $k=2$. We refer to these two clusters as \textit{predictable drivers} (purple crosses) and \textit{erratic drivers} (green triangles).
    Each of the panels from (g) to (r) shows the emissions' distribution of one pollutant (in grams) emitted by the vehicles in each cluster (the \textit{predictable drivers} on the left, and the \textit{erratic drivers} on the right) for a certain city. The mean of each distribution is indicated with the red diamond. The points in the tail of the distributions are not shown here as they would flatten the boxplots on the x-axis.}
    \label{fig:kmeans}
\end{figure}

\begin{figure}
    \centering
    \includegraphics[width=\textwidth]{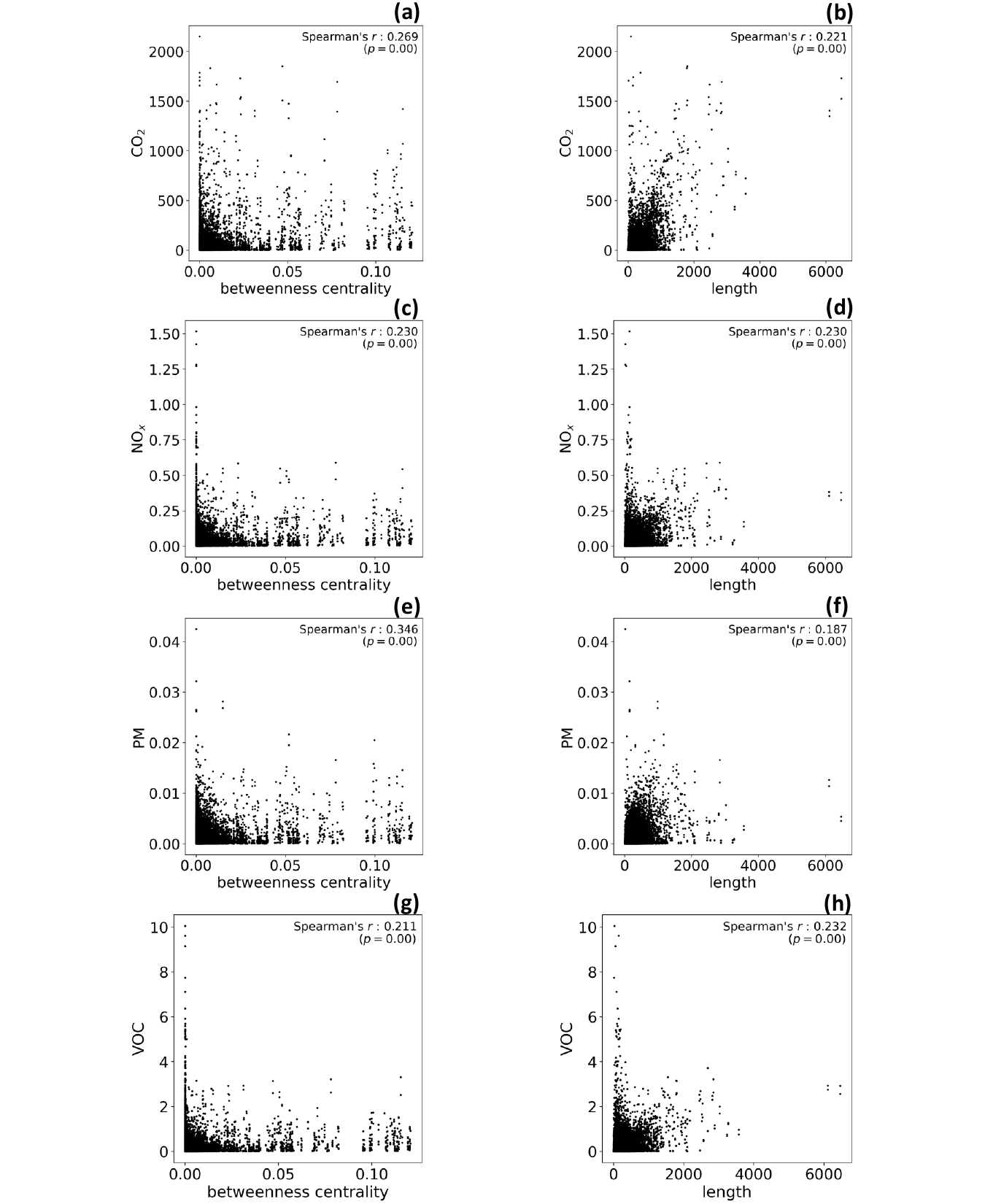}
    \caption{\textbf{Correlations between emissions and roads features for London.} 
    The scatter plots show the relation between the emissions of the four pollutants (rows) and two roads' features (columns): the betweenness centrality and length. 
    On the upper right corner of each figure we show the Spearman's correlation coefficient, and the corresponding $p$-value.}
    \label{fig:scatter_roads_london}
\end{figure}
\clearpage
\begin{figure}
    \centering
    \includegraphics[width=\textwidth]{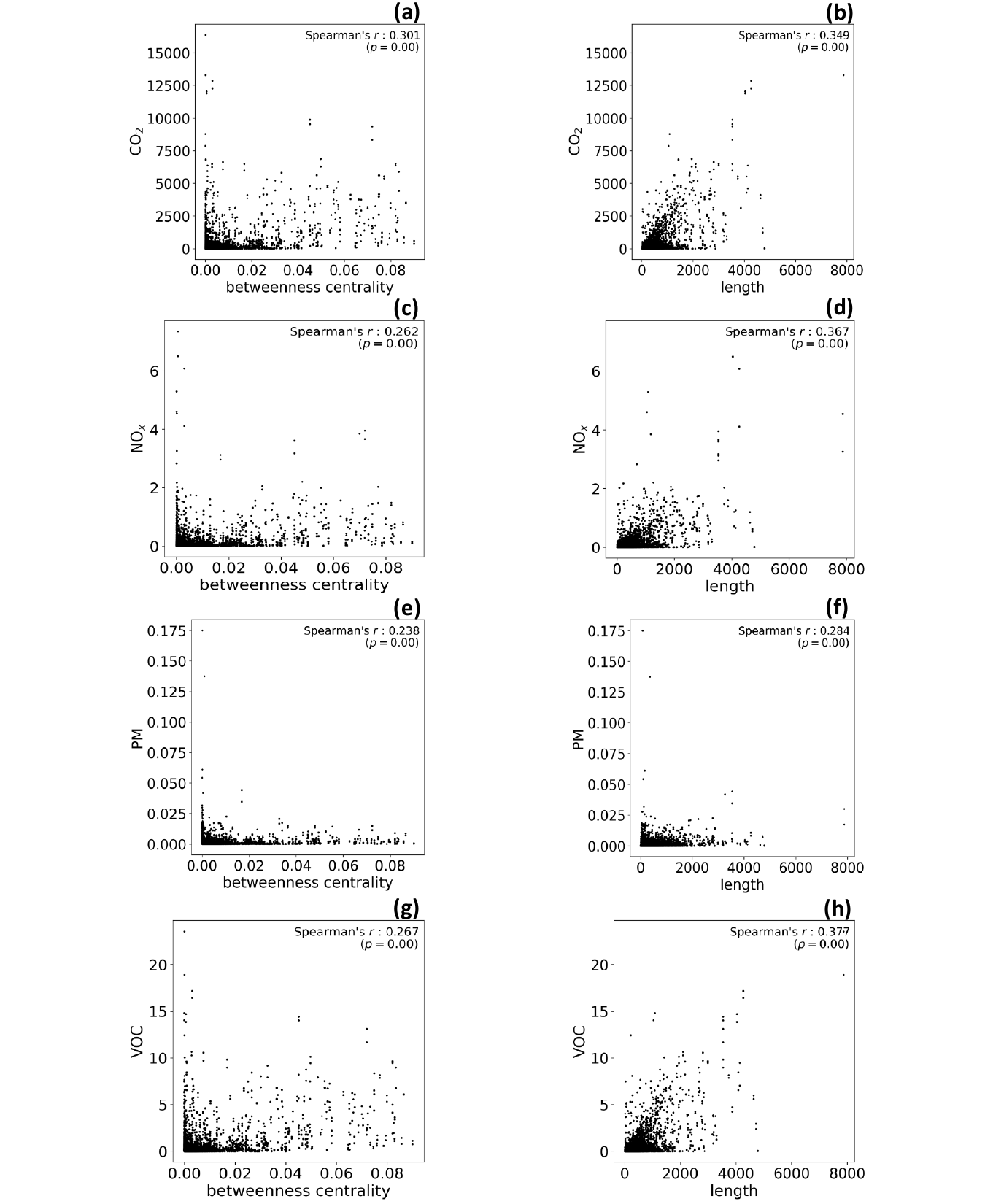}
    \caption{\textbf{Correlations between emissions and roads features for Rome.} 
    The scatter plots show the relation between the emissions of the four pollutants (rows) and two roads' features (columns): the betweenness centrality and length.
    On the upper right corner of each figure we show the Spearman's correlation coefficient, and the corresponding $p$-value.}
    \label{fig:scatter_roads_rome}
\end{figure}
\clearpage
\begin{figure}
    \centering
    \includegraphics[width=\textwidth]{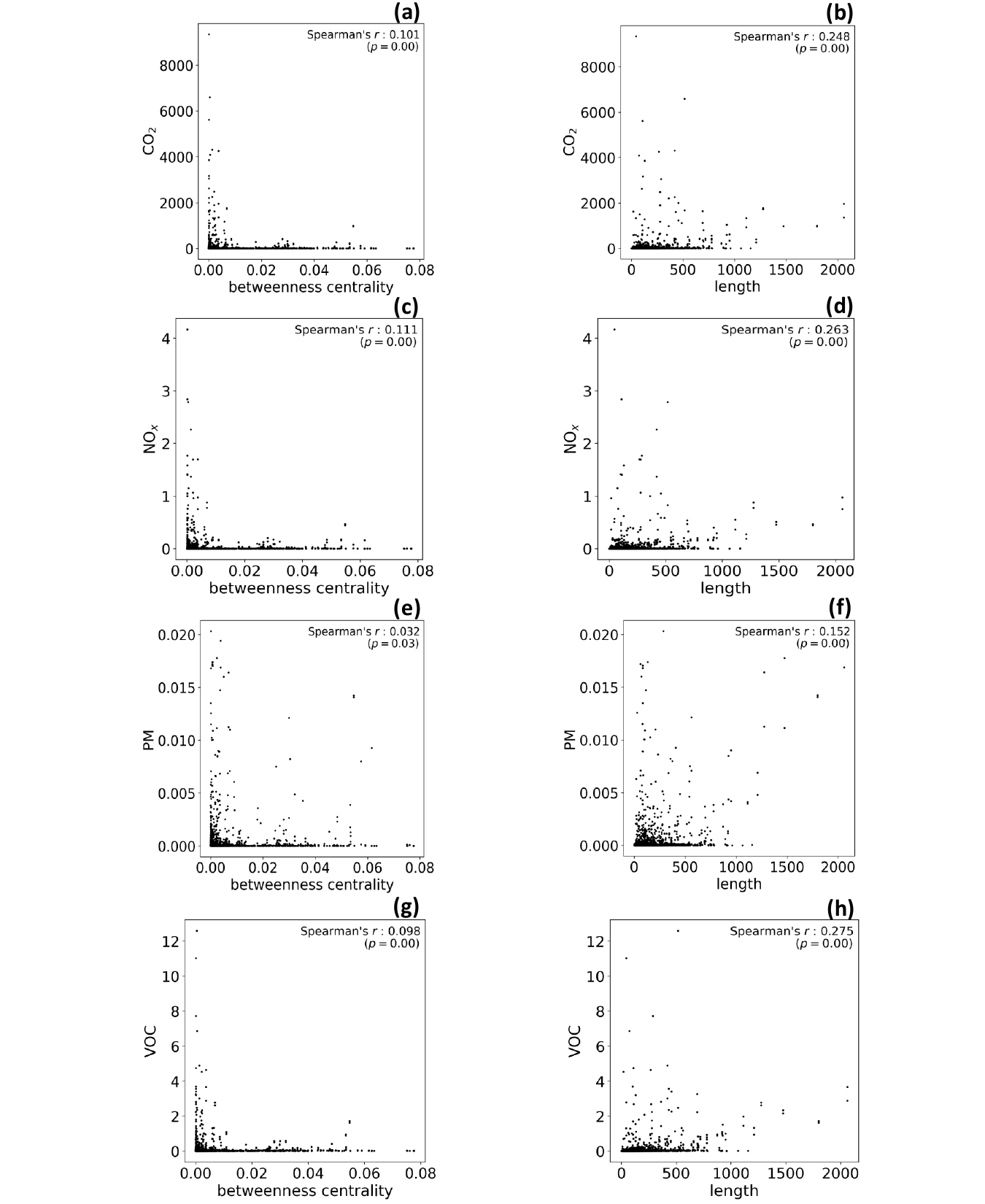}
    \caption{\textbf{Correlations between emissions and roads features for Florence.} 
    The scatter plots show the relation between the emissions of the four pollutants (rows) and two roads' features (columns): the betweenness centrality and length.
    On the upper right corner of each figure we show the Spearman's correlation coefficient, and the corresponding $p$-value.}
    \label{fig:scatter_roads_florence}
\end{figure}

\clearpage

\begin{figure}
    \centering
    \includegraphics[width=\textwidth]{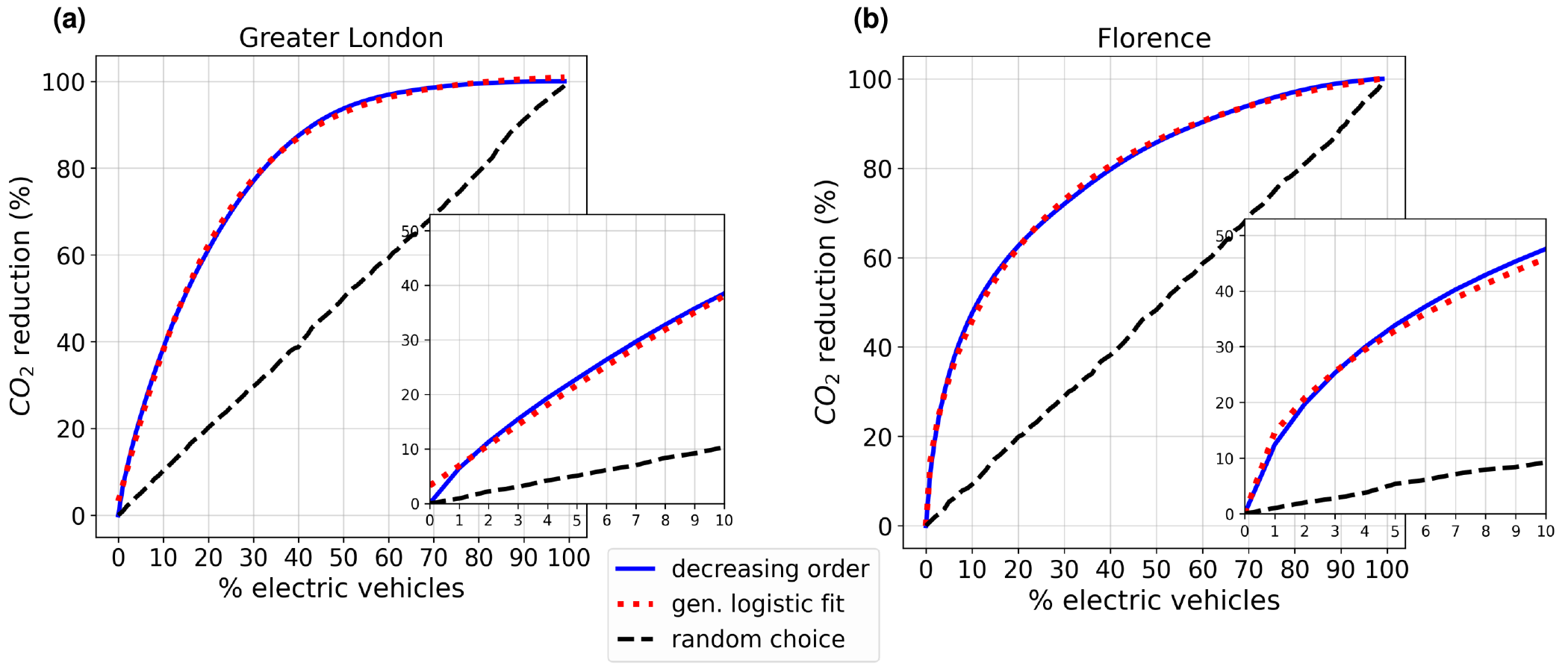}
    \caption{\textbf{CO$_2$ emissions reduction from vehicles electrification in Greater London and Florence}.
    Percentage reduction of CO$_2$ emissions corresponding to a certain share (0-100\%) of electric vehicles in (a) Greater London and (b) Florence.
    The inset plots zoom on the first tenth share of electric vehicles.
    The black dashed line is when the vehicles to be electrified are chosen at random.
    The blue solid line is when the vehicles to be electrified are chosen from the most polluting to the least polluting, and the dotted red line is its \textit{Generalised Logistic Function} (\textit{Richard's curve}) fit:
    $f(x) = \frac{\alpha}{(1 + \beta e^{-rx})^{1/\nu}}$, where $\alpha$ represents the upper asymptote, $\beta$ the growth range, $r$ the growth rate, and $\nu$ the slope of the curve.}
    \label{fig:electrification}
\end{figure}

\begin{figure}
    \centering
    \includegraphics[width=\textwidth]{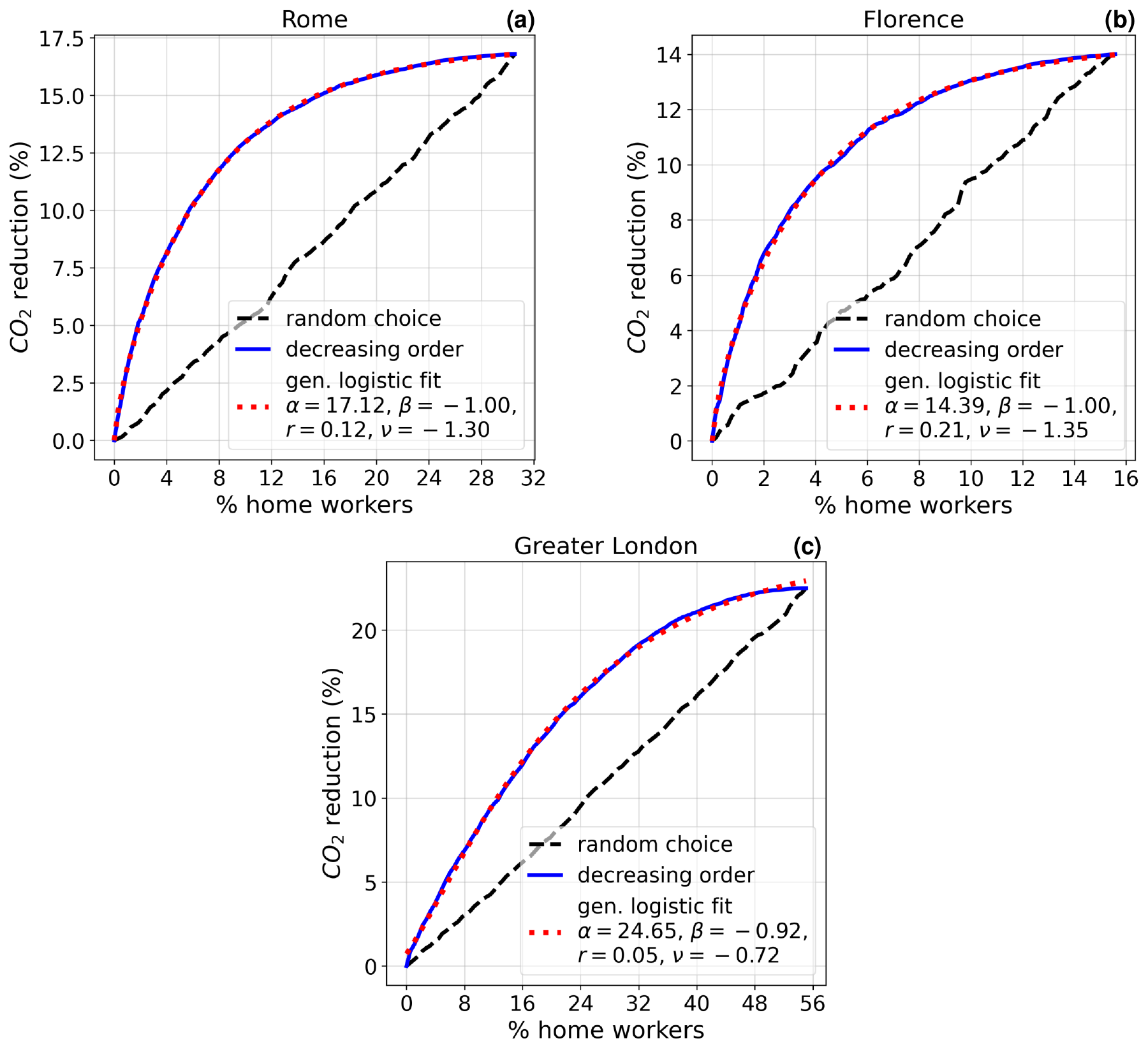}
    \caption{\textbf{CO$_2$ emissions reduction from remote working in Rome, Florence and Greater London}.
    Percentage reduction of CO$_2$ emissions corresponding to a certain share of vehicles that stop commuting in (a) Rome, (b) Florence, and (c) Greater London.
    The maximum share of vehicles that stop commuting in our simulation changes with the city: it is $31\%$ for Rome, $16\%$ for Florence, and $55\%$ for London, as these are the percentages of vehicles for which we can identify both the home and work locations and, thus, the commuting trajectories.
    The black dashed curve represents the case in which the vehicles that stop commuting are chosen at random.
    The blue solid curve represents the case in which the vehicles that stop commuting are chosen from the most polluting to the least polluting, and the dotted red curve is its \textit{Generalised Logistic Function} (\textit{Richard's curve}) fit: 
    $f(x) = \frac{\alpha}{(1 + \beta e^{-rx})^{1/\nu}}$, where $\alpha$ represents the upper asymptote, $\beta$ the growth range, $r$ the growth rate, and $\nu$ the slope of the curve.
    We show the estimated parameters in the legend.}
    \label{fig:homeworking}
\end{figure}


\clearpage
\section{Supplementary Tables}

\begin{table}[h!]
    \centering
    \footnotesize
    \begin{tabular}{| c | c | c | c |}
    \cline{2-4}
        
         \multicolumn{1}{c|}{} & \thead{Rome} & \thead{Greater London} & \thead{Florence}
         \\
         \cline{1-4}
        
        \thead{mean} &  207.7 & 72.9 & 261.3 \\
        \thead{std} & 245.6 & 144.0 & 400.4 \\
        \thead{min} &  0 & 0 & 0 \\
        \thead{25\%} &  86 & 60 & 130 \\
        \thead{50\%} &  146 & 60 & 211 \\
        \thead{75\%} &  271 & 60 & 324 \\
        \thead{90\%} &  425 & 90 & 469 \\
        \thead{max} &  69,712 & 37,005 & 56,490 \\
        \thead{mode} &  70 & 60 & 65 \\
        
        \cline{1-4}
    \end{tabular}
    \caption{\textbf{Descriptive statistics of the distributions of the sampling rate (in seconds) in the three cities before the filtering process.} 
    }
    \label{tab:stats_sampling_rate}
\end{table}

\begin{table}[h!]
    \centering
    \begin{tabular}{c | c | c | c | c | c |}
    \cline{2-6}
         & \multicolumn{5}{c|}{\bf road networks} \\
        \cline{2-6}
        
         & roads & crossroads & roads density & crossroads/$km^{2}$ & avg road length (std) \\
         \cline{2-6}
         
        London &  274,371 &  154,871 &  12,367 &  98.5 &  70.9 m (82.9) \\
        
        Rome &  87,029 &  50,705 &  6,759 &  39.5 &  99.8 m (153.0) \\
        
        Florence &  12,537 &  7,469 &  10,483 &  72.9 &  85.6 m (99.8) \\
        \cline{2-6}
    \end{tabular}
    \caption{\textbf{Summary statistics of the road networks.} 
    The number of roads and crossroads, their densities, and the mean road length (and standard deviation) of the road networks of Greater London, Rome, and Florence.
    The roads density is the total length of all the roads divided by the city's land area.
    The crossroads density (crossroads/$km^{2}$) is the number of crossroads divided by the land area of the city.}
    \label{tab:cities}
\end{table}

\begin{table}[h!]
    \centering
    \footnotesize
    \begin{tabular}{| c | c | c | c | c | c |}
    \cline{2-6}
        
         \multicolumn{1}{c|}{} & \thead{power law} & \thead{log-normal} & \thead{truncated \\[-0.2\normalbaselineskip] power law} & \thead{stretched \\[-0.2\normalbaselineskip] exponential} & \thead{exponential}
         \\
         \cline{1-6}
        
        \thead{power law} &  - & None & None & \makecell{power law \\[-0.2\normalbaselineskip] (R=2.59, p=0.01)} & \makecell{power law \\[-0.2\normalbaselineskip] (R=5.28, p=0.00)} \\
        
        \thead{log-normal} &  - & - & None & \makecell{log-normal \\[-0.2\normalbaselineskip] (R=2.56, p=0.01)} & \makecell{log-normal \\[-0.2\normalbaselineskip] (R=5.27, p=0.00)} \\
        
        \thead{truncated \\[-0.2\normalbaselineskip] power law} &  - & - & - & \makecell{tr. power law \\[-0.2\normalbaselineskip] (R=2.59, p=0.01)} & \makecell{tr. power law \\[-0.2\normalbaselineskip] (R=5.28, p=0.00)} \\
        
        \thead{stretched \\[-0.2\normalbaselineskip] exponential} &  - & - & - & - & \makecell{str. exponential \\[-0.2\normalbaselineskip] (R=5.34, p=0.00)} \\
        
        \cline{1-6}
    \end{tabular}
    \caption{\textbf{Results of log-likelihood ratio tests for comparing different models fitting the distribution of NO$_x$ emissions per vehicle in Rome.} 
    Each $(i,j)$ cell of the table shows the result of a log-likelihood ratio test used for comparing the goodness-of-fit of the model in row $i$ and the one in column $j$. If the test gives result in favour of a model, its name is shown together with the statistic $R$, that is negative if the result is in favour of the first model (the one on the row) or positive if in favour of the second (the one on the column), and the corresponding $p$-value.
    If the test gives no evidence in favour of one of the two models, the result is signed as None.
    In this case, there is no model that wins more comparisons than the others.}
    \label{tab:test_NOx_Rome}
\end{table}

\begin{table}[h!]
    \centering
    \footnotesize
    \begin{tabular}{| c | c | c | c | c | c |}
    \cline{2-6}
        
         \multicolumn{1}{c|}{} & \thead{power law} & \thead{log-normal} & \thead{truncated \\[-0.2\normalbaselineskip] power law} & \thead{stretched \\[-0.2\normalbaselineskip] exponential} & \thead{exponential}
         \\
         \cline{1-6}
        
        \thead{power law} &  - & None & None & None & \makecell{power law \\[-0.2\normalbaselineskip] (R=3.21, p=0.00)} \\
        
        \thead{log-normal} &  - & - & None & None & \makecell{log-normal \\[-0.2\normalbaselineskip] (R=3.32, p=0.00)} \\
        
        \thead{truncated \\[-0.2\normalbaselineskip] power law} &  - & - & - & None & \makecell{tr. power law \\[-0.2\normalbaselineskip] (R=3.27, p=0.00)} \\
        
        \thead{stretched \\[-0.2\normalbaselineskip] exponential} &  - & - & - & - & \makecell{str. exponential \\[-0.2\normalbaselineskip] (R=3.34, p=0.00)} \\
        
        \cline{1-6}
    \end{tabular}
    \caption{\textbf{Results of log-likelihood ratio tests for comparing different models fitting the distribution of PM emissions per vehicle in Rome.} 
    Each $(i,j)$ cell of the table shows the result of a log-likelihood ratio test used for comparing the goodness-of-fit of the model in row $i$ and the one in column $j$. If the test gives result in favour of a model, its name is shown together with the statistic $R$, that is negative if the result is in favour of the first model (the one on the row) or positive if in favour of the second (the one on the column), and the corresponding $p$-value.
    If the test gives no evidence in favour of one of the two models, the result is signed as None.
    In this case, there is no model that wins more comparisons than the others.}
    \label{tab:test_PM_Rome}
\end{table}

\begin{table}[h!]
    \centering
    \footnotesize
    \begin{tabular}{| c | c | c | c | c | c |}
    \cline{2-6}
        
         \multicolumn{1}{c|}{} & \thead{power law} & \thead{log-normal} & \thead{truncated \\[-0.2\normalbaselineskip] power law} & \thead{stretched \\[-0.2\normalbaselineskip] exponential} & \thead{exponential}
         \\
         \cline{1-6}
        
        \thead{power law} &  - & \makecell{log-normal \\[-0.2\normalbaselineskip] (R=-4.32, p=0.00)} & \makecell{tr. power law \\[-0.2\normalbaselineskip] (R=-4.40, p=0.00)} & \makecell{str. exponential \\[-0.2\normalbaselineskip] (R=-4.27, p=0.00)} & \makecell{power law \\[-0.2\normalbaselineskip] (R=7.99, p=0.00)} \\
        
        \thead{log-normal} &  - & - & None & None & \makecell{log-normal \\[-0.2\normalbaselineskip] (R=9.48, p=0.00)} \\
        
        \thead{truncated \\[-0.2\normalbaselineskip] power law} &  - & - & - & None & \makecell{tr. power law \\[-0.2\normalbaselineskip] (R=9.54, p=0.00)} \\
        
        \thead{stretched \\[-0.2\normalbaselineskip] exponential} &  - & - & - & - & \makecell{str. exponential \\[-0.2\normalbaselineskip] (R=9.56, p=0.00)} \\
        
        \cline{1-6}
    \end{tabular}
    \caption{\textbf{Results of log-likelihood ratio tests for comparing different models fitting the distribution of NO$_x$ emissions per road in London.} 
    Each $(i,j)$ cell of the table shows the result of a log-likelihood ratio test used for comparing the goodness-of-fit of the model in row $i$ and the one in column $j$. If the test gives result in favour of a model, its name is shown together with the statistic $R$, that is negative if the result is in favour of the first model (the one on the row) or positive if in favour of the second (the one on the column), and the corresponding $p$-value.
    If the test gives no evidence in favour of one of the two models, the result is signed as None.
    In this case, none of the log-normal, truncated power law and stretched exponential models wins more comparisons than the other two.}
    \label{tab:test_NOx_London}
\end{table}

\begin{table}[h!]
    \centering
    \footnotesize
    \begin{tabular}{| c | c | c | c | c | c |}
    \cline{2-6}
        
         \multicolumn{1}{c|}{} & \thead{power law} & \thead{log-normal} & \thead{truncated \\[-0.2\normalbaselineskip] power law} & \thead{stretched \\[-0.2\normalbaselineskip] exponential} & \thead{exponential}
         \\
         \cline{1-6}
        
        \thead{power law} &  - & \makecell{log-normal \\[-0.2\normalbaselineskip] (R=-2.87, p=0.00)} & \makecell{tr. power law \\[-0.2\normalbaselineskip] (R=-2.87, p=0.00)} & \makecell{str. exponential \\[-0.2\normalbaselineskip] (R=-2.89, p=0.00)} & \makecell{power law \\[-0.2\normalbaselineskip] (R=3.70, p=0.00)} \\
        
        \thead{log-normal} &  - & - & None & None & \makecell{log-normal \\[-0.2\normalbaselineskip] (R=5.06, p=0.00)} \\
        
        \thead{truncated \\[-0.2\normalbaselineskip] power law} &  - & - & - & None & \makecell{tr. power law \\[-0.2\normalbaselineskip] (R=5.10, p=0.00)} \\
        
        \thead{stretched \\[-0.2\normalbaselineskip] exponential} &  - & - & - & - & \makecell{str. exponential \\[-0.2\normalbaselineskip] (R=5.09, p=0.00)} \\
        
        \cline{1-6}
    \end{tabular}
    \caption{\textbf{Results of log-likelihood ratio tests for comparing different models fitting the distribution of PM emissions per road in London.} 
    Each $(i,j)$ cell of the table shows the result of a log-likelihood ratio test used for comparing the goodness-of-fit of the model in row $i$ and the one in column $j$. If the test gives result in favour of a model, its name is shown together with the statistic $R$, that is negative if the result is in favour of the first model (the one on the row) or positive if in favour of the second (the one on the column), and the corresponding $p$-value.
    If the test gives no evidence in favour of one of the two models, the result is signed as None.
    In this case, none of the log-normal, truncated power law and stretched exponential models wins more comparisons than the other two.}
    \label{tab:test_PM_London}
\end{table}

\begin{table}[h!]
    \centering
    \footnotesize
    \begin{tabular}{| c | c | c | c | c | c |}
    \cline{2-6}
        
         \multicolumn{1}{c|}{} & \thead{power law} & \thead{log-normal} & \thead{truncated \\[-0.2\normalbaselineskip] power law} & \thead{stretched \\[-0.2\normalbaselineskip] exponential} & \thead{exponential}
         \\
         \cline{1-6}
        
        \thead{power law} &  - & \makecell{log-normal \\[-0.2\normalbaselineskip] (R=-4.10, p=0.00)} & \makecell{tr. power law \\[-0.2\normalbaselineskip] (R=-4.56, p=0.00)} & \makecell{str. exponential \\[-0.2\normalbaselineskip] (R=-3.69, p=0.00)} & \makecell{power law \\[-0.2\normalbaselineskip] (R=10.11, p=0.00)} \\
        
        \thead{log-normal} &  - & - & None & None & \makecell{log-normal \\[-0.2\normalbaselineskip] (R=11.29, p=0.00)} \\
        
        \thead{truncated \\[-0.2\normalbaselineskip] power law} &  - & - & - & None & \makecell{tr. power law \\[-0.2\normalbaselineskip] (R=11.47, p=0.00)} \\
        
        \thead{stretched \\[-0.2\normalbaselineskip] exponential} &  - & - & - & - & \makecell{str. exponential \\[-0.2\normalbaselineskip] (R=11.46, p=0.00)} \\
        
        \cline{1-6}
    \end{tabular}
    \caption{\textbf{Results of log-likelihood ratio tests for comparing different models fitting the distribution of VOC emissions per road in London.} 
    Each $(i,j)$ cell of the table shows the result of a log-likelihood ratio test used for comparing the goodness-of-fit of the model in row $i$ and the one in column $j$. If the test gives result in favour of a model, its name is shown together with the statistic $R$, that is negative if the result is in favour of the first model (the one on the row) or positive if in favour of the second (the one on the column), and the corresponding $p$-value.
    If the test gives no evidence in favour of one of the two models, the result is signed as None.
    In this case, none of the log-normal, truncated power law and stretched exponential models wins more comparisons than the other two.}
    \label{tab:test_VOC_London}
\end{table}

\begin{table}[h!]
    \centering
    \begin{tabular}{c | c | c | c || c | c |}
    \cline{2-6}
         & \multicolumn{3}{ c||}{\bf mobility metrics} & \multicolumn{2}{c|}{\bf roads' features} \\
        \cline{2-6}
        
         & radius & entropy & travel time & betweenness centrality & length \\
         \cline{2-6}
         
        London &  0.02* &  -0.76 &  0.98 &   0.23 &  0.23 \\
        
        Rome &  0.36 &  -0.68 &  0.88 &  0.26 &  0.37 \\
        
        Florence &  0.13 &  -0.43 &  0.62 &  0.11 &  0.26 \\
        \cline{2-6}
    \end{tabular}
    \caption{\textbf{Correlations between vehicles' emissions of NO$_x$, mobility metrics, and road features.} Spearman's correlation coefficients between (left) NO$_x$ emissions per vehicle and vehicles' mobility metrics (their radius of gyration, uncorrelated entropy, maximum distance and distance travelled straight line), and (right) NO$_x$ emissions per road and roads' features (betweenness centrality and length).
    *p-value $>$ 0.05.}
    \label{tab:corrs_NOx}
\end{table}

\begin{table}[h!]
    \centering
    \begin{tabular}{c | c | c | c || c | c |}
    \cline{2-6}
         & \multicolumn{3}{ c||}{\bf mobility metrics} & \multicolumn{2}{c|}{\bf roads' features} \\
        \cline{2-6}
        
         & radius & entropy & travel time & betweenness centrality & length \\
         \cline{2-6}
         
        London &  0.02* &  -0.74 &  0.97 &   0.35 &  0.19 \\
        
        Rome &  0.28 &  -0.70 &  0.87 &  0.24 &  0.28 \\
        
        Florence &  -0.47 &  -0.66 &  0.74 &  0.03 &  0.15 \\
        \cline{2-6}
    \end{tabular}
    \caption{\textbf{Correlations between vehicles' emissions of PM, mobility metrics, and road features.} Spearman's correlation coefficients between (left) PM emissions per vehicle and vehicles' mobility metrics (their radius of gyration, uncorrelated entropy, maximum distance and distance travelled straight line), and (right) PM emissions per road and roads' features (betweenness centrality and length).
    *p-value $>$ 0.05.}
    \label{tab:corrs_PM}
\end{table}

\begin{table}[h!]
    \centering
    \begin{tabular}{c | c | c | c || c | c |}
    \cline{2-6}
         & \multicolumn{3}{ c||}{\bf mobility metrics} & \multicolumn{2}{c|}{\bf roads' features} \\
        \cline{2-6}
        
         & radius & entropy & travel time & betweenness centrality & length \\
         \cline{2-6}
         
        London &  0.03* &  -0.76 &  0.98 &   0.21 &  0.23 \\
        
        Rome &  0.48 &  -0.64 &  0.87 &  0.27 &  0.38 \\
        
        Florence &  0.09 &  -0.48 &  0.67 &  0.10 &  0.27 \\
        \cline{2-6}
    \end{tabular}
    \caption{\textbf{Correlations between vehicles' emissions of VOC, mobility metrics, and road features.} Spearman's correlation coefficients between (left) VOC emissions per vehicle and vehicles' mobility metrics (their radius of gyration, uncorrelated entropy, maximum distance and distance travelled straight line), and (right) VOC emissions per road and roads' features (betweenness centrality and length).
    *p-value $>$ 0.05.}
    \label{tab:corrs_VOC}
\end{table}

\begin{table}[h!]
    \centering
    \begin{tabular}{c | c | c | c | c || c |}
    \cline{2-6}
        
         & {\boldmath$\alpha$} & {\boldmath$\beta$} & {\boldmath$r$} & {\boldmath$\nu$} & {\boldmath$R^2$} \\
         \cline{2-6}
        
        London &  $101.66$ &  $-0.95$ &  $5.05 \times 10^{-2}$ &  $-0.86$ &  $0.99$ \\
        
        Rome &  $101.24$ &  $-0.99$ &  $3.96 \times 10^{-2}$ & $-1.56$ &  $0.99$ \\
        
        Florence &  $106.57$ &  $-0.99$ &  $2.20 \times 10^{-2}$ & $-1.91$ &  $0.99$ \\
        
        2$^{nd}$ Municipality (Rome) &  $99.8$ &  $-0.99$ &  $4.84 \times 10^{-2}$ &  $-1.55$ &  $0.99$ \\
        \cline{2-6}
    \end{tabular}
    \caption{\textbf{Estimated parameters of a \textit{Generalised Logistic Function} fitted to the reduction of the overall CO$_2$ emissions resulting from the electrification of the vehicles starting from the most polluting ones.} 
    The estimated parameters $\alpha$ (the upper asymptote), $\beta$ (the growth range), $r$ (the growth rate), and $\nu$ (the slope) obtained with non-linear least squares fitting of a \textit{Generalised Logistic Function} to the data are shown for London, Rome, Florence and a neighbourhood of Rome (the Second Municipality).
    We also show the $R^2$ for each model as a measure of its goodness of fit to the data.}
    \label{tab:logistic_fit_electrification}
\end{table}

\begin{table}[h!]
    \centering
    \begin{tabular}{c | c | c | c | c || c |}
    \cline{2-6}
        
         & {\boldmath$\alpha$} & {\boldmath$\beta$} & {\boldmath$r$} & {\boldmath$\nu$} & {\boldmath$R^2$} \\
         \cline{2-6}
        
        London &  $22.49$ &  $-0.92$ &  $0.05$ &  $-0.72$ &  $0.99$ \\
        
        Rome &  $17.12$ &  $-1.00$ &  $0.12$ & $-1.30$ &  $0.99$ \\
        
        Florence &  $14.39$ &  $-1.00$ &  $0.21$ & $-1.35$ &  $0.99$ \\
        \cline{2-6}
    \end{tabular}
    \caption{\textbf{Estimated parameters of a \textit{Generalised Logistic Function} fitted to the reduction of the overall CO$_2$ emissions resulting from the shift to remote working of the drivers of the vehicles starting from the most polluting ones.} 
    The estimated parameters $\alpha$ (the upper asymptote), $\beta$ (the growth range), $r$ (the growth rate), and $\nu$ (the slope) obtained with non-linear least squares fitting of a \textit{Generalised Logistic Function} to the data are shown for London, Rome, and Florence.
    We also show the $R^2$ for each model as a measure of its goodness of fit to the data.}
    \label{tab:logistic_fit_homeworking}
\end{table}

\begin{table}[h!]
    \centering
    \scriptsize
    \begin{tabular}{c | c | c | c | c | c | c | c | c |}
    \cline{2-9}
        
         & fuel type & acceleration & $f_1$ & $f_2$ & $f_3$ & $f_4$ & $f_5$ & $f_6$ \\
         \cline{2-9}
        
        CO$_2$ & petrol & all & $5.53\times 10^{-1}$ & $1.61\times 10^{-1}$ & $-2.89\times 10^{-3}$ & $2.66\times 10^{-1}$ & $5.11\times 10^{-1}$ & $1.83\times 10^{-1}$ \\
         & diesel & all & $3.24\times 10^{-1}$ & $8.59\times 10^{-2}$ & $4.96\times 10^{-3}$ & $-5.86\times 10^{-2}$ & $4.48\times 10^{-1}$ & $2.30\times 10^{-1}$ \\
         & LPG & all & $6.00\times 10^{-1}$ & $2.19\times 10^{-1}$ & $-7.74\times 10^{-3}$ & $3.57\times 10^{-1}$ & $5.14\times 10^{-1}$ & $1.70\times 10^{-1}$ \\
        \cline{2-9}
        NO$_x$ &  petrol & $\geq-0.5$ $m/s^2$ & $6.19\times 10^{-4}$ & $8.00 \times 10^{-5}$ & $-4.03 \times 10^{-6}$ & $-4.13\times 10^{-4}$ & $3.80\times 10^{-4}$ & $1.77\times 10^{-4}$ \\
         & petrol & $<-0.05$ $m/s^2$ & $2.17\times 10^{-4}$ & $0$ & $0$ & $0$ & $0$ & $0$ \\
         & diesel & $\geq-0.5$ $m/s^2$ & $2.41\times 10^{-3}$ & $-4.11\times 10^{-4}$ & $6.73 \times 10^{-5}$ & $-3.07\times 10^{-3}$ & $2.14\times 10^{-3}$ & $1.50\times 10^{-3}$ \\
         & diesel & $<-0.05$ $m/s^2$ & $1.68\times 10^{-3}$ & $-6.62 \times 10^{-5}$ & $9.00 \times 10^{-6}$ & $2.50\times 10^{-4}$ & $2.91\times 10^{-4}$ & $1.20\times 10^{-4}$ \\
         & LPG & $\geq-0.5$ $m/s^2$ & $8.92\times 10^{-4}$ & $1.61 \times 10^{-5}$ & $-8.06\times 10^{-7}$ & $-8.23 \times 10^{-5}$ & $7.60 \times 10^{-5}$ & $3.54 \times 10^{-5}$ \\
         & LPG & $<-0.05$ $m/s^2$ & $3.43\times 10^{-4}$ & $0$ & $0$ & $0$ & $0$ & $0$ \\
        \cline{2-9}
        PM & petrol & all & $0$ & $1.57\times 10^{-5}$ & $-9.21 \times 10^{-7}$ & $0$ & $3.75 \times 10^{-5}$ & $1.89 \times 10^{-5}$ \\
         & diesel & all & $0$ & $3.13\times 10^{-4}$ & $-1.84 \times 10^{-5}$ & $0$ & $7.50\times 10^{-4}$ & $3.78\times 10^{-4}$ \\
         & LPG & all & $0$ & $1.57 \times 10^{-5}$ & $-9.21 \times 10^{-7}$ & $0$ & $3.75 \times 10^{-5}$ & $1.89 \times 10^{-5}$ \\
        \cline{2-9}
        VOC & petrol & $\geq-0.5$ $m/s^2$ & $4.47\times 10^{-3}$ & $7.32 \times 10^{-7}$ & $-2.87 \times 10^{-8}$ & $-3.41 \times 10^{-5}$ & $4.94 \times 10^{-6}$ & $1.66 \times 10^{-6}$ \\
         & petrol & $<-0.05$ $m/s^2$ & $2.63\times 10^{-3}$ & $0$ & $0$ & $0$ & $0$ & $0$ \\
         & diesel & $\geq-0.5$ $m/s^2$ & $9.22 \times 10^{-5}$ & $9.09 \times 10^{-6}$ & $-2.29 \times 10^{-7}$ & $-2.20 \times 10^{-5}$ & $1.69 \times 10^{-5}$ & $3.75 \times 10^{-6}$ \\
         & diesel & $<-0.05$ $m/s^2$ & $5.25 \times 10^{-5}$ & $7.22 \times 10^{-6}$ & $-1.87 \times 10^{-7}$ & $0$ & $-1.02 \times 10^{-5}$ & $-4.22 \times 10^{-6}$ \\
         & LPG & $\geq-0.5$ $m/s^2$ & $1.44\times 10^{-2}$ & $1.74 \times 10^{-7}$ & $-6.82 \times 10^{-9}$ & $-8.11 \times 10^{-7}$ & $1.18 \times 10^{-6}$ & $3.96\times 10^{-7}$ \\
         & LPG & $<-0.05$ $m/s^2$ & $8.42\times 10^{-3}$ & $0$ & $0$ & $0$ & $0$ & $0$ \\
        \cline{2-9}
    \end{tabular}
    \caption{\textbf{Emission functions}.
    The emission functions for each pollutant, fuel type, and acceleration profile.}
    \label{tab:emission_factors}
\end{table}

\clearpage

\end{document}